\documentclass[11pt]{article}
\usepackage[dvips]{graphicx}
\usepackage{graphics}
\usepackage{graphicx}
\usepackage{amssymb}
\usepackage{amsmath}
\usepackage{amssymb}
\usepackage{overcite}
\usepackage{epsf}
\input{epsf}
\numberwithin{equation}{section}
\begin{document}
\begin{titlepage}
\title{\bf Statistical Mechanics  and the Physics of the  Many-Particle Model Systems\thanks{Physics of Particles and Nuclei, 2009, Vol. 40, No. 7, pp. 949-997. }}  
%
\author{
A. L. Kuzemsky 
\\
{\it Bogoliubov Laboratory of Theoretical Physics,} \\
{\it  Joint Institute for Nuclear Research,}\\
{\it 141980 Dubna, Moscow Region, Russia.}\\
{\it E-mail:kuzemsky@theor.jinr.ru} \\
{\it http://theor.jinr.ru/\symbol{126}kuzemsky}}
\date{}
\maketitle
\begin{abstract}
The development of methods of quantum statistical mechanics is considered in light of their applications
to quantum solid-state theory. We discuss fundamental problems of the physics of magnetic materials
and the methods of the quantum theory of magnetism, including the method of two-time temperature Green's
functions, which is widely used in various physical problems of many-particle systems with interaction. Quantum
cooperative effects and quasiparticle dynamics in the basic microscopic models of quantum theory of
magnetism: the Heisenberg model, the Hubbard model, the Anderson Model, and the spin-fermion model are
considered in the framework of novel self-consistent-field approximation. We present a comparative analysis
of these models; in particular, we compare their applicability for description of complex magnetic materials.
The concepts of broken symmetry, quantum protectorate, and quasiaverages are analyzed in the context of
quantum theory of magnetism and theory of superconductivity. The notion of broken symmetry is presented
within the nonequilibrium statistical operator approach developed by D.N. Zubarev. In the framework of the
latter approach we discuss the derivation of kinetic equations for a system in a thermal bath. Finally, the results
of investigation of the dynamic behavior of a particle in an environment, taking into account dissipative effects,
are presented.

%
%
\end{abstract}
%
%
%
\end{titlepage}
\newpage
\tableofcontents
\newpage

%
\section{  Introduction}
%
\noindent
The purpose of this review is to trace the development
of some methods of quantum statistical mechanics
formulated by N.N. Bogoliubov, and also to show
their effectiveness in applications to problems of quantum
solid-state theory, and especially to problems of
quantum theory of magnetism. It is necessary to stress,
that the path to understanding the foundations of the
modern statistical mechanics and the development of
efficient methods for computing different physical
characteristics of many-particle systems was quite
complex. The main postulates of the modern statistical
mechanics were formulated in the papers by J.P. Joule
(1818-1889), R. Clausius (1822-1888), W. Thomson
(1824-1907), J.C. Maxwell (1831-1879), L. Boltzmann
(1844-1906), and, especially, by J.W. Gibbs
(1839-1903). The monograph by Gibbs "Elementary
Principles in Statistical Mechanics Developed with
Special Reference to the Rational Foundations of Thermodynamics"~\cite{gib1,gib2}
remains one of the highest peaks of
modern theoretical science. A significant contribution
to the development of modern methods of equilibrium
and nonequilibrium statistical mechanics was made by
Academician N.N. Bogoliubov (1909-1992)~\cite{bog1,nnb84,bog2,nbog94,kuz07}.\\
Specialists in theoretical physics, as well as experimentalists,
must be able to find their way through theoretical
problems of the modern physics of many-particle
systems because of the following reasons. Firstly,
the statistical mechanics is filled with concepts, which
widen the physical horizon and general world outlook.
Secondly, statistical mechanics and, especially, quantum
statistical mechanics demonstrate remarkable efficiency
and predictive ability achieved by constructing
and applying fairly simple (and at times even crude)
many-particle models. Quite surprisingly, these simplified
models allow one to describe a wide diversity of
real substances, materials, and the most nontrivial
many-particle systems, such as quark-gluon plasma,
the DNA molecule, and interstellar matter. In systems
of many interacting particles an important role is
played by the so-called
correlation effects~\cite{acor04}, which
determine specific features in the behavior of most
diverse objects, from cosmic systems to atomic nuclei.
This is especially true in the case of solid-state physics.
Investigation of systems with strong inter-electron correlations,
complicated character of quasiparticle
states, and strong potential scattering is an extremely
important and topical problem of the modern theory of
condensed matter. Our time is marked by a rapid
advancement in design and application of new materials,
which not only find a wide range of applications in
different areas of engineering, but they are also connected
with the most fundamental problems in physics,
physical chemistry, molecular biology, and other
branches of science. The quantum cooperative effects,
such as
magnetism and superconductivity,
frequently
determine the unusual properties of these new materials. The same can be also said about other non-trivial
quantum effects like, for instance, the quantum Hall
effect, the Bose-Einstein condensation, quantum tunneling
and others. This research direction is developing
very rapidly, setting a fast pace for widening the
domain where the methods of quantum statistical
mechanics are applied. This review will support the
above statement by concrete examples.
\section{Quantum Statistical Mechanics and Solid State Physics}
The development of experimental techniques over
the recent years opened the possibility for synthesis and
investigations of a wide class of new substances with
unusual combination of properties~\cite{hum05,hand99,dshi,plan06,proz07,fred08,stef08}. Transition
and rare-earth metals and especially compounds containing
transition and rare-earth elements possess a
fairly diverse range of properties. Among those, one
can mention magnetically ordered crystals, superconductors,
compounds with variable valence and heavy
fermions, as well as substances which under certain
conditions undergo a metal-insulator transition, like
perovskite-type manganites, which possesses a large
magneto-resistance with a negative sign. These properties
find widest applications in engineering; therefore,
investigations of this class of substances should be classified
as the currently most important problems in the
physics of condensed matter.
A comprehensive description of materials and their
properties (as well as efficient predictions of properties
of new materials) is possible only in those cases, when
there is an adequate quantum-statistical theory based
on the information about the electron and crystalline
structures. The main theoretical problem of this direction
of research, which is the essence of the
quantum theory of magnetism~\cite{tyab,matt}, is investigations and improvements
of quantum-statistical models describing the
behavior of the above-mentioned compounds in order
to take into account the main features of their electronic
structure, namely, their dual \textbf{"band-atomic"} nature~\cite{b41,b151}. The construction of a consistent theory explaining
the electronic structure of these substances encounters
serious difficulties when trying to describe the collectivization-
localization duality in the behavior of electrons.
This problem appears to be extremely important,
since its solution gives us a key to understanding magnetic,
electronic, and other properties of this diverse
group of substances. The author of the present review
investigated the suitability of the basic models with
strong electron correlations and with a complex spectrum
for an adequate and correct description of the dual
character of electron states. A universal mathematical
formalism was developed for this investigation~\cite{kuz02}. It
takes into account the main features of the electronic
structure and allows one to describe the true quasiparticle
spectrum, as well as the appearance of the magnetically
ordered, superconducting, and dielectric (or
semiconducting) states. With a few exceptions, diverse physical phenomena
observed in compounds and alloys of transition and
rare-earth metals~\cite{b41,b151,kuz93}, cannot be explained in
the framework of the mean-field approximation, which
overestimates the role of inter-electron correlations in
computations of their static and dynamic characteristics.
The circle of questions without a precise and definitive
answer, so far, includes such extremely important
(not only from a theoretical, but also from a practical
point of view) problems as the adequate description of
quasiparticle dynamics
for quantum-statistical models
in a wide range of their parameter values. The source of
difficulties here lies not only in the complexity of calculations
of certain dynamic properties (such as, the
density of states, electrical conductivity, susceptibility,
electron-phonon spectral function, the inelastic scattering
cross section for slow neutrons), but also in the
absence of a well-developed method for a consistent
quantum-statistical analysis of a many-particle interaction
in such systems. A self-consistent field approach
was used in the papers~\cite{kuz02,b36,b91,b118,b127,b128,b161} for description of
various dynamic characteristics of strongly correlated
electronic systems. It allows one to consistently and
quite compactly compute quasiparticle spectra for
many-particle systems with strong interaction taking
into account damping effects. The correlation effects
and quasiparticle damping are the determining factors
in analysis of the normal properties of high-temperature
superconductors, and of the transition mechanism
into the superconducting phase. We also formulated a
general scheme for a theoretical description of electronic
properties of many-particle systems taking into
account strong inter-electron correlations~\cite{kuz02,b36,b91,b118,b128,b161}.
The scheme is a synthesis of the method of two-time
temperature Green's functions~\cite{tyab} and the diagram
technique. An important feature of this approach is a
clear-cut separation of the elastic and inelastic scattering
processes in many-particle systems (which is a
highly nontrivial task for strongly correlated systems).
As a result, one can construct a correct basic approximation
in terms of generalized mean fields (the elastic
scattering corrections), which allows one to describe
magnetically ordered or superconducting states of the
system. The residual correlation effects, which are the
source of quasiparticle damping, are described in
terms of the Dyson equation with a formally exact representation
for the mass operator. There is a general
agreement that for heavy-fermion compounds, the
model Hamiltonian is well established (the periodic
Anderson model or the periodic Kondo lattice), and the
main theoretical challenge in this case lies in constructing
accurate approximations. However, in the case of
high-temperature superconductors or perovskite-type
manganites, neither a model, nor adequate approximate
analytical methods for its solution are available. Thus,
the development and improvement of the methods of
quantum statistical mechanics still remains quite an
important direction of research~\cite{lou06}.
\section{Magnetic Properties of Substances and Models of Magnetic Materials}
It is widely accepted that the appearance of magnetically
ordered states in transition metals is to some
extent a consequence of the atom-like character of
$d$-states, but mostly it is the result of interatomic
exchange interactions. In order to better understand the
origin of quantum models of magnetic materials, we
discuss here briefly the physical aspects of the magnetic
properties of solid materials. The magnetic properties
of substances belong to the class of natural phenomena,
which were noticed a long time ago~\cite{matt,magic96}. Although
it is assumed that we encounter magnetic natural phenomena
less frequently than the electric ones, nevertheless
as was noticed by V. Weisskopf, "magnetism is
a striking phenomenon; when we hold a magnet in one
hand and a piece of iron in another, we feel a peculiar
force, some "force of nature", similar to the force of
gravity"~\cite{wei}. It is interesting to note that the experiment-based 
scientific approach began from investigations
of magnetic materials. This is the so called inductive
method, which insists on searching for truth about
the nature not in deductions, not in syllogisms and formal
logics, but in experiments with the natural substances
themselves. This method was applied for the
first time by William Gilbert (1544-1603), Queen Elizabeth's
physician. In his book, "\emph{On the magnet, magnetic
bodies, and on the great magnet, the Earth}"~\cite{gilb},
published in 1600, he described over 600 specially performed
experiments with magnetic materials, which
had led him to an extremely important and unexpected
for the time conclusion, that the Earth is a giant spherical
magnet. Investigations of Earth's and other planet's
magnetism is still an interesting and quite important
problem of modern science~\cite{bock,camp,zuber03}. Thus, it is with
investigations of the physics of magnetic phenomena,
that the modern experiment based science began. Note,
that although the creation of the modern scientific
methods is often attributed to Francis Bacon, Gilbert's
book appeared 20 years earlier than "The New Organon"
by Francis Bacon (1561-1626).\\
The key to understanding the nature of magnetism
became the discovery of a close connection between
magnetism and electricity~\cite{vons71}. For a long time the understanding
of magnetism's nature was based on the
hypothesis of how the magnetic force is created by
magnets. Andre Ampere (1775-1836) conjectured that
the principle behind the operation of an ordinary steel
magnet should be similar to an electric current passing
over a circular or spiral wire. The essence of his hypothesis
laid in the assumption, that each atom contains a
weak circular current, and if most of these atomic currents
are oriented in the same direction, then the magnetic
force appears. All subsequent developments of the
theory of magnetism consisted in the development and
refinement of Ampere's molecular currents hypothesis.\\
As an extension of this idea by Ampere a conjecture
was put forward that a magnet is an ensemble of  
elementary double poles, magnetic dipoles. The dipoles
have two magnetic poles which are inseparably linked.
In 1907, Pierre Weiss (1865-1940) proposed a phenomenological
picture of the magnetically ordered
state of matter. He was the first to perform a phenomenological
quantitative analysis of the magnetic phenomena
in substances~\cite{pweis}. Weiss's investigations were
based on the notion, introduced by him, of a molecular
field. Subsequently this approach was named the
molecular (or mean, or effective) field approximation,
and it is still being widely used even at the present time~\cite{smart}. 
The simplest microscopic model of a ferromagnet
in the molecular field's approximation is based on the
assumption that electrons form a free gas of magnetic
arrows (magnetic dipoles), which imitate Ampere's
molecular currents. In the simplest case it is assumed
that these "elementary magnets" could orient in space
either along a particular direction, or against it. In order
to find the thermodynamically equilibrium value of the
magnetization $\langle M \rangle$
as a function of temperature $T$, one
has to turn to general laws of thermodynamics. This is
especially important when we consider the behavior of
a system at finite temperatures. Finding the equilibrium
magnetization of a ferromagnet becomes quite a simple
task if we first succeed in writing down its energy
$ E(\langle M \rangle)$ as a function of magnetization. All we have to
do after that is to minimize the free energy
$ F(\langle M \rangle),$ 
which is defined by the following relationship~\cite{vons71}:
\begin{equation}\label{e1}
F(\langle M \rangle) = E(\langle M \rangle) - T S(\langle M \rangle).
\end{equation}
Here, $ S(\langle M \rangle)$ is the system entropy also written
down as a function of magnetization. It is important to
stress, that the problem of calculating the system
entropy cannot be solved in the framework of just thermodynamics.
In order to find the entropy one has to
turn to statistical mechanics~\cite{gib1,zub,kub1,kub2,min02}, which provides
a microscopic foundation to the laws of thermodynamics.
Note that derivations of equilibrium magnetization
$\langle M \rangle$ as a function of temperature
$T$, or, more generally,
investigations of relationships between the free energy
and the order parameter in magnetics and pyroelectrics
are still ongoing even at the present time~\cite{mar1,mar2,mar3,mar4}. Of
course, modern investigations take into account all the
previously accumulated experience.
In the framework of the P. Weiss approach one
investigates the appearance of a spontaneous magnetization
$\langle M \rangle \neq 0$
 for $ H = 0.$ This approach is based on the
following postulate for the behavior of the system's
energy
\begin{equation}\label{e2}
 E(\langle M \rangle) \simeq N I (\langle M \rangle)^{2}.
\end{equation}
This expression takes into account the interaction
between elementary magnets (arrows). Here,
$ I$ is the
energy of the Weiss molecular field per atomic magnetic
arrow. The question on the microscopic nature of
this field is beyond the framework of the Weiss
approach. The minimization of the free energy
$ F(\langle M \rangle)$ yields the following relationship
\begin{equation}\label{e3}
\langle M \rangle  = th (T_{C} \langle M \rangle / T),
\end{equation}
Where $ T_{C}$ is the Curie or the critical temperature. As the
temperature decreases below this critical value a spontaneous
magnetization appears in the system. The Curie
temperature was named in honor of Pierre Curie (1859-1906), who established the following law for the behavior
of susceptibility $ \chi$ in paramagnetic substances
\begin{equation}\label{e4}
\chi = \lim_{H \rightarrow 0} \frac{\langle M \rangle}{H} =  \frac{C}{T}.
\end{equation}
Depending on the actual material, the Curie parameter
$C$ obtains different (positive) values~\cite{vons71}. Note that
Pierre Curie performed thorough investigations of the
magnetic properties of iron back in 1895. In the process
of those experiments he established the existence of a
critical temperature for iron, above which the ferromagnetic
properties disappear. These investigations
laid a foundation for investigations of order-disorder
phase transitions, and other phase transformations in
gases, liquids, and solid substances. This research
direction created the core of the physics of critical phenomena,
which studies the behavior of substances in
the vicinity of critical temperatures~\cite{domb}.\\
Extensive investigations of spontaneous magnetization
and other thermal effects in nickel in the vicinity of
the Curie temperature were performed by Weiss and his
collaborators~\cite{pweis26}. They developed a technique for
measuring the behavior of the spontaneous magnetization
in experimental samples for different values of
temperature. Knowing the behavior of spontaneous
magnetization as a function of temperature one can
determine the character of magnetic transformations in
the material under investigation. Investigations of the
behavior of the magnetic susceptibility as a function of
temperature in various substances remain important
even at the present time~\cite{ww98,mkuz05,zw06}.\\
Within the P. Weiss approach we obtain the following
expression for the Curie temperature
\begin{equation}\label{e5}
T_{C} =   2 I  /k_{B}.
\end{equation}
In order to obtain a rough estimate for the magnitude
of $I$ take $ T_{C} = 1000 \, \textrm{K}.$ Then one obtains
$I \sim 10^{-13}$ erg/atom.
This implies that the only origin of the Weiss molecular
field can be the Coulomb interaction of electrical charges~\cite{tyab,vons71}.
Computations in the framework of the molecular
field method lead to the following formula for the magnetic
susceptibility
\begin{equation}\label{e6}
\chi =   N \mu_{B}^{2} \langle M \rangle /H = \frac{N \mu_{B}^{2}}{k_{B} (T - T_{C})},
\end{equation}
where $\mu_{B} = e \hbar / 2mc$ is the Bohr magneton (within the
Weiss approach this is the magnetic moment of the
magnetic arrows imitating Ampere's molecular currents).
The above expression for the susceptibility is
referred to as the Curie-Weiss law. Thus, the Weiss
molecular field, whose magnitude is proportional to the
magnetization, is given by
\begin{equation}\label{e7}
H_{W} =   k_{B} T_{C}  \langle M \rangle  /\mu_{B}.
\end{equation}
Researchers tried to find the answer to the question
on the nature of this internal molecular field in ferromagnets
for a long time. That is, they tried to figure out
which interaction causes the parallel alignment of electron
spins. As was stressed in the book~\cite{dorf}:  "At first
researchers tried to imagine this interaction of electrons
in a given atom with surrounding electrons as some
quasi-magnetic molecular field, acting on the electrons
of the given atom. This hypothesis served as a foundation
for the P. Weiss theory, which allowed one to
describe qualitatively the main properties of ferromagnets".
However, it was established that Weiss's molecular
field approximation is applicable neither for theoretical
interpretation, nor for quantitative description of
various phenomena taking place in the vicinity of the
Curie temperature. Although numerous attempts aiming
to improve Weiss's mean-field theory were undertaken,
none of them led to significant progress in this
direction.
Numerical estimates yield the value $ H_{W} = 10^{7}$ oersted
for the magnitude of the Weiss mean field. The
nonmagnetic nature of the Weiss molecular field was
established by direct experiments in 1927 (see the
books~\cite{vons71,dorf}). Ya.G. Dorfman performed the following
experiment. An electron beam passing though
nickel foil magnetized to the saturation level is falling
on a photographic plate. It was expected that if such a
strong magnetic field indeed exists in nickel, then the
electrons passing through the magnetized foil would
deflect. However, it turned out that the observed electron
deflection is extremely small. The experiment led
to the conclusion that, contrary to the consequences of
the Weiss theory, the internal fields of large intensity are
not present in ferromagnets. Therefore, the spin ordering
in ferromagnets is caused by forces of a nonmagnetic
origin. It is interesting that fairly recently, in 2001,
similar experiments were performed again~\cite{web01} (in a
substantially modified form, of course). A beam of
polarized "hot" electrons was scattered by thin ferromagnetic
nickel, iron, and cobalt films, and the polarization
of the scattered electrons was measured. The
concept of the Weiss exchange field $ \textbf{W}(x) \sim - J_{\alpha} \textbf{S}(x)$ was
used for theoretical analysis~\cite{web01,alb06}. The real part of
this field corresponds to the exchange interaction
between the incoming electrons and the electron density
of the film (the imaginary part is responsible for
absorption processes). The derived equations, describing
the beam scattering, resemble quite closely the corresponding
equations for the Faraday's rotation effect
in the light passing through a magnetized environment~\cite{web01,alb06}.
The theoretical consideration is based on using the
mean-field approximation, namely on the replacement
\begin{equation}\label{e8}
\textbf{W} \simeq  \langle \textbf{W}(x) \rangle = J_{\alpha}  \langle \textbf{S}(x) \rangle.
\end{equation}
The subsequent quite rigorous and detailed consideration~\cite{alb06}
aimed at deriving the effective quantum
dynamics of the field $W(x)$ showed, that this dynamics
is described by the Landau-Lifshitz equation~\cite{vons71}. The spatial
and temporal variations of the field $W(x)$ are
described by spin waves. The quanta of the Weiss
exchange field are magnons.\\
One has to note that in its original version, the Weiss
molecular field was assumed to be uniformly distributed
over the entire volume of the sample, and had the
same magnitude in all points of the substance. An
entirely different situation takes place in a special class
of substances called antiferromagnets. As the temperature
of antiferromagnets falls below a particular value,
a magnetically ordered state appears in the form of two
inserted into each other sublattices with opposite directions
of the magnetization. This special value of the
temperature became known as the Neel temperature,
after the founder of the antiferromagnetism theory
L. Neel (1904-2000). In order to explain the nature of
the antiferromagnetism (as well as of the ferrimagnetism)
L. Neel introduced a profound and nontrivial
notion of local molecular fields~\cite{neel}. However, there
was no a unified approach to investigations of magnetic
transformations in real substances. Moreover, a consistent
consideration of various aspects of the physics of
magnetic phenomena on the basis of quantum mechanics
and statistical physics was and still is an exceptionally
difficult task, which to the present days does not
have a complete solution~\cite{white06,magn06}. This was the reason
why the authors of the most complete, at that time,
monograph on the magnetism characterized the state of
affairs in the physics of magnetic phenomena as follows:
"Even recently the problems of magnetism
seemed to belong to an exceptionally unrewarding area
for theoretical investigations. Such a situation could be
explained by the fact that the attention of researchers
was devoted mostly to ferromagnetic phenomena,
because they played and still play quite an important
role in engineering. However, the theoretical interpretation
of the ferromagnetism presents such formidable
difficulties, that to the present day this area remains one
of the darkest spots in the entire domain of physics"~\cite{land29}.
The magnetic properties and the structure of matter
turned out to be interconnected subjects. Therefore,
a systematic quantum-mechanical examination of the
problem of magnetic substances was considered by
most researchers~\cite{dorf,ston26,vv32,bloch} as quite an important task.
Heisenberg, Dirac, Hund, Pauli, van Vleck, Slater, and
many other researchers contributed to the development
of the quantum theory of magnetism. As was noted by
D. Mattis~\cite{matt}, "\ldots by 1930, after four years of most exciting
and striking discoveries in the history of theoretical
physics, a foundation for the modern electron theory of
matter was laid down, after that an epoch of consolidation
and computations had began, which continues up
to the present day".\\
Over the last decades the physics of magnetic phenomena
became a vast and ramified domain of modern physical 
science~\cite{matt,vons71,white06,magn06,yos96,herr,pratt,barb88,casp89,ahar,chika,steve97,mied01,gigno02,gig02,boer03,spald03,zabel}. The rapid
development of the physics of magnetic materials was
influenced by introduction and development of new
physical methods for investigating the structural and
dynamical properties of magnetic substances~\cite{yzhu}.
These methods include magnetic neutron diffraction
analysis~\cite{ml71,tapa06}, NMR and EPR-spectroscopy, the
Mossbauer effect, novel optical methods~\cite{suga}, as well
as recent applications of synchrotron radiation~\cite{land97,msr01,msr06,sinh07}.
In particular, unique possibilities of the thermal
neutron's scattering methods~\cite{yzhu,ml71,tapa06,neut50} allow one to
obtain information on the magnetic and crystalline
structure of substances, on the distribution of magnetic
moments, on the spectrum of magnetic excitations, on
critical fluctuations, and on many other properties of
magnetic materials. In order to interpret the data
obtained via inelastic scattering of slow neutrons one
has to take into account electron-electron and electron-nuclear
interactions in the system, as well as the Pauli
exclusion principle. Here, we again face the challenge
of considering various aspects of the physics of magnetic
phenomena, consistently on the basis of quantum
mechanics and statistical physics. In other words, we
are dealing with constructing a consistent quantum theory
of magnetic substances. As was rightly noticed by
K. Yosida, "The question of electron correlations in
complex electronic systems is the beginning and the
end of all research on magnetism"~\cite{yosi}. Thus, the phenomena
of magnetism can be described and interpreted
consistently only in the framework of quantum statistical
theory of many interacting particles.
\section{Quantum Theory of Magnetism}
It is well known that "quantum mechanics is the key
to understanding magnetism"~\cite{vv79}. One of the first
steps in this direction was the formulation of "Hund's
rules" in atomic physics~\cite{pratt}. As was noticed by
D. Mattis~\cite{matt}, "The accumulated spectroscopic data
allowed Stoner (1899-1968) to attribute the correct
number of equivalent electrons to each atomic shell,
and Hund (1896-1997) to state his rules, related to the
spontaneous magnetic moments of a free atom or ion".
Hund's rules are empirical recipes. Their consistent
derivation is a difficult task. These rules are stated as
follows:\\
(1) The ground state of an atom or an ion with a $L-S$
coupling is a state with the maximal multiplicity $(2S + 1)$
for a given electron configuration.\\
(2) From all possible states with the maximal multiplicity,
the ground state is a state with the maximal
value of $L$ allowed by the Pauli exclusion principle.\\
Note that the applicability of these empirical rules is
not restricted to the case, when all electrons lie in a single
unfilled valency shell. A rigorous derivation of Hund's
rules is still missing. However, there are a few particular
cases which show their validity under certain restrictions~\cite{pratt,yaman}
(see a recent detailed analysis of this question in
the papers~\cite{mie93,frol05}). Nevertheless Hund's rules are very
useful and are widely used for analysis of various magnetic
phenomena. A physical analysis of the first Hund's
rule leads us to the conclusion, that it is based on the fact,
that the elements of the diagonal matrix of the electron-electron's Coulomb interaction contain the exchange's
interaction terms, which are entirely negative. This is the
case only for electrons with parallel spins. Therefore, the
more electrons with parallel spins involved, the greater
the negative contribution of the exchange to the diagonal
elements of the energy matrix. Thus, the first Hund's rule
implies that electrons with parallel spins "\emph{tend to avoid each other}" spatially. Here, we have a direct connection
between Hund's rules and the Pauli exclusion principle.\\
One can say that the Pauli exclusion principle
(1925) lies in the foundation of the quantum theory of
magnetic phenomena. Although this principle is merely
an empirical rule, it has deep and important implications~\cite{pauli05}.
W. Pauli (1900-1958) was puzzled by the
results of the ortho-helium terms analysis, namely, by
the absence in the term structure of the presumed
ground state, that is the $(1^{3}S)$ level. This observation
stimulated him to perform a general examination of
atomic spectra, with the aim to find out, if certain terms
are absent in other chemical elements and under other
conditions as well. It turned out, that this was indeed
the case. Moreover, the conducted analysis of term systems
had shown that in all the instances of missing
terms the entire sets of the quantum numbers were identical
for some electrons. And vice versa, it turned out
that terms always drop out in the cases when entire sets
of quantum numbers are identical. This observation
became the essence of the \textbf{Pauli exclusion principle}:\\
The sets of quantum numbers for two (or many)
electrons are never identical; two sets of quantum numbers,
which can be obtained from one another by permutations
of two electrons, define the same state.\\
In the language of many-electron wave functions
one has to consider permutations of spatial and spin
coordinates of electrons $i$ and $j$    in the case when both
the spin variables $\sigma_{i} = \sigma_{j} = \sigma_{0}$ and the spatial coordinates
$\vec{r}_{i} = \vec{r}_{j} = \vec{r}_{0} $ of these two electrons are identical.
Then, we obtain:
\begin{equation}\label{e9}
P_{ij} \psi (\vec{r}_{1}\sigma_{1}, \ldots \vec{r}_{i}\sigma_{i}, \ldots \vec{r}_{j}\sigma_{j}, \ldots) =   
\psi (\vec{r}_{1}\sigma_{1}, \ldots \vec{r}_{i}\sigma_{i}, \ldots \vec{r}_{j}\sigma_{j}, \ldots). 
\end{equation}
The Pauli exclusion principle implies that
\begin{equation}\label{e10}
P_{ij} \psi (\vec{r}_{1}\sigma_{1}, \ldots \vec{r}_{i}\sigma_{i}, \ldots \vec{r}_{j}\sigma_{j}, \ldots) =   
- \psi (\vec{r}_{1}\sigma_{1}, \ldots \vec{r}_{i}\sigma_{i}, \ldots \vec{r}_{j}\sigma_{j}, \ldots). 
\end{equation}
The above conditions are satisfied simultaneously
only in the case, when $\psi$ is equal to zero identically.
Therefore, we arrive at the following conclusion: electrons
are indistinguishable, that is, their permutations
must not change observable properties of the system.
The wave function changes or retains its sign under permutations
of two particles depending on whether these
indistinguishable particles are bosons or fermions. A
consequence of the Pauli exclusion principle is the \emph{Aufbau principle}, which leads 
to the periodicity in the
properties of chemical elements. The fact that not more
than one electron can occupy any single state leads also
to such fundamental consequences as the very existence
of solid bodies in nature. If the Pauli exclusion
principle was not satisfied, no substance could ever be
in a solid state. If the electrons would not have spin
(that is, if they were bosons) all substances would
occupy much smaller volumes (they would have higher
densities), but they would not be rigid enough to have
the properties of solid bodies.\\
Thus, the tendency of electrons with parallel spins
"to avoid each other" reduces the energy of electron-electron Coulomb interaction, and hence, lowers the
system energy. This property has many important
implications, in particular, the existence of magnetic
substances. Due to the presence of an internal unfilled
$nd$- or $nf$-shell, all free atoms of transition elements are
strong magnetic, and this is a direct consequence of
Hund's rules. When crystals are formed~\cite{matt,vons71,pratt,steve97}
the electronic shells in atoms reorganize, and in order to
understand clearly the properties of crystalline substances,
one has to know the wave function and the
energies of (previously) outer-shell electrons. At the
present time there are well-developed efficient methods
for computing electronic energy levels in crystals~\cite{cal,har,mar04}.
Speaking qualitatively, we have to find out how the
atomic wave's functions change when crystals are
formed, and how significantly they delocalize~\cite{b151}.
%
\subsection{The Method of Model Hamiltonians}
%
The method of model Hamiltonians proved to be
very efficient in the theory of magnetism. Without any
exaggeration one can say, that the tremendous successes
in the physics of magnetic phenomena were
achieved, largely, as a result of exploiting a few simple
and schematic model concepts for "the theoretical
interpretation of ferromagnetism"~\cite{pei80}. One can regard
the Ising model~\cite{iz2,iz3} as the first model of the quantum
theory of magnetism. In this model, formulated by
W. Lenz (1888-1957) in 1920 and studied by E. Ising
(1900-1998), it was assumed that the spins are
arranged at the sites of a regular one-dimensional lattice.
Each spin can obtain the values $\pm \hbar/2$:
\begin{equation}\label{e11}
\mathcal{H}  =  - \sum_{<ij>}  I_{ij} S^{z}_{i}S^{z}_{j}.
\end{equation}
This was one of the first attempts to describe the
magnetism as a cooperative effect. It is interesting that
the one-dimensional Ising model was solved by Ising in
1925, while the exact solution of the Ising model on a
two-dimensional square lattice was obtained by
L. Onsager (1903-1976)~\cite{onsag1,onsag2} only in 1944. However,
the Ising model oversimplifies the situation in real
crystals. W. Heisenberg (1901-1976)~\cite{wh28} and P. Dirac
(1902-1984)~\cite{dir29} formulated the Heisenberg model,
describing the interaction between spins at different sites
of a lattice by the following isotropic scalar function
\begin{equation}\label{e12}
\mathcal{H}  = -  \sum_{ij}  J(i-j) \vec S_{i} \vec S_{j} 
-g\mu_{B}H\sum_{i}S_{i}^{z}. 
\end{equation}
Here $J(i-j)$  (the "exchange integral") is the strength of
the exchange interaction between the spins located at the
lattice sites $i$ and $j$. It is usually assumed that $ J(i-j) = J(j-i)$  and $J(i - j = 0) = 0,$  which means that only the
inter-site interaction is present (there is no self-interaction).
The Heisenberg Hamiltonian (\ref{e12}) can be rewritten
in the following form:
\begin{equation}
\label{e13} \mathcal{H} = -  \sum_{ij} J(i-j) ( S^z_{i}S^z_{j} +
 S^+_{i}S^-_{j}).
\end{equation}
Here, $S^{\pm} = S^x \pm iS^y$ are the spin raising and lowering
operators. They satisfy the following set of commutation
relationships:
\begin{eqnarray} \nonumber
[S^{+}_{i},S^{-}_{j}]_{-} = 2S^{z}_{i} \delta_{ij}; \quad
[S^{+}_{i},S^{-}_{i}]_{+} = 2S(S + 1) - 2(S^{z}_{i})^{2};  \\
\nonumber [S^{\mp}_{i},S^{z}_{j}]_{-} = \pm
S^{\mp}_{i}\delta_{ij}; \quad S^{z}_{i} = S(S + 1) -
(S^{z}_{i})^{2} - S^{-}_{i}S^{+}_{i}; \\ \nonumber
(S^{+}_{i})^{2S+1} = 0, \quad (S^{-}_{i})^{2S+1} = 0. \nonumber
\end{eqnarray}
Note that in the isotropic Heisenberg model the $z$-component
of the total spin $S^z_{tot} = \sum_{i}S^z_{i}$ is a constant of
motion, that is  $ [H,S^z_{tot}] = 0$.\\
Thus, in the framework of the Heisenberg-Dirac-van Vleck model~\cite{vv32,wh28,dir29,vv36,kitt57}, describing the interaction
of localized spins, the necessary conditions for the
existence of ferromagnetism involve the following two
factors. Atoms of a "ferromagnet to be" must have a
magnetic moment, arising due to unfilled electron $d$- or
$f$-shells. The exchange integral $ J_{ij}$ related to the electron
exchange between neighboring atoms must be positive.
Upon fulfillment of these conditions the most energetically
favorable configurations in the absence of an
external magnetic field correspond to parallel alignment
of magnetic moments of atoms in small areas of
the sample (domains)~\cite{kitt57}. Of course, this simplified
picture is only schematic. A detail derivation of the
Heisenberg-Dirac-van Vleck model describing the
interaction of localized spins is quite complicated.
Because of a shortage of space we cannot enter into discussion
of this quite interesting topic~\cite{vv53,ara62,her75}. An
important point to keep in mind here is that magnetic
properties of substances are born by quantum effects,
the forces of exchange interaction~\cite{cars96}.\\
As was already mentioned above, the states with
antiparallel alignment of neighboring atomic magnetic
moments are realized in a fairly wide class of substances.
As a rule, these are various compounds of transition
and rare-earth elements, where the exchange
integral $ J_{ij}$ for neighboring atoms is negative. Such a
magnetically ordered state is called 
\textbf{antiferromagnetism}~\cite{neel,vv41,gort57,subl61,afmf62,bel65,smol74,smol75,zvezd85,nik89,tur01,bor05}. 
In 1948, L. Neel introduced the notion
of \textbf{ferrimagnetism}~\cite{neel48,wolf,pq90,neel88,neel91,neel40} to describe the properties
of substances in which spontaneous magnetization
appears below a certain critical temperature due to nonparallel
alignment of the atomic magnetic moment~\cite{pwa56,afmf62,bel65,smol74,smol75,zvezd85,nik89,tur01,bor05}.
These substances differ from antiferromagnets
where sublattice magnetizations $m_{A}$ and $m_{B}$
usually have identical absolute values, but opposite orientations.
Therefore, the sublattice magnetizations
compensate for each other and do not result in a macroscopically
observable value for magnetization. In ferrimagnetics
the magnetic atoms occupying the sites in
sublattices $A$ and $B$  differ both in the type and in the
number. Therefore, although the magnetizations in the
sublattices $A$ and $B$ are antiparallel to each other, there
exists a macroscopic overall spontaneous magnetization~\cite{afmf62,smol74,smol75,bor05,wolf}.\\
Later, substances possessing weak ferromagnetism
were investigated~\cite{afmf62,bel65,smol74,smol75,zvezd85,nik89,tur01,bor05}. It is interesting that originally
Neel used the term \emph{parasitic ferromagnetism}~\cite{neel49}
when referring to a small ferromagnetic moment,
which was superimposed on a typical antiferromagnetic
state of the $\alpha$ iron oxide $Fe_{2}O_{3}$ (hematite)~\cite{hemat}.
Later, this phenomenon was called canted antiferromagnetism,
or weak ferromagnetism~\cite{hemat,ahar62}. The
weak ferromagnetism appears due to antisymmetric
interaction between the spins $\vec{S}_{1}$ and $\vec{S}_{2}$ and which is proportional
to the vector product $\vec{S}_{1} \times \vec{S}_{2}.$ This interaction
is written in the following form
\begin{equation}\label{e14}
\mathcal{H}_{DM}  \sim  \vec{D}   \vec{S}_{1} \times \vec{S}_{2}.
\end{equation}
The interaction (\ref{e14}) is called the Dzyaloshinsky-Moriya interaction [127, 128]. 
Hematite is one of the
most well known minerals~\cite{hemat,ahar62,lin59,rem65,dun70}, which
is still being intensively studied~\cite{b26} even at the
present time~\cite{suber,sizeh,mdhnano,hnano}.\\
Thus, there exist a large number of substances and
materials that possess different types of magnetic
behavior: diamagnetism, paramagnetism, ferromagnetism,
antiferromagnetism, ferrimagnetism, and weak
ferromagnetism. We would like to note that the variety
of magnetism is not exhausted by the above types of
magnetic behavior; the complete list of magnetism
types is substantially longer~\cite{hurd}. As was already
stressed, many aspects of this behavior can be reasonably
well described in the framework of a very crude
Heisenberg-Dirac-van Vleck model of localized spins.
This model, however, admits various modifications
(see, for instance, the book~\cite{nag88}). Therefore, various
nontrivial generalizations of the localized spin models
were studied. In particular, a modification of the
Heisenberg model was investigated, where, in addition
to the exchange interaction between different sites, an
exchange interaction between the spins at the same site
was considered~\cite{b16}:
\begin{eqnarray}\label{e15}
\mathcal{H}   = -\mu_{B}H\sum_{<i\alpha>}  S^{z}_{i\alpha}  - \frac{1}{2}\sum_{i \neq j} \sum_{\alpha \beta} 
J(i\alpha;j\beta) \left( \lambda S^{+}_{i\alpha} S^{-}_{j\beta} + S^{z}_{i\alpha}S^{z}_{j\beta} \right) - \\ \nonumber
\frac{1}{2}\sum_{i} \sum_{\alpha \neq  \beta} J(i\alpha;i\beta) \left( \lambda S^{+}_{i\alpha} S^{-}_{i\beta} + 
S^{z}_{i\alpha}S^{z}_{i\beta} \right).
\end{eqnarray}
In the case when $J(i\alpha;i\beta) \gg J(i\alpha;j\beta)$, this model
Hamiltonian in some sense imitates Hund's rules.
Indeed, Hund's rules state that the triplet's spin state of
two electrons occupying one and the same site is energetically
more favorable than the singlet state. It is this
feature that is taken into account by the model (\ref{e15}). A
model of this type was used for description of composite
ferrites, which contain different types of atoms with
different spins (magnetic moments). In the limiting
case $J(i\alpha;i\beta)\neq 0; J(i\alpha;j\beta) \equiv 0$ the model (\ref{e15}) can be
considered as the simplest version of the Heisenberg
model~\cite{lul84}. In this case, the two-spin system is interpreted~\cite{lul84}
as the simplest one-dimensional periodic magnet
with the period $N = 2$. Despite the apparent
'shortages' model (\ref{e15}) has found numerous applications
for description of real substances~\cite{a11}, including
the composite $Cu(NO_{3})_{2}\cdot 2.5 H_{2}O$-type salts~\cite{ber63,tac70}, of
clusters~\cite{fur77,fur86}, as well as for improving mean-field
approximation by using various cluster methods~\cite{call63}.
%
\subsection{The Problem of Magnetism of Itinerant Electrons}
%
The Heisenberg model describing localized spins is
mostly applicable to substances where the ground
state's energy is separated from the energies of excited
current-type states by a gap of a finite width. That is, the
model is mostly applicable to semiconductors and
dielectrics~\cite{smol74,nag79}. However, the main strongly magnetic
substances, nickel, iron, and cobalt, are metals,
belonging to the transition group~\cite{vons71}. The development
of quantum statistical theory of transition metals
and of their compounds followed a more difficult path
than that of the theory of simple metals~\cite{rpei,rpei85,zi,mizu}. The
traditional physical picture of the metal state was based
on the notion of Bloch electron waves~\cite{rpei,rpei85,zi,mizu,b167}. However,
the role played by the inter-electron interaction
remained unclear within the conventional approach. On
the other hand, the development of the band theory of
magnetism~\cite{herr,vvmet57,mshi81,tmor88,kubl,mizia}, and investigations of the
electronic phase's transitions in transition and rare-earth
metal compounds gradually led to realization of
the determining role of electron correlations~\cite{mott64,raim2,jfr}. Moreover, in many cases inter-electron interaction
is very strong and the description in terms of the
conventional band theory is no longer applicable. Special
properties of transition  metals and of their
alloys and compounds are largely determined by the
dominant role of $d$-electrons. In contrast to simple metals,
where one can apply the approximation of quasi-free
electrons, the wave functions of $d$-electrons are
much more localized, and, as a rule, have to be
described by the tight-binding approximation~\cite{cal,har,tb97}. The main aim of the band theory of magnetism
and of related theories, describing phase ordering and
phenomena of phase transition in complex compounds
and oxides of transition and rare-earth metals, is to
describe in the framework of a unified approach both
the phenomena revealing the localized character of
magnetically active electrons, and the phenomena
where electrons behave as collectivized band entities~\cite{b151}.
A resolution of this apparent contradiction
requires a very deep understanding of the relationship
between the localized and the band description of electron
states in transition and rare-earth metals, as well as
in their alloys and compounds. The quantum statistical
theory of systems with strong inter-electron correlations
began to develop intensively when the main features
of early semi-phenomenological theories were formulated
in the language of simple model Hamiltonians.
Both the Anderson model~\cite{and61,and79}, which formalized
the Friedel theory of impurity levels, and the Hubbard
model~\cite{hub63,hub64d,hub64,hub65,hub66,hub66mb}, which formalized and developed early
theories by Stoner, Mott, and Slater, equally stress the role
of inter-electron correlations. The Hubbard Hamiltonian
and the Anderson Hamiltonian (which can be considered
as the local version of the Hubbard Hamiltonian)
play an important role in the electron's solid-state theory~\cite{kuz02}.
Therefore, as was noticed by E. Lieb~\cite{lieb69},
the Hubbard model is 'definitely the first candidate'
for constructing a '\textbf{more fundamental}' quantum theory
of magnetic phenomena than the 'theory based on the
Ising model'~\cite{lieb69} (see also the papers~\cite{kana63,lieb93,tas93,tas98,tas03}).\\
However, as it turned out, the study of Hamiltonians
describing strongly-correlated systems is an exceptionally
difficult many-particle problem, which requires
applications of various mathematical methods~\cite{lieb69,lieb93,tas93,tas98,tas03,loss93,yan96,stein97}.
In fact, with the exception of a few particular
cases, even the ground state of the Hubbard model is
still unknown. Calculation of the corresponding quasiparticle
spectra in the case of strong inter-electron correlations
also turned out to be quite a complicated problem.
As was quite rightly pointed out by J. Kanamori~\cite{kana82},
when one is dealing with "a metal state with the values
of parameters close to the critical point, where the
metal turns into a dielectric", then "the calculation of
excited states in such crystals becomes very difficult
(especially at low temperatures)". Therefore, in
contrast to quantum many-body systems with weak
interaction, the definition of such a notion as elementary
excitations for strongly-interacting electrons with
strong inter-electron correlations is quite a nontrivial
problem requiring special detailed investigations~\cite{kuz02,b36,b91,b118,b127,b128}.
At the same time, one has to keep in mind, that
the Anderson and Hubbard models were designed for
applications to real systems, where both the case of
strong and the case weak inter-electron correlations are
realized.\\  Often, a very important role is played by the
electron interaction with the lattice vibrations, the
phonons~\cite{fr66,mit69,blf}. Therefore, the number one necessity
became the development of a systematic self-consistent
theory of electron correlations applicable for a
wide range of the parameter values of the main model,
and the development of the electron-phonon's interaction
theory in the framework of a modified tight-binding
approximation of strongly correlated electrons, as well as
the examination of various limiting cases~\cite{b67,b102}. All
this activity allowed one to investigate the electric conductivity~\cite{b51,b65},
and the superconductivity~\cite{b56,b72} in
transition metals, and in their disordered alloys.
%
\subsection{The Anderson and Hubbard Models}
%
The Hamiltonian of the single-impurity Anderson
model~\cite{b128,and61,and79} is written in the following form:
\begin{equation} \label{e16}
\mathcal{H} =
\sum_{k\sigma}\epsilon_{k}c^{\dag}_{k\sigma}c_{k\sigma} + \sum_{\sigma}
E_{0\sigma}f^{\dag}_{0\sigma}f_{0\sigma} +
U/2\sum_{\sigma}n_{0\sigma}n_{0-\sigma} +
\sum_{k\sigma}V_{k}(c^{\dag}_{k\sigma}f_{0\sigma} +
f^{\dag}_{0\sigma}c_{k\sigma}).
\end{equation}
Here, $c^{\dag}_{k\sigma}$ and $f^{\dag}_{0\sigma}$ are the creation operators of conduction
electrons and of localized impurity electrons,
respectively, $\epsilon_{k}$ are the energies of conduction electrons,
$E_{0\sigma}$ is the energy level of localized impurity electrons,
and $U$ is the intra-atomic Coulomb interaction of
the impurity-site electrons;  $V_{k}$ is the $s-f$ hybridization.
One can generalize the Hamiltonian of the single-impurity
Anderson model to the periodic case:
\begin{equation} \label{e17}
\mathcal{H} =
\sum_{k\sigma}\epsilon_{k}c^{\dag}_{k\sigma}c_{k\sigma} +  
\sum_{i\sigma}E_{\sigma}f^{\dag}_{i\sigma}f_{i\sigma} +
U/2\sum_{i\sigma}n_{i\sigma}n_{i-\sigma} +
\sum_{kj\sigma}V_{kj}(c^{\dag}_{k\sigma}f_{j\sigma} +
f^{\dag}_{j\sigma}c_{k\sigma}).
\end{equation}
The above Hamiltonian is called the periodic
Anderson model.
The Hamiltonian of the Hubbard model~\cite{hub63} is
given by:
\begin{equation}\label{e18}
\mathcal{H} = \sum_{ij\sigma}t_{ij}a^{\dagger
}_{i\sigma}a_{j\sigma} + U/2\sum_{i\sigma}n_{i\sigma}n_{i-\sigma}.
\end{equation}
The above Hamiltonian includes the repulsion of the
single-site intra-atomic Coulomb $U$, and $t_{ij}$, the one-electron
hopping energy describing jumps from a $J$ site
to an $i$ site. As a consequence of correlations electrons
tend to "avoid one another". Their states are best modeled
by atom-like Wannier wave functions $[\phi({\vec r} -{\vec R_{j}})]$.
The Hubbard model's Hamiltonian can be characterized
by two main parameters: $U$, and the effective band
width of tightly bound electrons
$$\Delta = (N^{-1}\sum_{ij}\vert t_{ij}\vert^{2})^{1/2}.$$
The band energy of Bloch electrons $\epsilon(\vec k)$ is given by
$$\epsilon(\vec k)     = N^{-1}\sum_{\vec k}t_{ij} \exp[- i{\vec
k}({\vec R_{i}} -{\vec R_{j}}],$$
where $N$ is the total number of lattice sites. Variations
of the parameter $\gamma = \Delta/U$ allow one to study two interesting
limiting cases, the band regime ($\gamma \gg 1$) and the
atomic regime ($\gamma \rightarrow 0$).
Note that the single-band Hubbard model (\ref{e18}) is a
particular case of a more general model, which takes
into account the degeneracy of $d$-electrons. In this case
the second quantization is constructed with the aid of
the Wannier functions of the form $[\phi_{\lambda}(r-R_{i})]$, where
$\lambda$ is the band index ($\lambda$= 1,2,\ldots 5). The corresponding
Hamiltonian of the electron system is given by
\begin{equation} \label{e19a}
\mathcal{H} =
\sum_{ij\mu\nu\sigma}t^{\mu\nu}_{ij}a^{\dag}_{i\mu\sigma}a_{j\nu\sigma} +
\frac{1}{2}
\sum_{ij,mn}\sum_{\alpha\beta\gamma\delta\sigma\sigma'}
\langle i\alpha,j\beta|W|m\gamma,n\delta \rangle
a^{\dag}_{i\alpha\sigma}a^{\dag}_{j\beta\sigma'}a_{m\gamma\sigma'}a_{n\delta\sigma}.
\end{equation}
It can be rewritten in the following form
\begin{equation}\label{e20a}
\mathcal{H} = H_{1} + H_{2} + H_{3}. \end{equation}
The first term here represents the kinetic energy of
moving electrons
\begin{equation} \label{e21a}
H_{1} = \sum_{ij}
\sum_{\mu\nu\sigma} t^{\mu\nu}_{ij}a^{\dag}_{i\mu\sigma}a_{j\nu\sigma}.
\end{equation} 
The second term $H_{2}$ describes the single-center
Coulomb interaction terms:
\begin{eqnarray} \label{e22a}
H_{2} = \frac{1}{2} \sum_{i\mu\sigma}
U_{\mu\mu}n_{i\mu\sigma}n_{i\mu-\sigma} + \frac{1}{2}
\sum_{i\mu\nu}\sum_{\sigma\sigma'}
V_{\mu\nu}n_{i\mu\sigma}n_{i\nu\sigma'}(1 - \delta_{\mu\nu}) -\\
\nonumber
\frac{1}{2}\sum_{i\mu\nu\sigma}
I_{\mu\nu}n_{i\mu\sigma}n_{i\nu\sigma}(1 - \delta_{\mu\nu})
+
\frac{1}{2}\sum_{i\mu\nu\sigma}
I_{\mu\nu}a^{\dag}_{i\mu\sigma}a^{\dag}_{i\mu-\sigma}a_{i\nu-\sigma}a_{i\nu\sigma}
(1 - \delta_{\mu\nu}) -\\
\nonumber
\frac{1}{2} \sum_{i\mu\nu\sigma}
I_{\mu\nu}a^{\dag}_{i\mu\sigma}a_{i\mu-\sigma}a^{\dag}_{i\nu-\sigma}a_{i\nu\sigma}
(1 - \delta_{\mu\nu}).
\end{eqnarray}
Except for the integral of single-site repulsion $U_{\mu\mu}$,
which is also present in the single-band Hubbard
model, $H_{2}$ also contains three additional contributions
from the interorbital interaction. The last term $H_{3}$
describes the direct inter-site's exchange interaction:
\begin{equation}\label{e23a}
H_{3} = -\frac{1}{2}
\sum_{ij\mu}
\sum_{\sigma\sigma'}
J^{\mu\mu}_{ij}
a^{\dag}_{i\mu\sigma}a_{i\mu-\sigma'}a^{\dag}_{j\mu\sigma'}a_{j\mu\sigma}.
\end{equation}
It is usually assumed that
\begin{equation}\label{e24a}
U_{\mu\mu} = U; \quad V_{\mu\nu} = V; \quad I_{\mu\nu} = I; \quad
J^{\mu\mu}_{ij} = J_{ij}.  
\end{equation}
It is necessary to stress that the Hubbard model is
most closely connected with the \textbf{Pauli exclusion principle},
which in this case can be written as $n^{2}_{i\sigma} = n_{i\sigma}$. Thus,
the Anderson and the Hubbard models take into account
both the collectivized (band) and the localized behavior
of electrons. The problem of the relationship between the
collectivized and the localized description of electrons in
transition and rare-earth metals and in their compounds
is closely connected with another fundamental problem.
The case in point is the adequacy of the simple single-band
Hubbard model, which does not take into account
the interaction responsible for Hund's rules and the
orbital degeneracy for description of magnetic and some
other properties of matter. Therefore, it is interesting to
study various generalizations of the Anderson and the
Hubbard models. In a series of paper~\cite{b41,b151,b155} we
pointed out that the difference between these models is
most clearly visible when we consider dynamic (as
opposed to static) characteristics. Therefore, the
response of the systems to the action of external fields
and the spectra of excited quasiparticle states are of particular
interest. Introduction of additional terms in the
Anderson and the Hubbard model's Hamiltonians makes
the quasiparticle spectrum much more complicated,
leading to the appearance of new excitation branches,
especially in the optical region~\cite{b41,b151,b155}.
%
\subsection{The $s$-$d$ Exchange Model and the Zener model}
%
%
A generalized spin-fermion model, which is also
called the Zener model, or the $s$-$d$- ($d$-$f$)-model is of
primary interest in the solid-state theory. The Hamiltonian
of the $s$-$d$ exchange model~\cite{white06} is given by:
\begin{equation}\label{e19}
\mathcal{H} = H_{s} + H_{s-d},
\end{equation}
\begin{equation}\label{e20}
 H_{s} = \sum_{k\sigma}\epsilon_{k}c^{\dag}_{k\sigma}c_{k\sigma},
\end{equation}
\begin{equation} \label{e21}
H_{s-d} = J {\vec \sigma_{i}}{\vec S_{i}} = - J N^{-1/2}\sum_{kk'} 
\left (c^{\dag}_{k' \uparrow}c_{k \downarrow}S^{-}  + c^{\dag}_{k' \downarrow}c_{k \uparrow}S^{+} +
(c^{\dag}_{k' \uparrow}c_{k \uparrow} - c^{\dag}_{k' \downarrow}c_{k \downarrow})S^{z}  \right ).
\end{equation}
Here, $c^{\dag}_{k\sigma}$ and $c_{k\sigma}$ are the second-quantized operators
creating and annihilating conduction electrons. The
Hamiltonian (\ref{e19}) describes the interaction of the localized
spin of an impurity atom with a subsystem of
the host-metal conduction's electrons. This model is
used for description of the Kondo effect, which is
related to the anomalous behavior of electric conductivity
in metals containing a small amount of transition metal
impurities~\cite{white06,fis71,fis78,hew93}.\\
It is rather interesting to consider a generalized spin-fermion
model, which can be used for description of a
wider range of substances~\cite{white06,fis78,b143}. The Hamiltonian
of the generalized spin-fermion $d$-$f$ model~\cite{b143} is
given by:
\begin{equation} \label{e22}
\mathcal{H} = H_{d} + H_{d-f},  
\end{equation} 
\begin{equation} \label{e23}
H_{d} = \sum_{ij} \sum_{\sigma}
t_{ij}a^{\dag}_{i\sigma}a_{j\sigma} + \frac{1}{2} \sum_{i\sigma}
Un_{i\sigma}n_{i-\sigma}. 
\end{equation} 
The $H_{d-f}$ operator describes the interaction of a subsystem
of strongly localized $4f(5f)$-electrons with the
spin density of collectivized $d$-electrons.
\begin{equation} \label{e24}
 H_{d-f} = \sum_{i}J{\vec \sigma_{i}}{\vec S_{i}}
 = - J
N^{-1/2}\sum_{kq}\sum_{\sigma}[S^{-\sigma}_{-q}a^{\dagger
}_{k\sigma} a_{k+q-\sigma} + z_{\sigma}S^{z}_{-q}a^{\dagger
}_{k\sigma}a_{k+q\sigma}].
\end{equation}
The sign factors $z_{\sigma}$, introduced here for convenience,
are given by
\begin{equation}\label{e25}
z_{\sigma} = (+ ,-); - \sigma =  (\uparrow  ,  \downarrow); \quad
S^{-\sigma}_{-q} =  \begin{cases}
S^{-}_{-q},  & - \sigma = +, \\
S^{+}_{-q} & - \sigma = -.
\end{cases}
\end{equation}  
In the general case, the indirect exchange integral $J$
depends significantly on the wave vector $J(\vec k; \vec k+ \vec q)$
and attains the maximum value at the point $k=q=0$.
Note that the conduction electrons from the metal
$s$-band are also taken into account by the model, and
their role is the renormalization of the model parameters due to screening and other effects. 
Note that the
Hamiltonian of the $s$-$d$ model is a low-energy realization
of the Anderson model. This can be demonstrated
by applying the Schrieffer-Wolf canonical transformation~\cite{white06,hew93,wig83,gerd86}
to the latter model.
%
\subsection{Falicov-Kimball Model}
%
In 1969, Falicov and Kimball proposed a "\emph{simple}"
(in their opinion) model for description of the metal-insulator
transition in rare-earth metal compounds.
This model describes two subsystems, namely, the band
and the localized electrons and their interaction with
each other. The Hamiltonian of the Falicov-Kimball
model~\cite{fk69} is given by
\begin{equation} \label{fak1}
\mathcal{H} = H_{0} + H_{int},  
\end{equation} 
where
\begin{equation} \label{fak2}
 H_{0} = \sum_{k} \sum_{\nu \sigma} \epsilon_{\nu}(\vec{k})a^{\dag}_{\nu k \sigma}a_{\nu k\sigma} + 
\sum_{i} \sum_{\sigma} E b^{\dag}_{i\sigma}b_{i\sigma'}.
\end{equation} 
Here, $a^{\dag}_{\nu k \sigma}$  is the operator creating in the band $\nu$ an
electron in the state with the momentum $\vec{k}$ and the spin $\sigma$, 
and $b^{\dag}_{i\sigma}$  is the operator creating an electron (hole)
with the spin $\sigma$ in the Wannier state at the lattice site $\vec{R}_{i}.$
The energies $\epsilon_{\nu}(\vec{k})$ and $E$ are positive and such that
$\min [E + \epsilon_{\nu}(\vec{k})] > 0.$  It is assumed that due to screening
effects only intra-atomic interactions play a significant
role. Falicov and Kimball~\cite{fk69} took into account six
different types of intra-atomic interactions, and
described them by six different interaction integrals $G_{i}.$
In a simplified mean-field approximation the Hamiltonian
of the model (\ref{fak1}) was given by
\begin{equation} \label{fak3}
\mathcal{H} = N[\epsilon n_{a} + En_{b} - G n_{a}n_{b}],  
\end{equation} 
where $n_{b} = N^{-1} \sum_{i\sigma}  b^{\dag}_{i\sigma}b_{i\sigma}.$
Then, one can calculate the
free energy of the system, and to investigate the transition
of the first-order semiconductor-metal phase.
The Falicov-Kimball model together with its various
modifications and generalizations became very
popular~\cite{fk72,tk94,fkm94,fkm96,fkm97,fkm03,fkm05} in investigating various aspects of
the  theory of phase transitions, in particular, the metal-insulator
transition. It was also used in investigations of  
compounds with mixed valence, and as a crystallization
model. Lately, the Falicov-Kimball model was used in
investigations of electron ferroelectricity (EFE)~\cite{bat02}.
It also turned out that the behavior of a wide class of
substances can be described in the framework of this
model. This class includes, for instance, the compounds
$YbInCu_{4}$, $Eu Ni_{2}(Si_{1-x}Ge_{x})_{2}$, $NiI_{2}$, $Ta_{x}N.$  Thus, the Falicov-Kimball 
model is a microscopic model of the
metal-insulator phase's transition; it takes into account
the dual band-atomic behavior of electrons. Despite the
apparent simplicity, a systematic investigation of this
model, as well as of the Hubbard model, is very difficult,
and it is still intensively studied~\cite{fk72,tk94,fkm94,fkm96,fkm97,fkm03,fkm05}.
%
\subsection{The Adequacy of the Model Description}
%
%
As one can see, the Hamiltonians of $s-d$- and $d-f$- models
especially, clearly demonstrate the manifestation
of collectivized (band) and the localized behavior
of electrons. The Anderson, Hubbard, Falicov-Kimball,
and spin-fermion models are widely used for
description of various properties of the transition and
rare-earth metal compounds~\cite{b41,b151,kuz93,b36,b118,b127,b128,b143,b23,b31,b70,b80}.
In particular, they are applied for
description of various phenomena in the chemical adsorption
theory~\cite{chem83}, surface magnetism, in the theory
of the quantum diffusion in solid $He^{3}$, for description
of vacancy motion in quantum crystals, and the
properties of systems containing heavy fermions~\cite{white06,gerd86,hew93,kura,flo05}.
 The latter problem is especially
interesting and it is still an unsolved problem of the
physics of condensed matter. Therefore, development
of a systematic theory of correlation effects, and
description of the dynamics in the many-particle models
(\ref{e16}) - (\ref{e18}), (\ref{e19}) and (\ref{e22}),   were and still are very
interesting problems. All these models are different
\textbf{description languages}, different ways of describing
similar many-particle systems. They all try to give an
answer on the following questions: how the wave functions
of, formerly, valence electrons change, and how
large the effects of changes are; how strongly do they
delocalize? Their applicability in concrete cases
depends on the answers to those questions. On the
whole, applications of the above mentioned models
(and their combinations) allow one to describe a very
wide range of phenomena and to obtain qualitative, and
frequently quantitative, correct results. Sometimes (but
not always) very difficult and labor-intensive computations
of the electron band's structure add almost nothing
essential to results obtained in the framework of the
schematic and crude models described above.
In investigations of concrete substances, transition
and rare-earth metals and their compounds, actinides,
uranium compounds, magnetic semiconductors, and
perovskite-type manganites, most of the described
above models (or their combinations) are used to a
greater or lesser degree. This reflects the fact that the
electron states, which are of interest to us, have a dual
collectivized and localized character and can not be
described in either an entirely collectivized or entirely
localized form. As far back as 1960, C. Herring~\cite{cher60}, in
his paper on the $d$-electron states in transition metals,
stressed the importance of a "cocktail" of different
states. This is why efforts of many researchers are
directed towards building synthetic models, which take
into account the dual band-atomic nature of transition
and rare-earth metals and their compounds.\\  It was not
by accident, that E. Lieb~\cite{lieb69} made the following
statement: "Search for a model Hamiltonian describing 
collectivized electrons, which, at the same time, is
capable of describing correctly ferromagnetic properties,
is one of the main current problems of statistical
mechanics. Its importance can be compared to such
widely known recent achievements, as the proof of the
existence of extensive free energy for macroscopically
large systems" (see also~\cite{lieb93,lieb05,tasa07}). Solution of
this problem is a part of the more general task of a unified
quantum-statistical description of electrical, magnetic,
and superconducting properties of transition and
rare-earth metals, their alloys, and compounds. Indeed,
the dual band-atomic character of $d$- and, to some
extent, $f$-states manifests itself not only in various
magnetic properties, but also in superconductivity, as
well as in the electrical and thermal conduction processes.\\
The Nobel Prize winner K.G. Wilson noticed~\cite{wils}:
"There are a number of problems in science which
have, as a common characteristic, a complex microscopic
behavior that underlies macroscopic effects"
(see also~\cite{pnas00,pines05}). Eighty years since the formulation
of the Heisenberg model (in 1928), we still do
not have a complete and systematic theory, which
would allow us to give an unambiguous answer to the
question~\cite{mbspt}: "\emph{Why is iron magnetic}?"  Although
over the past decades the physics of magnetic phenomena
became a very extensive domain of modern physics,
and numerous complicated phenomena taking
place in magnetically ordered substances found a satisfactory
explanation, nevertheless recent investigations
have shown that there are still many questions that
remain without an answer. The model Hamiltonians
described above were developed to provide an understanding
(although only a schematic one) of the main
features of real-system behavior, which are of interest
to us. It is also necessary to stress, that the two types of
electronic states, the collectivized and the localized
ones, do not contradict each other, but rather are complementary
ways of quantum mechanical description of
electron states in real transition and rare-earth metals
and in their compounds. In some sense, all the Hamiltonians
described above can be considered as a certain
special extension of the Hubbard Hamiltonian that
takes into account additional crystal subsystems and
their mutual interaction.\\  The variety of the available
models reflects the diversity of magnetic, electrical, and
superconductivity properties of matter, which are of
interest to us. We would like to stress that \textbf{the creation
of physical models} is one of the essential features
of modern theoretical physics~\cite{pei80}. According to
Peierls, "various models serve absolutely different purposes
and their nature changes accordingly \ldots A common
element of all these different types of models is the
fact, that they help us to imagine more clearly the
essence of physical phenomena via analysis of simplified
situations, which are better suited for our intuition.
These models serve as footsteps on the way to the rational
explanation of real-world phenomena \ldots We can
take those models, turn them around, and most likely
we would obtain a better idea on the form and structure
of real objects, than directly from the objects themselves"~\cite{pei80}.
The development of the physics of magnetic
phenomena~\cite{mizia,kagzuk,rscom08} proves most convincingly
the validity of Peierls' conclusion.
%
\section{Theory of Many-particle Systems with Interactions}
%
The research program, which later became known
as the theory of many-particle systems with interaction,
began to develop intensively at the end of 1950s --
beginning of 60s~\cite{dth61}. Due to the efforts of numerous
researchers: F. Bloch, H. Fr\"{o}hlich, J. Bardeen,
N.N. Bogoliubov, H. Hugenholtz, L. Van Hove,
D. Pines, K. Brueckner, R. Feynman, M. Gell-Mann,
F. Dyson, R. Kubo, D. ter Haar, and many others, this
theory achieved significant successes in solving many
difficult problems of the physics of condensed matter~\cite{noz,pines,orl,maha}.
The book~\cite{lh02} contains an interesting details and the
story about the development of some aspects of
the theory of many-particle systems with interaction,
and about its applications to solid-state physics. For a
long time the perturbation theory (in its most diverse
formulations) remained the main method for theoretical
investigations of many-particle systems with interaction.
In the framework of that theory, the complete
Hamiltonian   of a macroscopic system under investigation
was represented as a sum of two parts, the
Hamiltonian of a system of noninteracting particles and
a weak perturbation:
\begin{equation} 
\label{e26} {\mathcal H} = H_{0} + V.
\end{equation}
In many practically important cases such approach
was quite satisfactory and efficient. Theory of many particle
systems found numerous applications to concrete
problems, for instance, in solid-state physics,
plasma, superfluid helium theory, to heavy nuclei, and
many others. It is intensive development of the theory
of many-particle systems that led to development of the
microscopic superconductivity theory~\cite{bar62,nnb58}.
Quite possibly, this was historically the first microscopic
theory based on a sound mathematical foundation~\cite{btsh58,nnb60,haag62,petr70}.
The development of the many-particle
systems theory led to adaptation of many methods from
quantum field theory to problems in statistical mechanics.
Among the most important adaptations are the
methods of Green's functions~\cite{js1,js2,js3,matsu}, and the diagram
technique~\cite{agd}. However, as the range of problems
under investigation widened, the necessity to go
beyond the framework of perturbation theory was felt
more and more acutely. This became a pressing necessity
with the beginning of theoretical investigations of
transition and rare-earth metals and their compounds,
metal-insulator transitions~\cite{mottmi}, and with the development
of the quantum theory of magnetism. This necessity
to go beyond the perturbation theory's framework
was felt by the founders of the Green's functions theory
themselves. Back in 1951 J. Schwinger wrote~\cite{js1}:\\
"\textbf{... it is desirable to avoid founding the formal theory
of the Green's functions on the restricted basis provided
by the assumption of expandability in powers of
coupling constants}."\\
Since the most important point of the theory of
many-particle systems with interaction is an adequate
and accurate treatment of the interaction, which can
change (sometimes quite significantly) the character of
the system behavior, in comparison to the case of noninteracting
particles, the above remark by J. Schwinger
seems to be quite farsighted. It is interesting to note,
that, apparently, admitting the prominent role of
J. Schwinger in development of the Green's functions
method, N.N. Bogoliubov in his paper~\cite{nnb54} uses the
term Green-Schwinger function (for an interesting
analysis of the origin of the Green's functions method
see the paper~\cite{ss05} and also the book~\cite{mehr}).
As far as the application of the Green's functions
method to the problems of statistical physics is concerned,
here, an essential progress was achieved after
reformulation of the original method in the form of the
two-time temperature Green's functions method.
%
\subsection{Two-time Temperature Green's Functions}
%
%
In statistical mechanics of quantum systems the
advanced and retarded two-time temperature Green's
functions (GF) were introduced by N.N. Bogoliubov
and S.V. Tyablikov~\cite{bt59}. In contrast to the causal GF,
the above function can be analytically continued to the
complex plane. Due to the convenient analytical property
the two-time temperature GF is a very widespread
method in statistical mechanics~\cite{nnb84,tyab,bt59,bbt61,zub60,mat64}. In
order to find the retarded and advanced GF we have to
use a hierarchy of coupled equations of motion together
with the corresponding spectral representations.
Let us consider a many-particle system with the
Hamiltonian ${\mathcal H} = H - \mu N$; here $\mu$
is the chemical potential
and $N$ is the operator of the total number of particles.
If $A(t)$ and $B(t')$   are some operators relevant to the
system under investigation, then their time evolution in
the Heisenberg representation has the following form
\begin{equation}
\label{e27} A(t) = \exp \left( \frac{i{{\mathcal H}}t}{\hbar} \right) A(0) \exp  \left( \frac{- i{\mathcal H}t}{\hbar} \right).
\end{equation}
The corresponding two-time correlation function is
defined as follows:
$$\langle A(t) B(t')\rangle = \textrm{Tr} ( \rho A(t) B(t')), \quad \rho = Z^{-1}\exp(- \beta {\mathcal H}).$$ 
This correlation function has the following property
\begin{eqnarray}
\label{e28}
\langle A(t) B(t')\rangle =  \\ \nonumber
Z^{-1} \textrm{Tr} \left ( \exp(- \beta {\mathcal H})  \exp \left( \frac{i{{\mathcal H}}t}{\hbar} \right) A(0) \exp \left( \frac{- i{\mathcal H}(t-t')}{\hbar} \right )
B(0) \exp \left( \frac{- i{\mathcal H}t'}{\hbar} \right) \right )  \\ \nonumber
= Z^{-1} \textrm{Tr} \left ( \exp(- \beta {\mathcal H}) \exp \left( \frac{ i{\mathcal H}(t-t')}{\hbar} \right ) A(0)  \exp \left( \frac{- i{\mathcal H}(t-t')}{\hbar} \right ) B(0)\right )\\ \nonumber
= \langle A(t-t') B(0)\rangle = \langle A(0) B(t'-t)\rangle.
\end{eqnarray}
Usually it is more convenient to use the following
compact notations $\langle A(t) B \rangle$ and  $\langle B A(t)  \rangle,$  where $t-t'$ is
replaced by  $t$. Since
\begin{equation}
\label{e29}
- \beta {\mathcal H}  + \frac{i{{\mathcal H}}t}{\hbar} = \frac{i{{\mathcal H}}(t + i \hbar \beta)}{\hbar} 
\end{equation}
these two correlation functions are related to each other.
Indeed, we have
\begin{eqnarray}
\label{e30}
\langle A(t) B \rangle =  \\ \nonumber
Z^{-1} \textrm{Tr} \left ( \exp(- \beta {\mathcal H})  \exp \left( \frac{i{{\mathcal H}}t}{\hbar} \right) A  \exp \left( \frac{- i{{\mathcal H}}t}{\hbar} \right) \exp( \beta {\mathcal H})
\exp(- \beta {\mathcal H})B \right )   \\ \nonumber
= 
Z^{-1} \textrm{Tr} \left ( \exp(- \beta {\mathcal H})B  \exp \left( \frac{ i{\mathcal H}(t + i \hbar \beta)}{\hbar} \right ) A \exp \left( \frac{- i{\mathcal H}(t + i \hbar \beta)}{\hbar} \right ) \right )  \\ \nonumber
=  \langle B A(t + i \hbar \beta) \rangle.
\end{eqnarray}
On can consider the correlation function $\langle  B A(t) \rangle$ as the
main one, because one can obtain the other function
$\langle   A(t) B \rangle$ by replacing the variable $t \rightarrow t_{1} = t + i \hbar \beta$ in $\langle  B A(t) \rangle.$
The spectral representation (Fourier transform over $\omega$)
of the function $\langle  B A(t) \rangle$ is defined as follows:
\begin{eqnarray}
\label{e31}
\langle  B A(t) \rangle =  \int^{ + \infty}_{ - \infty} d \omega
\exp [- \frac{i}{\hbar}\omega t] J (B, A;\omega),\\ \nonumber   
J (B, A;\omega) = \frac{1}{2 \pi \hbar}
\int^{ +\infty}_{ - \infty} d t \exp [\frac{i}{\hbar}\omega t] \langle  B A(t) \rangle. 
\end{eqnarray}
Equation (\ref{e31}) is the \textbf{spectral representation} of the
corresponding time correlation function. The quantities
$J (B, A;\omega)$ and $J (A, B;\omega)$   are the \textbf{spectral densities} (or
the spectral intensities). It is convenient to assume that
$\omega = \hbar \omega_{clas},$ where $\omega_{clas}$  is the classical angular frequency.
The time correlation function can be written down
in the following form
\begin{eqnarray}
\label{e32}
\langle  B A(t) \rangle =  Z^{-1} \sum_{nml} \langle n |B |m \rangle 
 \langle m |\exp [\frac{i}{\hbar}{\mathcal H} t]A \exp [\frac{- i}{\hbar}{\mathcal H} t] |l \rangle \langle l |\exp(- \beta {\mathcal H})|n \rangle 
\\ \nonumber  =
Z^{-1} \sum_{nm} \langle n |B |m \rangle \langle m |A |n \rangle \exp(- \beta \epsilon_{n})\exp\left( - \frac{i}{\hbar} ( \epsilon_{n} - \epsilon_{m})t\right),
\end{eqnarray}
where
$${\mathcal H}|n \rangle = \epsilon_{n}|n \rangle, 
\quad \exp [- \frac{i}{\hbar}{\mathcal H} t] |n \rangle = \exp(- \frac{i}{\hbar}  \epsilon_{n}t) |n \rangle.$$
Therefore, taking into account the identity
\begin{equation}
\label{e33}
\frac{1}{2 \pi \hbar}
\int^{ +\infty}_{ - \infty} d t \exp [- \frac{i}{\hbar}(\epsilon_{n} - \epsilon_{m} -\omega) t] = \delta(\epsilon_{n} - \epsilon_{m} -\omega),
\end{equation}
we obtain
\begin{eqnarray}
\label{e34}
J (B, A;\omega) = Z^{-1} \sum_{nm} \langle n |B |m \rangle \langle m |A |n \rangle \exp(- \beta \epsilon_{n})\delta(\epsilon_{n} - \epsilon_{m} - \omega).
\end{eqnarray}
Hence, the Fourier transform of the time correlation
function is given by
\begin{eqnarray}
\label{e35}
\langle  A(t) B  \rangle =  \langle  A B (-t) \rangle  = \int^{ + \infty}_{ - \infty} d \omega
\exp [\frac{i}{\hbar}\omega t] J (A, B; \omega) \\ \nonumber   
=  \int^{ +\infty}_{ - \infty} d \omega J (A, B; - \omega) \exp [- \frac{i}{\hbar}\omega t], 
\end{eqnarray}
where
\begin{eqnarray}
\label{e36}
J (A, B; - \omega) = 
Z^{-1} \sum_{nm} \langle m |A |n \rangle \langle n |B |m \rangle  \exp(- \beta \epsilon_{m})\delta(\epsilon_{m} - \epsilon_{n} + \omega) \\ \nonumber 
= Z^{-1} \sum_{nm} \langle n |B |m \rangle \langle m |A |n \rangle \exp(- \beta \epsilon_{n})\delta(\epsilon_{n} - \epsilon_{m} - \omega)\exp( \beta \omega).
\end{eqnarray}
It is easy to check, that the following identity holds
\begin{eqnarray}
\label{e37}
J (A, B; - \omega) = \exp( \beta \omega) J (B, A; \omega).
\end{eqnarray}
For the spectral density of a higher order correlation
function $\langle B [A(t),{\mathcal H}]_{-}  \rangle$ we obtain
\begin{eqnarray}
\label{e38}
J (B, [A,{\mathcal H}]_{-};  \omega) = \omega J (B, A; \omega),\\ \nonumber 
\omega J (A, B; \omega) = J (A, [{\mathcal H}, B]_{-};  \omega) = J ([A,{\mathcal H}]_{-}, B; \omega),\\ \nonumber 
\ldots \ldots \ldots \ldots \ldots \ldots \ldots \ldots \ldots \ldots \ldots \ldots \ldots \ldots \ldots
\end{eqnarray}
Now we introduce the retarded, advanced, and
causal GF:
\begin{eqnarray}
\label{e39} G^{r}(A,B; t-t') = \langle \langle A(t), B(t')\rangle \rangle^{r} = -i\theta(t -
t')\langle [A(t),B(t')]_{\eta}\rangle, \,  \eta = \pm,
 \\ \label{e40}
G^{a}(A,B; t-t') = \langle \langle A(t), B(t')\rangle \rangle^{a} =
i\theta(t' - t)\langle [A(t),B(t')]_{\eta}\rangle, \, \eta = \pm,  \\
G^{c}(A,B; t-t') = \langle \langle A(t), B(t')\rangle \rangle^{c} = i T \langle A(t)B(t')\rangle =   \\ \nonumber
i\theta(t - t')\langle A(t)B(t')\rangle + \eta i\theta (t'- t) \langle B(t')A(t)\rangle, \,
\eta = \pm. \label{e41}
\end{eqnarray}
Here $\langle \ldots \rangle$  is the average over the grand canonical ensemble,
$\theta(t)$ is the Heaviside step function;  the
square brackets denote either commutator or anticommutator
$(\eta = \pm)$:
\begin{equation} \label{e42}
[A,B]_{- \eta} = AB - \eta BA.
\end{equation}
An important ingredient for GF application is their
temporal evolution. In order to derive the corresponding
evolution's equation, one has to differentiate GF
over one of its arguments. Let us differentiate, for
instance, over the first one, the time $t$. The differentiation
yields the following equation of motion:
\begin{equation}
\label{e43}
id/dt G^{\alpha}(t,t') = \delta (t - t') \langle [A,B]_{\eta}\rangle + \langle \langle [A,{\mathcal H}]_{-}(t), B(t')\rangle \rangle^{\alpha}.
\end{equation}
Here, the upper index $\alpha = r,a,c$  indicates the type of
the GF: retarded, advanced, or causal, respectively.
Because this differential equation contains the delta
function in the inhomogeneous part, it is similar in its
form and structure to the defining equation of Green's
function from the differential equation theory~\cite{duffy}
(about the George Green (1793-1841) creative activity
see a detailed paper~\cite{lub94}). It is this similarity that
allows one to use the term Green's function for the
complicated object defined by Eqs. (\ref{e39}) - (\ref{e41}). It is necessary
to stress that the equations of motion for the
three GF: retarded, advanced, and causal, have the same
functional form. Only the temporal boundary conditions
are different there. The characteristic feature of all
equations of motion for GF is the presence of a higher
order GF (relative to the original one) in the right hand
side. In order to find the higher-order function, one has
to write down the corresponding equation of motion for
the GF $\langle \langle [A,{\mathcal H}](t), B(t')\rangle \rangle$, which will contain a new GF
of even higher order. Writing down consecutively the
corresponding equations of motion we obtain the hierarchy
of coupled equations of motion for GF. In principle,
one can write down infinitely many of such equations
of motion:
\begin{eqnarray}
\label{e44} (i)^{n}d^{n}/dt^{n} G(t,t') = \\ \nonumber
\sum_{k=1}^{n} (i)^{n-k} d^{n-k}/dt^{n-k} \delta (t - t')
\langle [[\ldots [A, \underbrace{{\mathcal H}]...{\mathcal H}]}_{k-1}, B]_{\eta} \rangle  \\  \nonumber +
\langle \langle [[\ldots [A, \underbrace{{\mathcal H}]_{-} \ldots {\mathcal H}]_{-}}_{n}(t), B(t')\rangle \rangle. \nonumber
\end{eqnarray}
The \textbf{infinite hierarchy of coupled equations of
motion for GF} is an obvious consequence of interaction
in many-particle systems. It reflects the fact that
none of the particles (or, no group of interacting particles)
can move independently of the remaining system.
The next task is the solution of the differential equation
of motion for GF. In order to do that one can use
the temporal Fourier transform, as well as the corresponding
boundary conditions, taking into account particular
features of the problem under consideration. The
spectral representation for GF, generalizing Eqs. (\ref{e31}) - (\ref{e34})
is given by
\begin{equation}
\label{e45}
 G^{r}(A,B; t-t') =
(2\pi \hbar)^{-1}\int_{\infty}^{\infty} dE G(A,B; E)
 \exp [-\frac{i}{\hbar}E( t - t')],
\end{equation}
\begin{equation}
 \label{e46}
 G(A,B; E) = \langle \langle A | B \rangle \rangle_{E} =
\int_{\infty}^{\infty} dt G(A,B; t)
 \exp (\frac{i}{\hbar}E t) .
\end{equation}
On substitution of Eq. (\ref{e45}) in Eqs. (\ref{e43}) and (\ref{e44}) one
obtains 
\begin{eqnarray}
\label{e47}
E G(A,B; E) = \langle [A,B]_{\eta}\rangle + \langle \langle [A,{\mathcal H}]_{-} | B \rangle \rangle_{E} ;  \\ \label{e48}
E^{n} G(A,B; E) = \sum_{k=1}^{n} E^{n-k}
\langle [[\ldots [A, \underbrace{{\mathcal H}] \ldots {\mathcal H}]}_{k-1}, B]_{\eta}\rangle \\  \nonumber +
\langle \langle [[\ldots [A, \underbrace{{\mathcal H}]_{-} \ldots {\mathcal H}}_{n}]_{-}| B \rangle \rangle_{E}. \nonumber
\end{eqnarray}
The above hierarchy of coupled equations of motion
for GF (\ref{e48}) is an extremely complicated and nontrivial
object for investigations. Frequently it is convenient to
rederive the same hierarchy of coupled equations of
motion for GF starting from differentiation over the
second time $t'$. The corresponding equations of motion
analogous to Eqs. (\ref{e47}) and (\ref{e48})  are given by
\begin{eqnarray}
\label{e49}
- E G(A,B; E) = - \langle [A,B]_{\eta}\rangle + \langle \langle A | [ B,{\mathcal H} ]_{-} \rangle \rangle_{E} ;  \\
\nonumber
(-1)^{n} E ^{n} G(A,B; E) = - \sum_{k=1}^{n} ( - 1 )^{n-k}E^{n-k}
\langle [A, [\ldots [B, \underbrace{{\mathcal H}] \ldots {\mathcal H}]}_{k-1}]_{\eta}\rangle \\
+ \langle \langle A |[\ldots [ B, \underbrace{{\mathcal H}]_{-} \ldots {\mathcal H}}_{n}]_{-} \rangle \rangle_{E}.
\label{e50}
\end{eqnarray}
The main problem is how to find solutions of the
hierarchy of coupled equations of motion for GF given
by either Eq. (\ref{e48}) or Eq.  (\ref{e50})? In order to approach this
difficult task one has to turn to the method of dispersion
relations, which, as was shown in the papers by
N.N. Bogoliubov and collaborators~\cite{nnb84,bt59,bbt61}, is
quite an effective mathematical formalism. The method
of retarded and advanced GF is closely connected with
the dispersion relations technique~\cite{nnb84}, which allows one
to write down the boundary conditions in the form of a
spectral representation for GF. The spectral representations
for correlation functions were used for the first
time in the paper~\cite{ca} by Callen and Welton (see also~\cite{bbk}) 
devoted to the fluctuation theory and the statistical
mechanics of irreversible processes. GF are combinations
of correlation functions
\begin{eqnarray}
\label{e51} F_{AB} (t-t') = \langle A (t) B(t') \rangle =  \langle A (t - t') B \rangle = \int^{ + \infty}_{ - \infty} d \omega
\exp [ \frac{i}{\hbar}\omega t] J (A, B; \omega), \\
\label{e52}
F_{BA} (t'-t) = \langle B(t') A(t) \rangle =  \langle  B A(t - t') \rangle = \int^{ + \infty}_{ - \infty} d \omega \exp [- \frac{i}{\hbar}\omega t] J (B, A; \omega).
\end{eqnarray}
Therefore, the spectral representations for two-time
temperature Green's functions can be written in the following
form
\begin{eqnarray}
\label{e53}
\langle \langle A|B \rangle \rangle_{{\mathcal E}} = 
\int^{ + \infty}_{ - \infty} d \omega \frac{J (B, A; \omega)(\exp( \beta \omega) - \eta)}{{\mathcal E} - \omega}
= \int^{ + \infty}_{ - \infty} d \omega \frac{J' (B, A; \omega)}{{\mathcal E} - \omega},
\end{eqnarray}
where 
\begin{eqnarray}
\label{e54}
J' (B, A; \omega) = (\exp( \beta \omega) - \eta) J (B, A; \omega)
\end{eqnarray}
and ${\mathcal E}$ is the complex energy ${\mathcal E} = {\rm Re} {\mathcal E} + i {\rm Im} {\mathcal E}. $ \\
Hence,
\begin{eqnarray}\label{e55}
\int^{ + \infty}_{ - \infty} d \omega  \biggl( J (B, A; \omega)\exp( \beta \omega) - \eta J (B, A; \omega) \biggr ) \nonumber \\
= \int^{ + \infty}_{ - \infty} d \omega  \biggl( J (B, A; - \omega) - \eta J (B, A; \omega) \biggr ) = \langle  A B  - \eta B A \rangle.
\end{eqnarray}
Therefore, we obtain the following equation
\begin{eqnarray}\label{e56}
\langle \langle A|B \rangle \rangle_{{\mathcal E}} =  \langle  A B  - \eta B A \rangle + \langle \langle [A,{\mathcal H}]_{-}|B \rangle \rangle_{{\mathcal E}}.
\end{eqnarray}
One should note that the two-time temperature
Green's functions are not defined for $t = t'$; moreover,
$\langle \langle A (t) B(t') \rangle \rangle^{r} = 0$ for $t < t'$ и $\langle \langle A (t) B(t') \rangle \rangle^{a} = 0$ для $t > t'.$
Using the following representations for the step-function
$\theta (t):$ 
\begin{eqnarray}
\label{e57}
\theta (t) = \exp (- \varepsilon t)(\varepsilon \rightarrow 0, \varepsilon > 0 ),\quad t > 0; \quad \theta (t) = 0,\quad t < 0.
\end{eqnarray}
we can rewrite the Fourier transform of the retarded
(advanced) GF in the following form
\begin{eqnarray}
\label{e58}
\lim_{\varepsilon \rightarrow 0} \langle \langle A|B \rangle \rangle_{E \pm i \varepsilon}= G^{r(a)}(A, B; E).
\end{eqnarray}
It is clear that the two functions, $G^{r}(A, B; E) $ and $G^{a}(A, B; E)$, are functions of a real variable $E$; they are
defined as limiting values of the Green's function
$\langle \langle A|B \rangle \rangle_{{\mathcal E}}$ in the upper and lower half-plane, respectively.
According to the Bogoliubov-Parasiuk theorem~\cite{tyab,bt59,bbt61,zub60} the function
\begin{eqnarray}
\label{e59}
\langle \langle A|B \rangle \rangle_{{\mathcal E}} = 
\int^{ + \infty}_{ - \infty} d \omega \frac{J (B, A; \omega)(\exp( \beta \omega) - \eta)}{{\mathcal E} - \omega}
\end{eqnarray}
is an analytic function in the complex ${\mathcal E}$-plane; this
function coincides with $G^{r}(A, B; E)$ everywhere in the
upper half-plane, and with $G^{a}(A, B; E)$ everywhere in
the lower half-plane. It has singularities on the real
axis; therefore, one has to make a cut along the real
axis. Note that $G^{r(a)}(A, B; t)$ is a generalized function in
the Sobolev-Schwartz sense~\cite{tyab,bt59,bbt61,zub60}. The function
$G(A, B; {\mathcal E})$ is an analytic function in the complex
plane with the cut along the real axis. It has two
branches; one is defined in the upper half-plane, the
other in the lower half-plane for complex values of ${\mathcal E}$:
\begin{eqnarray}
\label{e60}
\langle \langle A|B \rangle \rangle_{{\mathcal E}} = 
\begin{cases}
G^{r}(A, B; {\mathcal E}), & {  \textrm{если}}  \quad  {\mathcal E} > 0 ,\cr
 G^{a}(A, B; {\mathcal E}), &  {  \textrm{если}}  \quad  {\mathcal E} < 0.
\end{cases}
\end{eqnarray}
The corresponding Fourier transform is given by
\begin{eqnarray}
\label{e61}
 G^{r(a)}(A, B; t) = (2\pi \hbar)^{-1}\int_{- \infty}^{+ \infty} dE G^{r(a)}(A,B; E)
 \exp [-\frac{i}{\hbar}Et] = \nonumber \\ (2\pi \hbar)^{-1}\int_{- \infty}^{+ \infty} dE \exp [-\frac{i}{\hbar}Et]
\int^{ + \infty}_{ - \infty} d \omega \frac{J' (B, A; \omega)}{E - \omega \pm i\varepsilon  }. 
\end{eqnarray}
Here, $J' (B, A; \omega)$ can be written down as follows $(\varepsilon \rightarrow 0)$
\begin{eqnarray}
\label{e62}
J' (B, A; \omega) = - \frac{1}{2 \pi i}\left ( \langle \langle A|B \rangle \rangle_{\omega + i\varepsilon  } - 
\langle \langle A|B \rangle \rangle_{\omega - i\varepsilon  }\right ).
\end{eqnarray}
Therefore, the spectral representations for the
retarded and the advanced GF are determined by the
following relationships:
\begin{eqnarray}
\label{e63}
G^{r} (A,B; E) = \langle \langle A | B \rangle \rangle^{r}_{\omega + i\varepsilon} = \\ \nonumber
  \int^{ + \infty}_{ - \infty}
\frac {d\omega}{ E - \omega  + i\varepsilon}
 J' (B, A; \omega)    \\ \nonumber
= P  \int^{ + \infty}_{ - \infty} d\omega \frac {J' (B, A; \omega)}{ E - \omega } - i \pi J' (B, A; E),\\
\label{e64}
G^{a} (A,B; E) = \langle \langle A | B \rangle \rangle^{a}_{\omega - i\varepsilon} =  \\
 \int^{ + \infty}_{ - \infty}
\frac {d\omega}{ E - \omega  - i\varepsilon}
 J' (B, A; \omega)   \nonumber \\
= P  \int^{ + \infty}_{ - \infty} d\omega \frac {J' (B, A; \omega)}{ E - \omega } + i \pi J' (B, A; E). \nonumber
\end{eqnarray}
In the derivation of the above equations we made
use of the following relationship~\cite{tyab,bt59,bbt61,zub60}
\begin{eqnarray}
\label{e65}
\lim_{\varepsilon \rightarrow 0} \frac{1}{x \pm i \varepsilon} \rightarrow P \frac{1}{x} \mp i \pi \delta(x).
\end{eqnarray}
Here, $P (1/x)$ indicates that one has to take the principal
value when calculating integrals. As a result we obtain
the following fundamental relationship for the spectral
density
\begin{eqnarray}
\label{e66}
J (B, A; E) = - \frac{1}{2 \pi i}  \frac{G^{r} (A,B; E) - G^{a} (A,B; E)}{\exp( \beta E) - \eta}.  
\end{eqnarray}
Thus, once we know the Green's function $G^{r(a)} (A,B; E)$
we can find $J (B, A; E)$, and then calculate the corresponding correlation function. 
Using the relationship (\ref{e66}), one can obtain the following dispersion relationships:
\begin{eqnarray}
\label{e67}
{\rm Re} G^{r(a)} (A,B; E) = \mp \frac{1}{\pi} P \int^{ + \infty}_{ - \infty} d\omega \frac{{\rm Im} G^{r(a)} (A,B; E)}{E - \omega}.
\end{eqnarray}
The most important practical consequence of the
spectral representations for the retarded and advanced
GF is the so called spectral theorem:
\begin{eqnarray}
\label{e68}
\langle B(t') A(t) \rangle = \\
\nonumber
- \frac {1}{\pi} \int^{ + \infty}_{ - \infty} dE
\exp [\frac{i}{\hbar} E (t-t')] [\exp ( \beta E) - \eta ]^{-1} {\rm Im} G_{AB} (E + i\varepsilon), \\
\label{e69}
 \langle A(t) B(t')\rangle = \\ \nonumber - \frac {1}{\pi} \int^{ +
\infty}_{ - \infty} dE \exp  (\beta E)  \exp [\frac{i}{\hbar}E
(t-t')] [\exp ( \beta E) - \eta ]^{-1} {\rm Im} G_{AB} (E +
i\varepsilon).
\end{eqnarray}
Equations (\ref{e68}) and (\ref{e69}) are of a fundamental importance
for the entire method of two-time temperature
GF. They allow one to establish a connection between
statistical averages and the Fourier transforms of
Green's functions, and are the basis for practical applications
of the entire formalism for solutions of concrete
problems~\cite{nnb84,tyab,bt59,bbt61,zub60}.
%
 \subsection{The Method of Irreducible Green's Functions}    
%
When working with infinite hierarchies of equations
for GF the main problem is finding the methods for
their efficient decoupling, with the aim of obtaining a
closed system of equations, which determine the GF. A
decoupling approximation must be chosen individually
for every particular problem, taking into account its
character. This "individual approach" is the source of
critique for being too \emph{ad hoc}, which sometimes
appear in the papers using the causal GF and diagram
technique. However, the ambiguities are also present in
the diagram technique, when the choice of an appropriate
approximation is made there. The decision, which
diagrams one has to sum up, is obvious only for a narrow
range of relatively simple problems. In the papers~\cite{ze62,ze66,np70,np73}
devoted to Bose-systems, and in the papers
by the author of this review~\cite{kuz02,b36,b91,b118,b128,b143,b101,b132,b142}  
devoted to Fermi systems it was shown that for a
wide range of problems in statistical mechanics and
theory of condensed matter one can outline a fairly systematic
recipe for constructing approximate solutions
in the framework of irreducible Green's functions
method. Within this approach one can look from a unified
point of view at the main problems of fundamental
characters arising in the method of two-time temperature
GF. \\ The method of irreducible Green's functions is
a useful reformulation of the ordinary Bogoliubov-
Tyablikov method of equations of motion. The constructive
idea can be summarized as follows. During
calculations of single-particle characteristics of the system
(the spectrum of quasiparticle excitations, the density
of states, and others) it is convenient to begin from
writing down GF (\ref{e39}) as a formal solution of the Dyson
equation. This will allow one to perform the necessary
decoupling of many-particle correlation functions in
the mass operator. This way one can to control the
decoupling procedure conditionally, by analogy with
the diagrammatic approach. The method of irreducible
Green's functions is closely related to the Mori-Zwanzig's
projection method~\cite{zwa60,zwan01,mori65,hmori65,ichi,ze81,howl}, which essentially
follows from Bogoliubov's idea about the reduced
description of macroscopic systems~\cite{nnb46}. In this
approach the infinite hierarchy of coupled equations for
correlation functions is reduced to a few relatively simple
equations that effectively take into account the
essential information on the system under consideration,
which determine the special features of this concrete
problem. \\ It is necessary to stress that the structure
of solutions obtained in the framework of irreducible
GF method is very sensitive to the order of equations
for GF~\cite{kuz02,b91} in which irreducible parts are separated.
This in turn determines the character of the approximate
solutions constructed on the basis of the exact representation.
In order to clarify the above general description, let
us consider the equations of motion (\ref{e47}) for the
retarded GF (\ref{e39}) of the form $ \langle \langle A(t), A^{\dagger}(t')\rangle \rangle $ 
\begin{equation}
\label{e70} \omega G(\omega) = \langle [A, A^{\dagger}]_{\eta} \rangle + \langle \langle [A,
H]_{-} \vert A^{\dagger}\rangle \rangle_{\omega}. 
\end{equation}
The irreducible (\textbf{ir}) GF is defined by
\begin{equation}
\label{e71} ^{(ir)}\langle \langle[A, H]_{-}\vert A^{\dagger}\rangle \rangle = \langle \langle [A, H]_{-}
- zA\vert A^{\dagger}\rangle \rangle.
\end{equation}
The unknown constant $z$ is found from the condition
\begin{equation}
\label{e72} \langle [^{(ir)}[A, H]_{-}, A^{\dagger}]_{\eta} \rangle = 0.
\end{equation}
In some sense the condition (\ref{e72}) corresponds to the
orthogonality conditions within the Mori formalism~\cite{zwa60,zwan01,mori65,hmori65,ichi,ze81,howl}.
 It is necessary to stress, that instead of finding
the irreducible part of GF $(^{(ir)}\langle \langle [A, H]_{-}\vert A^{\dagger}\rangle \rangle)$ , one can
absolutely equivalently consider the irreducible operators
$(^{(ir)}[A, H]_{-}) \equiv ([A, H]_{-})^{(ir)}$. Therefore, we will use
both the notation ($ ^{(ir)}\langle \langle A  \vert B \rangle \rangle$) and $\langle \langle (A)^{(ir)}  \vert B \rangle \rangle$), whichever
is more convenient and compact. Equation (\ref{e72})
implies
\begin{equation}
\label{e73} z = \frac{\langle [[A, H]_{-}, A^{\dagger}]_{\eta} \rangle}{\langle [A,
A^{\dagger}]_{\eta}\rangle} =
 \frac{M_{1}}{M_{0}}.
\end{equation}
Here, $M_{0}$ and $M_{1}$   are the zero and first moments of the
spectral density~\cite{tyab,bt59,bbt61,zub60}. Green's function is
called \textbf{irreducible} (i.e. impossible to reduce to a desired, simpler, or smaller form or amount) if it cannot 
be turned into a lower order GF via decoupling. The well-known objects in
statistical physics are irreducible correlation functions
(see, e.g. papers~\cite{zwan01,pokr}). In the framework of the diagram
technique~\cite{agd} the irreducible vertices are a set of
graphs, which cannot be cut along a single line. The
definition (\ref{e71}) translates these notions to the language
of retarded and advanced Green's functions. We
attribute all the mean-field renormalizations that are
separated by Eq. (\ref{e71}) to GF within a generalized mean field
approximation
\begin{equation}
\label{e74} G^{0}(\omega) = \frac{\langle [A,
A^{\dagger}]_{\eta}\rangle}{(\omega - z)}.
\end{equation}
For calculating GF (\ref{e71}), $\,  ^{(ir)}\langle \langle [A, H]_{-}(t),A^{\dagger}(t')\rangle \rangle,$   we
make use of differentiation over the second time $t'$.
Analogously to Eq. (\ref{e71}) we separate the irreducible
part from the obtained equation and find
\begin{equation} \label{e75} 
G(\omega) = G^{0}(\omega) + G^{0}(\omega) P(\omega) G^{0}(\omega).
\end{equation}
Here, we introduced the scattering operator
\begin{equation}
\label{e76} P = (M_{0})^{-1}\biggl (\langle \langle([A,
H]_{-})^{(ir)}   \vert ([A^{\dagger}, H]_{-})^{(ir)} \rangle \rangle \biggr ) (M_{0})^{-1}.
\end{equation}
In complete analogy with the diagram technique one
can use the structure of Eq. (\ref{e75}) to define the mass
operator $M$:
\begin{equation} \label{e77} P = M + M G^{0} P.
\end{equation}
As a result we obtain the exact Dyson equation (we
did not perform any decoupling yet) for two-time temperature
GF:
\begin{equation} \label{e78} G =
G^{0} + G^{0} M G.
\end{equation}
According to Eq. (\ref{e77}), the mass operator $M$ (also
known as the self-energy operator) can be expressed in
terms of the \textbf{proper} (called \emph{connected} within the diagram
technique) part of the many-particle irreducible
GF. This operator describes inelastic scattering processes,
which lead to damping and to additional renormalization
of the frequency of self-consistent quasiparticle
excitations. One has to note that there is quite a
subtle distinction between the operators $P$ and $M$. Both
operators are solutions of two different integral equations
given by Eqs. (\ref{e77}) and (\ref{e78}), respectively. However,
only the Dyson equation (\ref{e78}) allows one to write
down the following formal solution for the GF:
\begin{equation}
\label{e79} G = [ (G^{0})^{-1} - M ]^{-1}.
\end{equation}
This fundamental relationship can be considered as
an alternative form of the Dyson equation, and as the
\textbf{definition} of the mass operator under the condition that
the GF within the generalized mean-field approximation,
$G^{0}$, was appropriately defined using the equation
\begin{equation}
\label{e80} G^{0}G^{-1} + G^{0}M = 1.
\end{equation}
In contrast, the operator $P$ does not satisfy Eq. (\ref{e80}).
Instead we have
\begin{equation}
\label{e81} (G^{0})^{-1} - G^{-1} = P G^{0}G^{-1}.
\end{equation}
Thus, it is \textbf{the functional structure} of Eq. (\ref{e79}) that
determines the essential differences between the operators
$P$ and $M$. To be absolutely precise, the definition
(\ref{e77}) has a symbolic character. It is assumed there that
due to the similar structure of equations (\ref{e39}) - (\ref{e41})
defining all three types of GF, one can use the causal GF
at all stages of calculation, thus confirming the sensibility
of the definition (\ref{e77}). Therefore, one should rather
use the phrase "an analogue of the Dyson equation".
Below we will omit this stipulation, because it will not
lead to misunderstandings. One has to stress that the
above definition of irreducible parts of the GF (irreducible
operators) is nothing but a general scheme. The
specific way of introducing the irreducible parts of the
GF depends on the concrete form of the operator $A$ on
the type of the Hamiltonian, and on the problem under
investigation. \\ Thus, we managed to reduce the derivation
of the complete GF to calculation of the GF in the
generalized mean-field approximation and with the
generalized mass operator. The essential part of the
above approach is that the approximate solutions are
constructed not via decoupling of the equation-of-motion
hierarchy, but via choosing the functional form
of the mass operator in an appropriate self-consistent
form. That is, by looking for approximations of the
form $ M \approx F[G]$. Note that the exact functional structure
of the one-particle GF (\ref{e79}) is preserved in this
approach, which is quite an essential advantage in comparison
to the standard decoupling schemes.
%
\section{The Generalized Mean Fields}
%
%
Apparently, the mean field concept was originally
formulated for many-particle systems (in an implicit
form) in Van der Waals (1837-1923) Ph.D. thesis "On
the Continuity of Gaseous and Liquid States". This
classical paper was published in 1874 and became
widely known~\cite{kip85}. At first, Van der Waals expected
that the volume correction to the equation of state
would lead only to an obvious reduction of the available
space for the molecular motion by an amount $b$ equal to
the overall volume of the molecules. However, the
actual situation turned out to be much more complicated.
It was necessary to take into account both corrections,
the volume correction $b$, and the pressure correction
$a/V^2$, which led to the Van der Waals equation~\cite{nov75}.
Thus, Van der Waals realized that "the range of
attractive forces contains many neighboring molecules".
The development of this approach led to the
insight, that one can try to describe the complex many-particle
behavior of gases, liquids, and solids in terms
of a single particle moving in an average (or effective)
field created by all the other particles, considered as
some homogeneous (or inhomogeneous) environment.
That is, the many-particle behavior was reduced to
effective (or renormalized) behavior of a single particle
in a medium (or a field). Later, these ideas were
extended to the physics of magnetic phenomena, where
magnetic substances were considered as some kind of a
peculiar liquid. That was the origin of the terminology
magnetically soft and hard materials. Beginning from
1907 the Weiss molecular-field approximation~\cite{pweis}
became widespread in the theory of magnetic phenomena~\cite{smart},
and even at the present time it is still being
used efficiently~\cite{ivan07}. Nevertheless, back in 1965 it
was noticed that~\cite{hbcal65}
\begin{quote}
"The Weiss molecular field theory plays an enigmatic
role in the statistical mechanics of magnetism".
\end{quote}
In order to explain the concept of the molecular
field on the example of the Heisenberg ferromagnet
one has to transform the original many-particle
Hamiltonian (\ref{e12}) into the following reduced one-particle
Hamiltonian
\begin{equation} 
\mathcal{H}  = -   2 \mu_{0} \mu_{B} \vec{\textbf{S}} \cdot \vec{\textbf{h}}^{(mf)}.  \nonumber
\end{equation}
This transformation is achieved with the help of the
identity
$$\vec{S} \cdot \vec{S'} = \vec{S}\cdot \langle \vec{S'}\rangle + \langle \vec{S} \rangle \cdot \vec{S'} - 
\langle \vec{S} \rangle \cdot \langle \vec{S'}\rangle + C.$$
Here, the constant $C = (\vec{S} - \langle \vec{S} \rangle) \cdot (\vec{S'} - \langle \vec{S'} \rangle)$ describes
spin correlations. The usual molecular-field approximation
is equivalent to discarding the third term in the
right hand side of the above equation, and using the
approximation $C \sim \langle C \rangle = 
\langle \vec{S} \cdot \vec{S'} \rangle - \langle \vec{S} \rangle \cdot \langle \vec{S'}\rangle.$ for the
constant $C$. \\ Let us consider this point in more detail. It
is instructive to trace the evolution of the mean or concept
of the molecular field for different systems. The
list of some papers, which contributed to the development
of the mean-field concept, is presented in Table 1.
%
%
\protect\begin{table} \label{tab1}
\caption{ \textbf{The development of the mean-field concept}}
\begin{tabular}{|l|c|r|} \hline \hline 
 Mean-field type  &Author&Year\cr \hline \hline  
A homogeneous molecular field  in dense gases &J.D. Van der Waals&1873 \\
\hline  
A homogenous quasi-magnetic   mean-field\\ in magnetics&P.Weiss&1907 \\ 
\hline  
A mean-field in atoms:\\ the Thomas-Fermi model 
&L.H.Thomas, E.Fermi&1926-28 \\ 
\hline
A homogeneous mean-field \\ in many-electron atoms&D.Hartree, V.A. Fock&1928-32 \\
\hline 
A molecular field in ferromagnets
&Ya. G. Dorfman, F.Bloch, &1927 - 1930 \\
\hline
Inhomogeneous (local) mean-fields\\ in antiferromagnets&L.Neel&1932 \\ 
\hline 
A molecular field, taking into account\\ the cavity reaction in polar substances&L.Onsager&1936 \\
\hline
The Stoner model of band magnetics& E.Stoner&1938 \\
\hline
Generalized mean-field approximation\\ in many-particle systems&T.Kinoshita, Y. Nambu&1954 \\ 
\hline
The BCS-Bogoliubov mean-field\\ in superconductors &N.N. Bogoliubov&1958 \\ 
\hline
The Tyablikov decoupling for ferromagnets & S. V. Tyablikov&1959 \\ 
\hline
The mean-field theory for the Anderson model& P.W.Anderson&1961 \\ 
\hline
The density functional theory for electron gas&W.Kohn&1964 \\ 
\hline
The Callen decoupling for ferromagnets&H.B.Callen&1963 \\ 
\hline
The alloy analogy (mean-field)\\ for the Hubbard model&J.Hubbard&1964 \\ 
\hline
The generalized H-F approximation\\ for the Heisenberg model&Yu.A. Tserkovnikov, Yu.G. Rudoi&1973-1975 \\
\hline 
A generalized mean-field approximation\\ for ferromagnets  &N.M. Plakida&1973 \\
\hline
A generalized mean-field approximation\\ for the Hubbard model&A.L. Kuzemsky&1973-2002 \\
\hline
A generalized mean-field approximation\\ for antiferromagnets&A.L. Kuzemsky, D. Marvakov&1990 \\ 
\hline
A generalized random-phase approximation\\ in the theory of ferromagnets& A.Czachor, A.Holas&1990 \\ 
\hline
A generalized mean-field approximation\\ for band antiferromagnets&A.L. Kuzemsky&1999 \\
\hline
The Hartree-Fock-Bogoliubov mean-field\\ in Fermi systems &N.N. Bogoliubov, Jr.&2000 \\ 
\hline
\end{tabular}
%
\end{table}
%
%
A brief look at that table allows one to notice a certain
tendency. Earlier molecular-field concepts
described the mean-field in terms of some functional of
the average density of particles $\langle n \rangle$ (or, using the magnetic
terminology, the average magnetization $\langle M \rangle$), that
is, as $F[\langle n \rangle, \langle M \rangle]$. Using the modern language, one can
say that the interaction between the atomic spins $S_{i}$
and their neighbors can be equivalently described by
effective (or mean) field $ h^{(mf)}$. As a result one can write
down
$$ M_{i} = \chi_{0} [ h_{i}^{(ext)} + h_{i}^{(mf)} ]. $$
The mean field $ h^{(mf)}$ can be represented in the form
(in the case $T >  T_{c}$)
\begin{equation}
\label{e82} h^{(mf)} = \sum_{i} J(R_{ji})\langle S_{i} \rangle.
\end{equation}
Here, $ h^{ext}$ is the external magnetic field, $\chi_{0}$ is the system's
response function, and $ J(R_{ji})$ is the interaction
between the spins. In other words, in the mean-field
approximation a many-particle system is reduced to the
situation, where the magnetic moment at any site aligns
either parallel or anti-parallel to the overall magnetic
field, which is the sum of the applied external field and
the molecular field. \\ Note that only the "\textbf{averaged}" interaction
with $i$ neighboring sites is taken into account,
while the fluctuation effects are ignored. We see that the
mean-field approximation provides only a rough
description of the real situation and overestimates the
interaction between particles. Attempts to improve the
homogeneous mean-field approximation were undertaken
along different directions~\cite{beh03}. \\ An extremely
successful and quite nontrivial approach was developed
by L. Neel~\cite{neel}, who essentially formulated the concept
of \textbf{local mean fields} (1932). Neel assumed that the sign
of the mean-field could be both positive and negative.
Moreover, he showed that below some critical temperature
(the Neel temperature) the energetically most
favorable arrangement of atomic magnetic moments is
such, that there is an equal number of magnetic
moments aligned against each other. This novel magnetic
structure became known as the \textbf{antiferromagnetism}~\cite{afm56}.
It was established that the antiferromagnetic
interaction tends to align neighboring spins
against each other. In the one-dimensional case this corresponds
to an alternating structure, where an "\textbf{up}" spin
is followed by a "\textbf{down}" spin, and vice versa. Later it
was conjectured that the state made up from two
inserted into each other sublattices is the ground state of
the system (in the classical sense of this term). Moreover,
the mean-field sign there alternates in the "chessboard"
(staggered) order. \\ The question of the true antiferromagnetic
ground state is not completely clarified
up to the present time~\cite{ander52,tau67,casp80,misg02,kurk05}. This is related to the
fact that, in contrast to ferromagnets, which have a
unique ground state, antiferromagnets can have several
different optimal states with the lowest energy. The
Neel ground state is understood as a possible form of
the system's wave function, describing the antiferromagnetic
ordering of all spins~\cite{kurk05}. Strictly speaking,
the ground state is the thermodynamically equilibrium
state of the system at zero temperature. Whether the
Neel state is the ground state in this strict sense or not,
is still unknown. It is clear though, that in the general
case, the Neel state is not an eigenstate of the Heisenberg
antiferromagnet's Hamiltonian. On the contrary,
similar to any other possible quantum state, it is only
some linear combination of the Hamiltonian eigenstates.
Therefore, the main problem requiring a rigorous
investigation is the question of Neel state's~\cite{kurk07}
stability. In some sense, only for infinitely large lattices,
the Neel state becomes the eigenstate of the Hamiltonian
and the ground state of the system. Nevertheless,
the sublattice structure is observed in experiments on
neutron scattering~\cite{ml71}, and, despite certain worries~\cite{vons71},
the actual existence of sublattices~\cite{subl61} is beyond
doubt.\\
Once Neel's investigations were published, the
effective mean-field concept began to develop at a
much faster pace. An important generalization and
development of this concept was proposed in 1936 by
L. Onsager~\cite{onsag36} in the context of the polar liquid theory.
This approach is now called \textbf{the Onsager reaction field}
approximation. It became widely known, in particular,
in the physics of magnetic phenomena~\cite{wysin1,wysin2,wysin3,medv07}.
In 1954, Kinoshita and Nambu~\cite{namb54} developed a systematic method
for description of many-particle systems in the framework
of an approach which corresponds to the \textbf{generalized
mean-field} concept. Later, various schemes of
"effective mean-field theory taking into account correlations"
were proposed (see the review~\cite{kuz02}). One can
show in the framework of the variation principle~\cite{tyab,fey81,sold}
that various mean-field approximations can be
described on the basis of the Bogoliubov inequality~\cite{nnb84}:
\begin{eqnarray}
\nonumber
 F =  - \beta^{-1} \ln (\textrm{Tr} e^{-\beta H } ) \leq \\ -
\beta^{-1} \ln (\textrm{Tr} e^{-\beta H^{mf}}) + \frac {\textrm{Tr} e^{ - \beta
H^{mf} } ( H - H^{mf})}{\textrm{Tr} e^{- \beta H^{mf} }}. \label{e83}
\end{eqnarray}
Here, $F$ is the free energy of the system under consideration,
whose calculation is extremely involved in the
general case. The quantity $H^{mf}$ is some trial Hamiltonian
describing the effective-field approximation. The
inequality (\ref{e83}) yields an upper bound for the free
energy of a many-particle system. One should note that
the BCS-Bogoliubov superconductivity theory~\cite{bar62,nnb58,btsh58,nnb60}
is formulated in terms of a trial (approximating)
Hamiltonian, which is a quadratic form with respect to
the second-quantized creation and annihilation operators,
including the terms responsible for anomalous (or
non-diagonal) averages. For the single-band Hubbard
model the BCS-Bogoliubov functional of generalized
mean fields can be written in the following form~\cite{kuz02}
\begin{equation} \label{e84}
\Sigma^{c}_{\sigma} =  U 
\begin{pmatrix}   \langle a^{\dagger}_{i-\sigma}
a_{i-\sigma} \rangle & - \langle a_{i\sigma} a_{i-\sigma} \rangle \cr
- \langle a^{\dagger}_{i-\sigma} a^{\dagger}_{i\sigma} \rangle &
- \langle a^{\dagger}_{i\sigma} a_{i\sigma} \rangle \cr  
\end{pmatrix}. 
\end{equation}
The anomalous (or nondiagonal) mean values in this
expression fix the vacuum state of the system exactly in
the BCS-Bogoliubov form. A detailed analysis of
Bogoliubov's approach to investigations of (Hartree-
Fock-Bogoliubov) mean-field type approximations for
models with a four-fermion interaction is given in the
papers~\cite{nbog94,bog90}.\\
There are many different approaches to construction
of generalized mean-field approximations; however, all
of them have a special-case character. The method of
irreducible Green's functions allows one to tackle this
problem in a more systematic fashion. In order to clarify
this statement let us consider as an example two
approaches for linearizing GF equations of motion.
Namely, the Tyablikov approximation~\cite{tyab} and the
Callen approximation~\cite{cal63} for the isotropic Heisenberg model (\ref{e12}). 
We begin from the equations of
motion (\ref{e43}) for GF of the form $\langle \langle S^{+}|S^{-} \rangle \rangle $:
\begin{eqnarray}
\label{e85}
\omega \langle \langle S_{i}^{+}|S_{j}^{-} \rangle \rangle_{\omega} =  
\nonumber 2 \langle S^{z} \rangle \delta_{ij} + \sum_{g} J (i-g)
\langle \langle S^{+}_{i}S^{z}_{g} - S^{+}_{g}S_{i}^{z}|S_{j}^{-} \rangle \rangle_{\omega}.
\end{eqnarray}
Within the Tyablikov approximation the second order
GF is written in terms of the first-order GF as follows~\cite{tyab}:
\begin{equation}
\label{e86} \langle \langle S^{+}_{i}S_{g}^{z}|S^{-}_{j} \rangle \rangle \simeq
\langle S^{z} \rangle \langle \langle S^{+}_{i}|S^{-}_{j} \rangle \rangle.
\end{equation}
It is well know, that the Tyablikov approximation
(\ref{e86}) corresponds to the random phase approximation
for a gas of electrons. The spin-wave's excitation spectrum
does not contain damping in this approximation:
\begin{equation} \label{e87}
E(q) = \sum_{g} J(i-g) \langle S^{z} \rangle\exp [i(\vec R_{i} - \vec R_{g})
\vec q ] = 2 \langle S^{z} \rangle (J_{0} - J_{q}).
\end{equation}
This is due to the fact that the Tyablikov approximation
does not take into account the inelastic quasiparticle's
scattering processes. One should also mention
that within the Tyablikov approximation the exact commutation
relations $[S^{+}_{i},S^{-}_{j}]_{-} = 2S^{z}_{i} \delta_{ij}$ are replaced by
approximate relationships of the form $ [S^{+}_{i},S^{-}_{j}]_{-} \simeq 2 \langle S^{z} \rangle  \delta_{ij}.$
Despite being simple, the Tyablikov approximation
is widely used in different problems even at the
present time~\cite{medv08}.\\
Callen proposed a modified version of the Tyablikov
approximation, which takes into account some correlation
effects. The following linearization of equations-of-
motion is used within the Callen approximation~\cite{cal63}:
\begin{equation} \label{e88}
\langle \langle S^{z}_{g}S_{f}^{+}|B \rangle \rangle \rightarrow \langle S^{z} \rangle \langle \langle S^{+}_{f}|B \rangle \rangle -
\alpha \langle S^{-}_{g}S^{+}_{f}\rangle \langle \langle S^{+}_{g}|B \rangle \rangle.
\end{equation}
Here, $ 0 \leq \alpha \leq 1$. In order to better understand Callen's
decoupling idea one has to take into account that the
spin $1/2$ operator $ S^{z}$ can be represented in the form $ S^{z}_{g} = S - S^{-}_{g}S^{+}_{g}$ or $S^{z}_{g} =
{1 \over 2} ( S^{+}_{g}S^{-}_{g} - S^{-}_{g}S^{+}_{g})$.
Therefore, we have
$$
S^{z}_{g}  = \alpha S + \frac {1 - \alpha}{2} S^{+}_{g}S^{-}_{g}
- \frac {1 + \alpha}{2} S^{-}_{g}S^{+}_{g}.$$  
The operator $S^{-}_{g}S^{+}_{g}$   is the   "deviation" of the quantity
$\langle S^{z} \rangle$ from $S$. In the low-temperature domain that "deviation"
is small and $\alpha \sim 1$. Analogously, the operator ${1 \over 2} (S^{+}_{g}S^{-}_{g} - S^{-}_{g}S^{+}_{g})$
is the "deviation" of the quantity $\langle S^{z} \rangle$ from  $0$.
Therefore, when $\langle S^{z} \rangle$ approaches zero one can
expect that $\alpha \sim 0$. Thus, the Callen approximation has
an interpolating character. Depending on the choice of
the value for the parameter $\alpha$, one can obtain both positive
and negative corrections to the Tyablikov approximation,
or even almost vanishing corrections. The particular
case $\alpha = 0$ corresponds to the Tyablikov approximation.\\
We would like to stress that the Callen
approach is by no means rigorous. Moreover, it has
serious drawbacks~\cite{kuz02}. However, one can consider this
approximation as the first serious attempt to construct
an approximating interpolation scheme in the framework
of the GF's equations-of-motion method. In contrast
to the Tyablikov approximation, the spectrum of
spin-wave excitations within the Callen approximation
is given by
\begin{equation} \label{e89}
E(q) = 2 \langle S^{z} \rangle \bigl ((J_{0} - J_{q}) +\frac {\langle S^{z} \rangle}{NS^2}
\sum_{k} [ J(k) - J(k - q)] N(E(k)) \bigr ).
\end{equation}
Here, $N(E(k))$ is the Bose's distribution function
$ N(E(k)) = [\exp ( E(k)\beta) - 1 ]^{-1}$
Equation (\ref{e89}) clearly
shows how the Callen approximation improves Tyablikov's
approximation. From a general point of view,
one has to find the form of the effective self-consistent
generalized mean-field functional. That is, to find
which averages determine that field
$$F = \{\langle S^{z} \rangle, \langle S^{x} \rangle, \langle S^{y} \rangle, \langle S^{+}S^{-} \rangle,
\langle S^{z}S^{z} \rangle, \langle S^{z}S^{+}S^{-} \rangle, \ldots \}.$$
Later many approximate schemes for decoupling
the hierarchy of equations for GF were proposed~\cite{tyab},
improving the Tyablikov and Callen decouplings. Various
approaches generalizing the random phase's
approximation in the ferromagnetism theory for wide
ranges of temperature were considered in the paper~\cite{hol90}
by Czachor and Holas.
%
\subsection{Heisenberg Antiferromagnet and Anomalous Averages}
%
%
In order to illustrate the scheme of the irreducible
GF method we are going to consider now the Heisenberg
antiferromagnet. Note that a systematic microscopic
theory of antiferromagnetism has not been built
yet. In the framework of the model of localized spins
the appearance of the antiferromagnetic phase is usually
associated with the first divergence of the generalized
spin's susceptibility, if the exchange integral
between the nearest neighbors is negative. The first
divergence appears at $\vec{Q} = \pi/a (\vec{a} + \vec{b} + \vec{c}).$
Which means that when transiting from one atomic plane to
another along the vector the phase of the magnetization
vectors changes by $\pi$. Generally speaking, in crystals
with a complicated structure the exchange interaction
may be different for different pairs of neighbors. In
this case, we have a large variety of antiferromagnetic
configurations. The simplest and the most frequently
used model of localized spins of antiferromagnetic phenomena
is the Heisenberg model of two-sublattice antiferromagnets.
Let us consider now the calculation of the
renormalized quasiparticle spectrum of magnetic excitations
in the framework of the irreducible GF method~\cite{b100}.
The Hamiltonian of the system is given by
\begin{equation}
\label{e90} {\mathcal H} = - \frac {1}{2} \sum_{ij}\sum_{\alpha \alpha'}
J^{\alpha \alpha'}(i-j) \vec S_{i \alpha} \vec S_{j \alpha'} = -
\frac {1}{2} \sum_{q}\sum_{\alpha \alpha'} J_{q}^{\alpha \alpha'}
\vec S_{q \alpha} \vec S_{-q \alpha'}.
\end{equation}
Here, $S_{i \alpha}$ is the spin operator at the site $i$ подрешетки  of the sublattice $\alpha$,
and $J^{\alpha \alpha'}(i-j)$ is the exchange integral between
the spins at the sites $R_{i\alpha}$ and $R_{j\alpha'}$;  the indexes $\alpha, \alpha'$
assume two values $(a)$ and $(b)$. It is assumed that all the
atoms in a sublattice $\alpha$ are identical and have the spin
$S_{\alpha}$. It is convenient to rewrite the Hamiltonian (\ref{e90}) in
the following form:
\begin{equation}
\label{e91} H = - \frac {1}{2} \sum_{q}\sum_{\alpha \alpha'}
I_{q}^{\alpha \alpha'} ( S^{+}_{q \alpha} S^{-}_{-q \alpha'} +
S^{z}_{q\alpha} S^{z}_{-q\alpha'}),
\end{equation}
where $$ I^{\alpha \alpha'}_{q} = 1/2( J^{\alpha \alpha'}_{q} +
J^{\alpha' \alpha}_{-q}).$$  
Let us again consider the equations of motion (\ref{e43})
for the Green's function of the form$\langle \langle S^{+}|S^{-} \rangle \rangle. $ In contrast
to Heisenberg's ferromagnet model, for the two-sublattice
antiferromagnet we have to use the matrix
GF of the form
\begin{equation} \label{e92}
\hat G(k;\omega) = 
\begin{pmatrix}
   \langle \langle S^{+}_{ka} \vert S^{-}_{-ka} \rangle \rangle &
\langle \langle S^{+}_{ka} \vert S^{-}_{-kb}\rangle \rangle \cr \langle \langle S^{+}_{kb}\vert
S^{-}_{-ka}\rangle \rangle& \langle \langle S^{+}_{kb}\vert S^{-}_{-kb}\rangle \rangle \cr
\end{pmatrix}. 
\end{equation}
Here, the GF on the main diagonal are the usual or
normal GF, while the off-diagonal GF describe contributions
from the so-called anomalous terms, analogous
to the anomalous terms in the BCS-Bogoliubov superconductivity
theory (\ref{e84}). The anomalous (or off-diagonal)
average values in this case select the vacuum state
of the system precisely in the form of the two-sublattice
Neel state. The Dyson equation (\ref{e78}) is derived with the
help of irreducible operators of the form
\begin{eqnarray}
\label{e93}
(S^{ab}_{kq})~^{(ir)} = S^{ab}_{kq} - A^{ab}_{q}S^{+}_{ka} + A^{ba}_{k-q}S^{+}_{kb},  \\
\label{e94} (S^{z}_{q\alpha})^{~(ir)} = S^{z}_{q\alpha} -
N^{1/2} <S^{z}_{\alpha}>\delta_{q,0},
\end{eqnarray}
where $S^{ab}_{kq} = (S^{+}_{k-q,a}S^{z}_{qb} - S^{+}_{qb}S^{z}_{k-q,a})$. On performing
standard transformations one can obtain the Dyson
equation in the matrix form:
\begin{equation}
\label{e95} \hat G(k,\omega) = \hat G_{0}(k,\omega) + \hat
G_{0}(k,\omega) \hat M(k,\omega) \hat G(k,\omega).
\end{equation}
Here, $\hat G_{0}(k,\omega)$ is the GF within the generalized mean field
approximation
\begin{eqnarray}
\label{e96} \hat G_{0} =  
\begin{pmatrix}
G_{0}^{aa}(k,\omega) & G_{0}^{ab}(k,\omega)\cr
G_{0}^{ba}(k,\omega) &G_{0}^{bb}(k,\omega)\cr  
\end{pmatrix}
= \frac
{2<S^z_a>}{\textrm{det} \hat \Omega} 
\begin{pmatrix} (\omega -
\omega_{aa})&\omega_{ab} \cr \omega_{ab} & ( \omega -
\omega_{bb}) \cr  
\end{pmatrix},
\end{eqnarray}
where
$$ \textrm{det} \hat \Omega = (\omega - \omega_{aa})(\omega - \omega_{bb})
- \omega_{aa}\omega_{ab}.$$   
The poles of the GF (\ref{e96}) determine the spectrum of
magnetic excitations in the generalized mean-field
approximation (the elastic scattering corrections):
$$ \textrm{det} \hat \Omega = 0.$$
As a result we obtain
\begin{equation}
\label{e97} \omega_{\pm}(k) = \pm \sqrt {(\omega^{2}_{aa}(k) -
\omega^{2}_{ab}(k))},
\end{equation}
\begin{equation}
\label{e98} \omega(k) = I z \langle S^{z}_{a} \rangle \Bigl [ 1 - \frac
{1}{N^{1/2}\langle S^{z}_{a} \rangle} \sum_{q} \gamma_{q} A^{ab}_{q} \Bigr ]
\sqrt {( 1 - \gamma^{2}_{k} )},
\end{equation}
where $I_{q} = zI\gamma_{q}$, $ \gamma_{k} = 1/z \sum_{i} \exp (ikR_{i})$   and $z$  is the number
of nearest neighbors. The first term in (\ref{e98}) corresponds
to the Tyablikov approximation. The second
term describes the corrections of elastic scattering
within the generalized mean-field approximation. Note
that the quantity which determines these corrections,
is given by
\begin{equation}
\label{e99} A^{ab}_{q} = \frac { 2 \langle (S^{z}_{-qa})^{(ir)}
(S^{z}_{qb})^{(ir)} \rangle + \langle S^{-}_{-qa}S^{+}_{qb} \rangle} {
2N^{1/2} \langle S^{z}_{a} \rangle}.
\end{equation}
This expression contains anomalous averages
$ \langle S^{-}_{-qa}S^{+}_{qb} \rangle$,
which characterize the Neel ground state.
%
\subsection{Many-particle Systems with Strong and Weak Electron Correlations}
%
%
%
The efficiency of the method of the irreducible
Green's functions for description of normal and superconducting
properties of systems with a strong interaction
and complicated character of the electron spectrum
was demonstrated in the papers~\cite{kuz02,b36,b91,b118,b101}. Let us
consider the Hubbard model (\ref{e18}). The properties of this
Hamiltonian are determined by the relationship
between the two parameters: the effective band's width $\Delta$
and the electron's repulsion energy $U$. Drastic transformations
of the metal-dielectric phase transition's
type take place in the system as the ratio of these
parameters changes. Note that, simultaneously, the
character of the system description must change as
well, that is, we always have to describe our system by
the set of relevant variables.
In the case of weak correlation~\cite{kuz02,b36,b91,b118,b101} the
corresponding set of relevant variables contains the
ordinary second-quantized Fermi operators and $a^{\dagger}_{i\sigma}$ и $a_{i\sigma}$,
as well as the number of particles operator $n_{i\sigma} = a^{\dagger}_{i\sigma}a_{i\sigma}.$ 
These operators have the following properties:
$$a^{\dagger}_{i}\Psi^{(0)} = \Psi^{(1)}_{i} ; \quad a_{i}\Psi^{(1)} =
\Psi^{(0)},$$ $$a_{i}\Psi^{(0)} = 0, \quad  a_{j}\Psi^{(1)}_{i} =
0 \quad ( i \not= j).$$ 
Here $\Psi^{(0)}$ and $\Psi^{(1)}$
 describe the vacuum and the single-particle
states, respectively [159]. In order to find the
low-lying excited quasiparticle states of the many-electron
system with the Hamiltonian (\ref{e18}), one has to pass
to the vector space of Bloch states
$$a_{{\vec k}\sigma}= N^{-1/2}\sum_{i}\exp(-i{\vec k}{\vec R_{i}}) a_{i\sigma}.$$
In this representation the Hamiltonian (\ref{e18}) is given by
\begin{equation}
\label{e100} \mathcal{H} =
\sum_{k\sigma}\epsilon(k) a^{\dagger}_{k\sigma}a_{k\sigma} +
U/2N \sum_{pqrs} \sum_{\sigma} a^{\dagger}_{p+r-q\sigma} a_{p\sigma} a^{\dagger}_{q-\sigma} a_{r-\sigma}.
\end{equation}
Let us now consider the one-particle electron's GF
of the form
\begin{equation}
\label{e101} G_{k\sigma}(t - t') = \langle \langle a_{k\sigma},
a^{\dagger}_{k\sigma} \rangle \rangle = -i\theta(t - t') \langle [a_{k\sigma}(t),
a^{\dagger}_{k\sigma}(t')]_{+} \rangle.
\end{equation}
The corresponding equation of motion (\ref{e43}) for
$G_{k\sigma}(\omega)$ is given by
\begin{equation}
\label{e102} (\omega - \epsilon_{k})G_{k\sigma}(\omega) = 1 +
U/N\sum_{pq} \langle \langle a_{k+p\sigma}a^{\dagger}_{p+q-\sigma}a_{q-\sigma}
\vert a^{\dagger}_{k\sigma} \rangle \rangle_{\omega}.
\end{equation}
In line with Eq.(\ref{e71}) we introduce the irreducible GF
\begin{eqnarray}
\label{e103}
^{(ir)} \langle \langle a_{k+p\sigma}a^{\dagger}_{p+q-\sigma}a_{q-\sigma}
\vert a^{\dagger}_{k\sigma}\rangle \rangle_{\omega} =\nonumber\\
 \langle \langle  a_{k+p\sigma}a^{\dagger}_{p+q-\sigma}a_{q-\sigma} \vert
a^{\dagger}_{k\sigma} \rangle \rangle_{\omega} -\delta_{p,
0} \langle n_{q-\sigma} \rangle G_{k\sigma}.
\end{eqnarray}
The irreducible (\textbf{ir}) GF in Eq. (\ref{e103}) is defined in
such a way that it can not be transformed to a lower
order GF by arbitrary pairings of second-quantized fermion
operators. Next, according to Eqs. (\ref{e71}) - (\ref{e79}) we
find
\begin{eqnarray}
\label{e104}
G_{k\sigma}(\omega) = G^{MF}_{k\sigma}(\omega) + \\
\nonumber G^{MF}_{k\sigma}(\omega)
U/N\sum_{pq}{}^{(ir)} \langle \langle a_{k+p\sigma}a^{\dagger}_{p+q-\sigma}a_{q-\sigma}
\vert a^{\dagger}_{k\sigma} \rangle \rangle_{\omega}. \nonumber
\end{eqnarray}
The following notation were introduced here
\begin{equation}
\label{e105} G^{MF}_{k\sigma}(\omega) = (\omega -
\epsilon(k\sigma))^{-1} ; \, \epsilon(k\sigma) = \epsilon(k)
+U/N\sum_{q} \langle n_{q-\sigma}\rangle.
\end{equation}
Below, for simplicity we consider only paramagnetic
solutions, where $\langle n_{\sigma}\rangle = \langle n_{-\sigma} \rangle$. According to
Eqs. (\ref{e71}) - (\ref{e79}) we obtain
\begin{equation}
\label{e106} G_{k\sigma}(\omega) = G^{MF}_{k\sigma}(\omega) +
G^{MF}_{k\sigma}(\omega)
P_{k\sigma}(\omega)G^{MF}_{k\sigma}(\omega).
\end{equation}
The operator $P$ is given by
\begin{eqnarray}
\label{107} P_{k\sigma}(\omega) =
\frac{U^{2}}{N^{2}}\sum_{pqrs}D^{(ir)}_{k\sigma} (p, q\vert r, s,
;\omega) =  \\ \nonumber
\frac{U^{2}}{N^{2}}\sum_{pqrs}\biggl( \, ^{(ir)}\langle \langle a_{k+p\sigma}a^{\dagger}_{p+q-\sigma}
a_{q-\sigma}\vert
a^{\dagger}_{r-\sigma}a_{r+s-\sigma}a^{\dagger}_{k+s\sigma} \rangle \rangle^{(ir)}_{\omega} \biggr).
\end{eqnarray}
The proper part of the operator $P$ is given by
\begin{eqnarray}
\label{e108} D^{(ir)}_{k\sigma}(p, q\vert r, s;\omega) =
 L^{(ir)}_{k\sigma}(p, q\vert r, s;\omega)
\nonumber\\
+ \frac{U^{2}}{N^{2}}\sum_{r's'p'q'}L^{(ir)}_{k\sigma}(p, q\vert
r's';\omega) G^{MF}_{k\sigma}(\omega)D^{(ir)}_{k\sigma}(p',
q'\vert r, s;\omega).
\end{eqnarray}
Here, $L^{(ir)}_{k\sigma}(p, q\vert r, s;\omega)$ is the proper part of the GF
$D^{(ir)}_{k\sigma}(p, q \vert r, s;\omega)$
Therefore, we obtain
\begin{equation}
\label{e109} G_{k\sigma} = G^{MF}_{k\sigma}(\omega) +
G^{MF}_{k\sigma}(\omega) M_{k\sigma}( \omega)G_{k, \sigma}(\omega).
\end{equation}
Equation (\ref{e109}) is the desired Dyson equation for
two-time temperature GF $ G_{k\sigma}(\omega)$. It has the following
formal solutions, cf. (\ref{e79}):
\begin{equation}
\label{e110} G_{k\sigma}(\omega) = [\omega -\epsilon(k\sigma)
-M_{k\sigma}( \omega)]^{-1}.
\end{equation}
The mass operator $M$ is given by
\begin{eqnarray}
\label{e111} M_{k\sigma}(\omega) =  \frac{U^{2}}{N^{2}}
\sum_{pqrs}L^{(ir)}_{k\sigma}
(p, q \vert r, s;\omega) =  \\
 \frac{U^2}{N^2}{} \sum_{pqrs}  \biggl( \,^{(ir)}\langle \langle a_{k+p\sigma}a^{\dagger}_{p+q-\sigma}
a_{q-\sigma} \vert
a^{\dagger}_{r-\sigma}a_{r+s-\sigma}a^{\dagger}_{k+s\sigma} \rangle \rangle^{(ir)}\biggr )^{(p)}. \nonumber
\end{eqnarray}
As was shown in the papers~\cite{kuz02,b36,b91,b118,b101}, an
approximation to the mass operator $M$ can be calculated
as follows:
\begin{eqnarray}
\label{e112}
M_{k\sigma}( \omega) \simeq  
 \frac{U^2}{N^2} \sum_{pq} \int
\frac{d\omega_{1}d\omega_{2}d\omega_{3}}{\omega + \omega_{1} -
\omega_{2} - \omega_{3}} \times  \nonumber \\
~\Bigl [n(\omega_{2})n(\omega_{3}) + n(\omega_{1})\Bigl (1 -
n(\omega_{2}) - n(\omega_{3})\Bigr ) \Bigr]
g_{p+q-\sigma}(\omega_{1})g_{k+p\sigma}(\omega_{2})g_{q-\sigma}(\omega_{3}).
\end{eqnarray}
Here,
$$n(\omega) = [\exp(\beta\omega) + 1]^{-1}; \,
g_{k\sigma}(\omega) = -{1 \over \pi} \textrm{Im} G_{k\sigma}(\omega + i\varepsilon).$$
Equations (\ref{e110}) and (\ref{e112})  are a self-consistent system
of equations for calculating the one-particle GF
$G_{k\sigma}(\omega).$  As the first iteration one can substitute the
expression
\begin{equation}
\label{e113} g_{k\sigma}(\omega) \approx \delta(\omega -
\epsilon(k\sigma)).
\end{equation}
in the right hand side of Eq.(\ref{e112}). The substitution
yields
\begin{equation}
\label{e114} M_{k\sigma}( \omega) = \frac{U^2}{N^2} \sum_{pq}
\frac{n_{p+q-\sigma}(1 - n_{k+p\sigma} - n_{q-\sigma}) +
n_{k+p\sigma} n_{q-\sigma}}{\omega + \epsilon(p+q\sigma) -
\epsilon(k+p\sigma) - \epsilon(q\sigma)}.
\end{equation}
Equation (\ref{e114}) describes the renormalization of the
electron spectrum due to the inelastic electron's scattering
processes. All elastic scattering corrections have
already been taken into account by the electron
energy's renormalization, see Eq. (\ref{e105}). Thus, the
investigation of the Hubbard model in the weak coupling
limit is relatively easy.\\
The most challenging case is the solution of the
Hubbard model when the electron correlations are
strong, but are finite. In this limit it is convenient to consider
the one-particle GF in the Wannier representations
\begin{equation}
\label{e115} G_{ij\sigma}(t - t') = \langle \langle a_{i\sigma}(t);
a^{\dagger}_{j\sigma}(t') \rangle \rangle.
\end{equation}
In the case of strong correlation, the algebra of relevant
operators must be chosen according to specific
features of the problem under investigation. It is convenient
to use the Hubbard operators~\cite{hub64}:
\begin{eqnarray}
\label{e116} d_{i\alpha\sigma} =
n^{\alpha}_{i-\sigma}a_{i\sigma}, (\alpha = \pm);\quad
n^{+}_{i\sigma} = n_{i\sigma},\quad n^{-}_{i\sigma} = (1 -
n_{i\sigma});
\nonumber\\
\sum n^{\alpha}_{i\sigma} = 1; \quad
n^{\alpha}_{i\sigma}n^{\beta}_{i\sigma} =
\delta_{\alpha\beta}n^{\alpha}_{i\sigma}; \quad \sum_{\alpha}
d_{i\alpha\sigma} = a_{i\sigma}.
\end{eqnarray}
The new operators $d_{i\alpha \sigma}$ and $d^{\dagger}_{j\beta\sigma}$  have complicated
commutation relations, namely
$$[d_{i\alpha \sigma}, d^{\dagger}_{j\beta \sigma}]_{+} = \delta_{ij}
\delta_{\alpha \beta}n^{\alpha}_{i-\sigma}.$$ 
The advantages of using these operators become
clear when we consider their equations of motion:
\begin{eqnarray}
\label{e117} [d_{i\alpha \sigma}, H]_{-} = E_{\alpha}d_{i\alpha
\sigma} + \sum_{ij}t_{ij}(n^{\alpha}_{i-\sigma}a_{j\sigma} +
\alpha a_{i\sigma}
b_{ij-\sigma}),  \nonumber\\
b_{ij\sigma} = (a^{\dagger}_{i\sigma}a_{j\sigma} -
a^{\dagger}_{j\sigma}a_{i\sigma}).
\end{eqnarray}
According to Hubbard~\cite{hub64}, the contributions to
the above equation describe the "alloy analogy" corrections and the resonance broadening corrections. Using
the Hubbard operators one can write down GF (\ref{e115}) in
the following form
\begin{equation}
\label{e118} G_{ij \sigma}(\omega) = \sum_{\alpha
\beta} \langle \langle d_{i\alpha \sigma} \vert d^{\dagger}_{j\beta
\sigma} \rangle \rangle_{\omega} = \sum_{\alpha \beta} F^{\alpha \beta}_{
ij\sigma}(\omega).
\end{equation}
The equation of motion for the auxiliary GF $F$
\begin{equation}
\label{e119} F^{\alpha \beta}_{ij\sigma}(\omega) =
\begin{pmatrix}
\langle \langle d_{i+\sigma} \vert
d^{\dagger}_{j+\sigma}\rangle \rangle_{\omega}&\langle \langle d_{i+\sigma} \vert
d^{\dagger}_{j-\sigma} \rangle \rangle_{\omega}
 \cr
\langle \langle d_{i-\sigma} \vert
d^{\dagger}_{j+\sigma} \rangle \rangle_{\omega}&\langle \langle d_{i-\sigma} \vert
d^{\dagger}_{j-\sigma} \rangle \rangle_{\omega}\cr
\end{pmatrix} 
\end{equation}
is now given by
\begin{equation}
\label{e120} ({\bf E}{\bf F}_{ij\sigma}(\omega) -{\bf I}
\delta_{ij})_{\alpha \beta} = \sum_{l\not=
i}t_{il} \langle \langle n^{\alpha}_{i-\sigma}a_{l\sigma} + \alpha a_{i\sigma}
b_{il-\sigma} \vert d^{\dagger}_{j\beta \sigma} \rangle \rangle_{\omega}.
\end{equation}
Here, we used the following notation:
\begin{equation}
\label{e121} {\bf E} = 
\begin{pmatrix}
(\omega- E_{+})&0\cr 0&(\omega -
E_{-})\cr 
\end{pmatrix} ; \,
{\bf I} = 
\begin{pmatrix}
n^{+}_{-\sigma}&0\cr
0&n^{-}_{-\sigma} \cr
\end{pmatrix}. 
\end{equation}
The determination of the irreducible parts of the GF
is more involved:
\begin{eqnarray}
\label{e122} {\bf D}^{(ir)}_{il, j}(\omega) =
\begin{pmatrix}
\langle \langle Z_{11}\vert d^{\dagger}_{j+\sigma} \rangle \rangle_
{\omega}&\langle \langle Z_{12}\vert d^{\dagger}_{j-\sigma} \rangle \rangle_{\omega}\cr
\langle \langle Z_{21}\vert d^{\dagger}_{j+\sigma} \rangle \rangle_{\omega}&\langle \langle Z_{22}\vert
d^{\dagger}_{j-\sigma} \rangle \rangle_{\omega}\cr 
\end{pmatrix}
- \nonumber\\
\sum _{\alpha'}\left ( {A^{+\alpha'}_{il}\brack A^{-\alpha'}_{il}}[F^{\alpha'+}_{
ij\sigma} \ F^{\alpha'-}_{ij\sigma}] - {B^{+\alpha'}_{li}\brack
B^{-\alpha'}_{
li}}[F^{\alpha'+}_{lj\sigma} \ F^{\alpha'-}_{lj\sigma}]\right ).
\end{eqnarray}
In order to make the equations more compact we
have introduced the following notation:
$$Z_{11} = Z_{12} = n^{+}_{i-\sigma}a_{l\sigma} + a_{i\sigma}b_{il-\sigma};
\ Z_{21} = Z_{22} = n^{-}_{i-\sigma}a_{l\sigma} -
a_{i\sigma}b_{il-\sigma}.$$ 
One has to stress that the definition (\ref{e122}) plays the
central role in this method. The coefficients $A$ and  $B$   are
found from the orthogonality condition (\ref{e72})
\begin{equation}
\label{e123} \langle [({\bf D}^{(ir)}_{il, j})_{\alpha \beta},
d^{\dagger}_{j\beta \sigma}]_{+} \rangle = 0.
\end{equation}
Next, the exact Dyson equation is derived according
to Eqs.(\ref{e70}) - (\ref{e79}). Its mass operator is given by
\begin{equation}
\label{e124}      {\bf M}_{q\sigma}(\omega)   = \left ( {\bf P}_{q\sigma}(\omega) \right )^{p} =  \left ( {\bf
I}^{-1}[\sum_{lm}t_{il}t_{mj} \langle \langle {\bf D}^{(ir)}_{il, j}|{\bf
D}^{(ir) \dagger}_{i, mj} \rangle \rangle_{\omega}]_{q}{\bf I}^{-1} \right )^{p}.
\end{equation}
The GF in the generalized mean-field's approximation has the following very 
complicated functional structure~\cite{kuz02,b36,b91,b118,b101}:
\begin{equation}
\label{e125} G^{MF}_{k\sigma}( \omega) = \frac{\omega -
(n^{+}_{-\sigma}E_{-} + n^{-}_{-\sigma}E_{+}) - \lambda(k)}{
(\omega -E_{+} - n^{-}_{-\sigma}\lambda_{1}(k))(\omega - E_{-} -
 n^{+}_{-\sigma}
\lambda_{2}(k)) -
n^{-}_{-\sigma}n^{+}_{-\sigma}\lambda_{3}(k)\lambda_{4}(k)}.
\end{equation}
Here, the quantities $\lambda_{i}(k)$ are the components of the
generalized mean field, which cannot be reduced to the
functional of the mean particle's densities. The expression
for GF (\ref{e125}) can be written down in the form of the
following \textbf{generalized two-pole solution}
\begin{eqnarray}
\label{e126} G^{MF}_{k\sigma}( \omega) = \frac
{n^{+}_{-\sigma}(1 + cb^{-1})}{a - db^{-1}c} + \frac
{n^{-}_{-\sigma}(1 + da^{-1})}{b - ca^{-1}d} \approx
\nonumber\\
\frac {n^{-}_{-\sigma}}{\omega - E_{-} -
n^{+}_{-\sigma}W^{-}_{k-\sigma}} + \frac {n^{+}_{-\sigma}}{\omega
- E_{+} - n^{-}_{-\sigma}W^{\dagger}_{k-\sigma}},
\end{eqnarray}
where
\begin{eqnarray}
\label{e127}
n^{+}_{-\sigma}n^{-}_{-\sigma}W^{\pm}_{k-\sigma} =
N^{-1}\sum_{ij} t_{ij}
\exp[-ik(R_{i} -R_{j})] \times  \\
\left ((\langle a^{\dagger}_{i-\sigma}n^{\pm}_{i\sigma}a_{j-\sigma} \rangle +
\langle a_{i-\sigma}
n^{\mp}_{i\sigma}a^{\dagger}_{j-\sigma} \rangle)    \right . + \nonumber \\
 \left . (\langle n^{\pm}_{j-\sigma}n^{\pm}_{i-\sigma} \rangle +
\langle a_{i\sigma}a^{\dagger}_{i-\sigma}
a_{j-\sigma}a^{\dagger}_{j\sigma} \rangle -
\langle a_{i\sigma}a_{i-\sigma}a^{\dagger}_{j-\sigma}
a^{\dagger}_{j\sigma} \rangle) \right). \nonumber
\end{eqnarray}
Green's function (\ref{e126}) is the most general solution
of the Hubbard model within the generalized mean field 
approximation. Equation (\ref{e127}) is nothing else
but the explicit expression for the generalized mean field.
As we see, this mean field is not a functional of
the mean particle's densities. The solution (\ref{e126}) is
more general than the solution "Hubbard III"~\cite{hub64} and
the two-pole solution from the papers~\cite{roth1,roth2} by
Roth. It was shown in the papers~\cite{kuz02,b36,b91,b118,b101} by
the author of this review, that the solution "Hubbard I"~\cite{hub63}
is a particular case of the solution (\ref{e126}), which
corresponds to the additional approximation
\begin{equation}
\label{e128} n^{+}_{-\sigma}n^{-}_{-\sigma}W^{\pm}(k) \approx
N^{-1}\sum_{ij}t_{ij} {\exp[-ik(R_{i} -
R_{j})]} \langle n^{\pm}_{j-\sigma}n^{\pm}_{i-\sigma}\rangle.
\end{equation}
Assuming $\langle n_{j-\sigma}n_{i-\sigma}\rangle \approx n^{2}_{-\sigma},$ we obtain the approximation
"Hubbard I"~\cite{hub63}. Thus, we have shown that in
the cases of systems of strongly correlated particles
with a complicated character of quasiparticle spectrums
the generalized mean fields can have quite a nontrivial
structure, which is difficult to establish by using
any kind of independent considerations. The method of
irreducible GF allows one to obtain this structure in the
most general form.
%
\subsection{Superconductivity Equations}
%
The nontrivial structure of the generalized mean fields
in many-particle systems is vividly revealed in
the description of the superconductivity phenomenon.
Let us now briefly consider this topic following the
papers~\cite{kuz02,b91,b67,b47}. We describe our system by
the following Hamiltonian:
\begin{equation}
\label{e129}
 H = H_{e} + H_{i} + H_{e-i}.
\end{equation}
Here, the operator $H_{e}$ is the Hamiltonian of the crystal's
electron subsystem, which we describe by the
Hubbard Hamiltonian (\ref{e18}). The Hamiltonian of the ion
subsystem and the operator describing the interaction
of electrons with the lattice are given by
\begin{eqnarray}
\label{e130} H_{i} = \frac {1}{2} \sum_{n} \frac {P^{2}_{n}}{2M}
+
{1\over 2} \sum_{mn \alpha \beta} \Phi^{\alpha\beta}_{nm}u^{\alpha}_{n}u^{\beta}_{m}, \\
\label{e131} H_{e-i} = \sum_{\sigma} \sum_{n,i \not = j}
V^{\alpha}_{ij} ( \vec R^{0}_{n})
a^{\dagger}_{i\sigma}a_{j\sigma}u^{\alpha}_{n},
\end{eqnarray}
where
\begin{equation}
\label{e132} \sum_{n} V^{\alpha}_{ij} ( \vec
R^{0}_{n})u^{\alpha}_{n} = \frac {\partial t_{ij}(\vec
R_{ij}^{0})}{\partial R_{ij}^{0}} (\vec u_{i} - \vec u_{j}).
\end{equation}
Here, $P_{n}$ is the momentum operator, $M$ is the ion
mass, and $u_{n}$ is the ion displacement relative to its equilibrium
position at the lattice site $R_{n}$. Using more convenient
notations one can write down the operator
describing the interaction of electrons with the lattice as
follows~\cite{b91,b67,b47}
\begin{equation}
\label{e133} H_{e-i} = \sum_{\nu\sigma}\sum_{kq} V^{\nu}(\vec
k, \vec k + \vec q)Q_{\vec q\nu}a^{\dagger}_{k+q\sigma}
a_{k\sigma},
\end{equation}
where
\begin{equation}
\label{e134} V^{\nu}(\vec k, \vec k + \vec q) = \frac{2iq_{0}}{(
N M )^{1/2}}\sum_{\alpha} t(\vec a_{\alpha})e^{\alpha}_{\nu}(\vec
q)[\sin \vec a_{\alpha} \vec k - \sin \vec a_{\alpha} (\vec k -
\vec q)].
\end{equation}
Here, $q_{0}$ is the Slater coefficient~\cite{kuz02,b67,b47}, describing
the exponential decay of the $d$-electrons' wave
function. The quantities $\vec e_{\nu}(\vec q)$
are the phonon-mode's
polarization vectors. The Hamiltonian of the ion subsystem can be rewritten in the following form
\begin{equation}
\label{e135} H_{i} = \frac{1}{2} \sum_{q\nu}
(P^{\dagger}_{q\nu}P_{q\nu} + \omega^{2}(\vec q
\nu)Q^{\dagger}_{q\nu}Q_{q\nu})
\end{equation}
Here, $P_{q\nu}$ and $Q_{q\nu}$   are the normal coordinates, $\omega(q\nu)$
are the acoustic phonons' frequencies.\\
Consider now the generalized one-electron GF of
the following form:
\begin{eqnarray}
\label{e136}
 G_{ij} (\omega) =
\begin{pmatrix} 
G_{11}&G_{12}\cr 
G_{21}&G_{22}\cr 
\end{pmatrix}    =   \begin{pmatrix}
\langle \langle a_{i\sigma}\vert a^{\dagger}_{j\sigma} \rangle \rangle & \langle \langle a_{i\sigma}\vert
a_{j-\sigma} \rangle \rangle \cr \langle \langle a^{\dagger}_{i-\sigma}\vert
a^{\dagger}_{j\sigma} \rangle \rangle & \langle \langle a^{\dagger}_{i-\sigma}\vert
a_{j-\sigma} \rangle \rangle \cr  \nonumber
\end{pmatrix}.
\end{eqnarray}
As was already discussed above, the off-diagonal
entries of the above matrix select the vacuum state of
the system in the BCS-Bogoliubov form, and they are
responsible for the presence of anomalous averages.
The corresponding equations of motion are given by
\begin{eqnarray}
\label{e137} \sum_{j}(\omega \delta_{ij} - t_{ij})\langle \langle a_{j\sigma}
\vert a^{\dagger}_{i'\sigma} \rangle \rangle = \delta_{ii'} + \\  \nonumber
U \langle \langle a_{i\sigma} n_{i-\sigma} \vert a^{\dagger}_{i'\sigma} \rangle \rangle +
\sum_{nj} V_{ijn} \langle \langle a_{j\sigma} u_{n} \vert a^{\dagger}_{i'\sigma} \rangle \rangle, \\
\label{e138} \sum_{j}(\omega \delta_{ij} +
t_{ij}) \langle \langle a^{\dagger}_{j-\sigma} \vert a^{\dagger}_{i'\sigma} \rangle \rangle =
\\  \nonumber
-U \langle \langle a^{\dagger}_{i-\sigma} n_{i\sigma} \vert
a^{\dagger}_{i'\sigma} \rangle \rangle + \sum_{nj} V_{jin}
\langle \langle a^{\dagger}_{j-\sigma} u_{n} \vert a^{\dagger}_{i'\sigma} \rangle \rangle.
\end{eqnarray}
Following the general scheme of the irreducible GF
method, see Eqs. (\ref{e71}) - (\ref{e79}), we introduce the irreducible
GF as follows
\begin{eqnarray}
\label{e139}
(^{(ir)}\langle \langle a_{i\sigma}a^{\dagger}_{i-\sigma}a_{i-\sigma} \vert
a^{\dagger}_{i'\sigma}\rangle \rangle_ {\omega} )   =
\langle \langle a_{i\sigma}a^{\dagger}_{i-\sigma}a_{i-\sigma}\vert
a^{\dagger}_{i'\sigma} \rangle \rangle_{\omega} - \\ \nonumber
- \langle n_{i-\sigma}\rangle G_{11} + \langle a_{i\sigma}a_{i-\sigma}\rangle
\langle \langle a^{\dagger}_{i-\sigma} \vert a^{\dagger}_{i'\sigma} \rangle \rangle_{\omega},\\
\nonumber
(^{(ir)}\langle \langle a^{\dagger}_{i\sigma}a_{i\sigma}a^{\dagger}_{i-\sigma}
\vert a^{\dagger}_{i'\sigma} \rangle \rangle_ {\omega} )  =
\langle \langle a^{\dagger}_{i\sigma}a_{i\sigma}a^{\dagger}_{i-\sigma}\vert
a^{\dagger}_{i'\sigma} \rangle \rangle_{\omega} - \\ \nonumber
- \langle n_{i\sigma}\rangle G_{21} +
\langle a^{\dagger}_{i\sigma}a^{\dagger}_{i-\sigma}\rangle \langle \langle a_{i\sigma} \vert
a^{\dagger}_{i'\sigma} \rangle \rangle_{\omega}. \nonumber
\end{eqnarray}
Therefore, instead of the algebra of the normal state's
operator $(a_{i\sigma}$, $a^{\dagger}_{i\sigma}$, $n_{i\sigma})$ , for description of superconducting
states, one has to use a more general algebra,
which includes the operators 
$(a_{i\sigma}$, $a^{\dagger}_{i\sigma}$, $n_{i\sigma}$,
$a^{\dagger}_{i\sigma}a^{\dagger}_{i-\sigma}$, and
$a_{i-\sigma}a_{i\sigma}) $. 
The self-consistent system of
superconductivity equations follows from the Dyson
equation
\begin{equation}
\label{e140} \hat G_{ii'}(\omega) = \hat G^{0}_{ii'}(\omega) +
\sum_{jj'} \hat G^{0}_{ij} (\omega) \hat M_{jj'}( \omega) \hat
G_{j'i'} (\omega).
\end{equation}
Green's function in the generalized mean-field's
approximation, $G^{0}$, and the mass operator $ M_{jj'}$ are
defined as follows
\begin{equation}
\label{e141} \sum_{j}(\omega \tau_{0}\delta_{ij} - t_{ij}
\tau_{3} - \Sigma^{c}_{i\sigma}) G^{0}_{ji'} =
\delta_{ii'}\tau_{0},
\end{equation}
\begin{equation}
\label{e142} M_{kk'} = \sum_{jj'} (\langle \langle ( \rho_{kj} \tau_{3}
\psi_{j} )^{(ir)} \vert (\psi^{\dagger}_{j'} \tau_{3}
\rho_{j'k'})^{(ir)} \rangle \rangle)^{(p)}_{\omega},
\end{equation}
\begin{eqnarray}
\label{e143} \hat M_{ii'} (\omega) = \\ \sum_{jj'} 
\begin{pmatrix}
(^{(ir)}\langle \langle a_{j \uparrow} \rho_{ij\uparrow} \vert
\rho_{j'i'\uparrow} a^{\dagger}_{j'\uparrow} \rangle \rangle^{(ir)})^{(p)} &
(^{(ir)}\langle \langle a_{j \uparrow} \rho_{ij\uparrow} \vert
\rho_{j'i'\downarrow} a_{j'\downarrow} \rangle \rangle^{(ir)})^{(p)} \cr
(^{(ir)}\langle \langle a^{\dagger}_{j \downarrow} \rho_{ji\downarrow} \vert
\rho_{j'i'\uparrow} a^{\dagger}_{j'\uparrow}
\rangle \rangle^{(ir)})^{(p)} & (^{(ir)}\langle \langle a^{\dagger}_{j \downarrow} \rho_{ji\downarrow} \vert
\rho_{i'j'\downarrow} a_{j'\downarrow} \rangle \rangle^{(ir)})^{(p)} \cr 
\end{pmatrix}.
\nonumber
\end{eqnarray}
The mass operator (\ref{e143}) describe the processes of
inelastic electron scattering on lattice vibrations.
The elastic processes are described by the quantity $\Sigma^{c}_{i\sigma},$  
see Eq. (\ref{e84}). An approximate expression for the
mass operator (\ref{e143}) follows from the following trial
solution:
\begin{equation}
\label{e144} \langle \rho_{j'i'\sigma}(t)a^{\dagger}_{j'\sigma}(t)
a_{j\sigma} \rho_{ij\sigma}\rangle^{(ir)} \approx \langle \rho_{j'i'\sigma}(t)
\rho_{ij\sigma}\rangle \langle a^{\dagger}_{i'\sigma}(t)a_{j\sigma}\rangle.
\end{equation}
This approximation corresponds to the standard
approximation in the superconductivity theory, which
in the diagram-technique language is known as neglecting
vertex corrections, that is, neglecting electron correlations
in the propagation of fluctuations of charge
density. Taking into account this approximation, one
can write down the mass operator (\ref{e143}) in the following
form
\begin{equation}
\label{e145} \hat M_{ii'} (\omega) = \hat M^{1}_{ii'} (\omega) +
\hat M^{2}_{ii'} (\omega).
\end{equation}
The first term,$M^{1}$, has the form typical for an interacting
electron-phonon system
\begin{eqnarray}
\label{e146} M^{1}_{ii'}( \omega) = \sum_{nn'} \sum_{jj'}
V_{ijn} V_{j'i'n'}{1 \over 2} \int_{-\infty}^{+\infty} \frac
{d\omega_{1}d{\omega}_{2}}{\omega - \omega_{1} - \omega_{2}}
\left(\cot \frac{\beta \omega_{1}}{2} + \tan \frac{\beta \omega_{2}}{2} \right) \times \nonumber\\
\left(-{1 \over \pi} \textrm{Im} \langle \langle u_{n} \vert u_{n'}\rangle \rangle_{\omega_{2}} \right)  \left(-{1 \over
\pi} \tau_{3} \textrm{Im} \langle \langle \psi_{j} \vert
\psi^{\dagger}_{j'}\rangle \rangle_{\omega_{1}}\tau_{3} \right).
\end{eqnarray}
The second term $M^{2}_{ii'}$  has a more complicated structure
\begin{equation}
\label{e147} M^{2}_{ii'} = {U^2 \over 2}
\int_{-\infty}^{+\infty} \frac {d\omega_{1}d{\omega}_{2}}{\omega
- \omega_{1} - \omega_{2}} \left(\cot \frac{\beta \omega_{1}}{2} + \tan \frac{\beta \omega_{2}}{2} \right)
 \begin{pmatrix} m_{11}&m_{12}\cr
m_{21}&m_{22}\cr 
\end{pmatrix},
\end{equation}
where
\begin{eqnarray}
\label{e148}
m_{11} = \left(-{1 \over \pi} \textrm{Im} \langle \langle n_{i\downarrow} \vert
n_{i'\downarrow}\rangle \rangle_{\omega_{2}} \right)  \left(-{1 \over \pi} \textrm{Im}
\langle \langle a_{i\uparrow} \vert a^{\dagger}_{i'\uparrow}\rangle \rangle_{\omega_{1}} \right),
 \\
m_{12} = \left({1 \over \pi} \textrm{Im} \langle \langle n_{i\downarrow} \vert
n_{i'\uparrow}\rangle \rangle_{\omega_{2}} \right)  \left(-{1 \over \pi} \textrm{Im} \langle \langle a_{i\uparrow}
\vert a^{\dagger}_{i'\downarrow}\rangle \rangle_{\omega_{1}} \right),
\nonumber \\
m_{21} = \left({1 \over \pi} \textrm{Im} \langle \langle n_{i\uparrow} \vert
n_{i'\downarrow}\rangle \rangle_{\omega_{2}} \right)  \left(-{1 \over \pi} \textrm{Im} \langle \langle
a_{i\downarrow} \vert a^{\dagger}_{i'\uparrow}\rangle \rangle_{\omega_{1}} \right),
\nonumber \\
m_{22} = \left(-{1 \over \pi} \textrm{Im} \langle \langle n_{i\uparrow} \vert
n_{i'\uparrow}\rangle \rangle_{\omega_{2}} \right) \left(-{1 \over \pi}
\textrm{Im} \langle \langle a_{i\downarrow} \vert
a^{\dagger}_{i'\downarrow}\rangle \rangle_{\omega_{1}} \right). \nonumber
\end{eqnarray} 
The definition (\ref{e139}) and Eqs. (\ref{e140}) - (\ref{e148}) allowed
us to perform a systematic derivation of superconductivity
equations for transition metals~\cite{kuz02,b91,b67,b47}
and disordered binary alloys~\cite{b56,b72} in the strong coupling
approximation. Thus, it is the adequate
description of the generalized mean-field in superconductors,
taking into account anomalous mean values,
which allowed us to construct compactly and self-consistently,
the superconductivity equations in the strong coupling
approximation.
%
\subsection{Magnetic Polaron Theory}
%
%
%
To obtain a clear idea of the fundamental importance
of the complex structure of mean fields let us
investigate the problem of the magnetic polaron~\cite{b81,b160}
in magnetic semiconductors~\cite{nag79}. That is, in substances
which have a subsystem of itinerant carriers and
a subsystem of local magnetic moments~\cite{b161,b81,b160}.
Usually the model of  $s-d$ exchange (\ref{e25}) is used for
description of magnetic semiconductors. It is important
to keep in mind that there are different spin and charge
degrees of freedom in that model, which are described
by the operators: %
$ a_{k \sigma}, \quad a^{\dagger}_{k \sigma}, \quad n_{k \sigma} = a^{\dagger}_{k \sigma}a_{k \sigma};$ \,
$ S^{+}_{k}, \quad S^{-}_{-k} = ( S^{+}_{k} )^{\dagger};$ 
$ b_{k\sigma} =  \sum_{q} ( S^{-\sigma}_{-q} a_{q+k-\sigma} +
z_{\sigma}S^{z}_{-q} a_{q+k\sigma}); $ 
and
$\sigma^{+}_{k} = \sum_{q}
a^{\dagger}_{k\uparrow}a_{k+q\downarrow} ;\quad \sigma^{-}_{k} =
\sum_{q} a^{\dagger}_{k\downarrow}a_{k+q\uparrow}. $ 
The complete
algebra of \textbf{relevant operators} is given by
$$\{a_{i\sigma},\quad S^{z}_{i},\quad S^{-\sigma}_{i},\quad S^{z}_{i} a_{i\sigma},\quad S^{-\sigma}_{i} a_{i-\sigma}\}.$$
Three additional GFs arise upon calculating the one-electron
GF, because of the interaction between the
subsystems. In order to describe correctly the spin and
charge degrees of freedom in magnetic semiconductors,
as well as their interaction, the original GF must
have the following matrix form:
\begin{equation}\label{e149}
{\tiny
\begin{pmatrix}
 \langle \langle a_{i \sigma}  \vert a^{\dagger}_{j \sigma'} \rangle \rangle & \langle \langle a_{i \sigma} \vert
S^{z}_{j} \rangle \rangle & \langle \langle a_{i \sigma}  \vert S^{\sigma'}_{j} \rangle \rangle &
\langle \langle a_{i \sigma}  \vert a^{\dagger}_{j \sigma'} S^{z}_{j} \rangle \rangle & \langle \langle a_{i \sigma}  \vert a^{\dagger}_{j -\sigma'} S^{\sigma'}_{j} \rangle \rangle \cr
\langle \langle S^{z}_{i} \vert a^{\dagger}_{j \sigma'}\rangle \rangle & \langle \langle S^{z}_{i} \vert S^{z}_{j} \rangle \rangle & \langle \langle S^{z}_{i} \vert S^{\sigma'}_{j} \rangle \rangle &
\langle \langle S^{z}_{i} \vert a^{\dagger}_{j \sigma'}S^{z}_{j} \rangle \rangle & \langle \langle S^{z}_{i} \vert a^{\dagger}_{j -\sigma'}S^{\sigma'}_{j} \rangle \rangle \cr
\langle \langle S^{-\sigma}_{i} \vert a^{\dagger}_{j \sigma'}\rangle \rangle & \langle \langle S^{-\sigma}_{i} \vert S^{z}_{j} \rangle \rangle & \langle \langle S^{-\sigma}_{i}\vert S^{\sigma'}_{j} \rangle \rangle&
\langle \langle S^{-\sigma}_{i} \vert a^{\dagger}_{j \sigma'}S^{z}_{j} \rangle \rangle & \langle \langle S^{-\sigma}_{i} \vert a^{\dagger}_{j -\sigma'}S^{\sigma'}_{j} \rangle \rangle \cr
\langle \langle S^{z}_{i} a_{i\sigma}  \vert a^{\dagger}_{j\sigma'} \rangle \rangle & \langle \langle S^{z}_{i}a_{i\sigma} \vert
S^{z}_{j}  \rangle \rangle & \langle \langle S^{z}_{i}a_{i \sigma}  \vert S^{\sigma'}_{j} \rangle \rangle &
\langle \langle S^{z}_{i}a_{i \sigma}  \vert a^{\dagger}_{j \sigma'} S^{z}_{j} \rangle \rangle & \langle \langle S^{z}_{i}a_{i \sigma}
\vert a^{\dagger}_{j -\sigma'} S^{\sigma'}_{j}\rangle \rangle \cr
\langle \langle S^{-\sigma}_{i}a_{i -\sigma} \vert a^{\dagger}_{j \sigma'} \rangle \rangle &\langle \langle S^{-\sigma}_{i}a_{i -\sigma} \vert S^{z}_{j} \rangle \rangle & \langle \langle S^{-\sigma}_{i}a_{i -\sigma}\vert S^{\sigma'}_{j} \rangle \rangle&
\langle \langle S^{-\sigma}_{i} a_{i -\sigma} \vert a^{\dagger}_{j \sigma'}S^{z}_{j} \rangle \rangle & \langle \langle S^{-\sigma}_{i}a_{i -\sigma} \vert a^{\dagger}_{j -\sigma'}S^{\sigma'}_{j} \rangle \rangle \cr 
\end{pmatrix}}.
\end{equation}
The functional structure of GF (\ref{e149}) shows that
there are two regimes of quasiparticle dynamics: the
scattering regime and the regime, where the electron-magnon's
bound states (the magnetic polaron) are
formed. To somewhat simplify our task we will use the
following reduced algebra of relevant operators 
$(a_{k \sigma},a^{\dagger}_{k \sigma}, b_{k\sigma},b^{\dagger}_{k\sigma})$. 
In this case, however, we will need a
separate consistent consideration of the dynamic in the
localized spin's subsystem~\cite{b81,b160}. For this purpose
we use GF
\begin{eqnarray}\label{e150} \mathcal G^{+-}(k;t - t') =
\langle \langle S^{+}_{k}(t),S^{-}_{-k}(t')\rangle \rangle.  
\end{eqnarray}
Now, the relevant matrix's GF for the problem of
magnetic dynamics is given by
\begin{equation}\label{e151}
\hat {\mathcal G} (k;\omega) =
\begin{pmatrix} 
\langle \langle S^{+}_{k}\vert S^{-}_{-k} \rangle \rangle & \langle \langle S^{+}_{k}\vert \sigma^{-}_{-k}\rangle \rangle \cr 
\langle \langle \sigma^{+}_{k} \vert S^{-}_{-k} \rangle \rangle &\langle \langle \sigma^{+}_{k} \vert \sigma^{-}_{-k} \rangle \rangle \cr 
\end{pmatrix}.
\end{equation}
The Dyson equation for GF (\ref{e151})
\begin{equation}
\label{e152} \hat{\mathcal G} = \hat{\mathcal G_{0}} + \hat{\mathcal G_{0}} \hat M \hat{\mathcal G}
\end{equation}
determines GF $\hat {\mathcal G_{0}}$   in the generalized mean-field
approximation, and the mass operator $\hat M$~\cite{b160}. For
description of the charge-carriers subsystem we use the
GF in the form
\begin{eqnarray}\label{e153} g_{k\sigma}(t - t') =
\langle \langle a_{k \sigma}(t),a^{\dagger}_{k \sigma}(t')\rangle \rangle.
\end{eqnarray}
The Dyson equation for this GF is given by~\cite{b160}
\begin{equation}
\label{e154}
  g_{k\sigma}(\omega) =
g_{k\sigma}^{0}(\omega) +
g_{k\sigma}^{0}(\omega) M_{k\sigma}(\omega) g_{k\sigma}(\omega).
\end{equation}
Equations (\ref{e152}) and (\ref{e154})  allow one to investigate
self-consistently, the spin and the charge's quasiparticle
dynamics in the system. In contrast to the scattering
regime, for the one-electron GF (\ref{e153}) in the bound state's
formation regime we find the following expression
for the GF in the generalized mean-field's approximation
\begin{equation} \label{e155}
 \langle \langle a_{k \sigma}  \vert a^{\dagger}_{k \sigma} \rangle \rangle^{0} =
 (\textrm{det} \hat \Omega)^{-1} = (\omega  - \varepsilon(k\sigma) - I^{2} N^{-1} \chi^{b}_{k \sigma} (\omega))^{-1},
\end{equation}
where
\begin{eqnarray}\label{e156}
\chi^{b}_{k \sigma} (\omega) =
\sum_{q} \{ \frac{ \langle S^{-\sigma}_{-q}S^{\sigma }_{q} \rangle}{(1 - 
I \Lambda_{k\sigma}(\omega))(\omega + z_{\sigma}\omega(q) - \varepsilon(k+q -\sigma)) }
\nonumber \\
+ \frac{(1 + I \Lambda_{k\sigma}(\omega)) \langle (S^{z}_{-q})~^{ir}(S^{z}_{q})~^{ir} \rangle }{(1 - 
I \Lambda_{k\sigma}(\omega)) (\omega - \varepsilon(k+q \sigma))} \},
\end{eqnarray}
\begin{equation} \label{e157}
 \Lambda_{k\sigma}(\omega) = \frac{1}{N} \sum_{q}  \frac{1}{(\omega + z_{\sigma}\omega(q) - \varepsilon(k+q -\sigma))}.
\end{equation}
The quantity  $\chi^{b}_{k \sigma} (\omega)$ plays the role of the generalized
susceptibility for spin-electron bound states. It is this
property that distinguishes the bound-state regime from
the scattering regime, where instead of the
electron-spin susceptibility  $\chi^{b}_{k \sigma} (\omega)$ appears $\chi ^{s}_{0}(k,\omega)$
\begin{eqnarray}\label{e158}
\chi ^{s}_{0}(k,\omega) = N^{-1} \sum_{p} \frac {
(f_{p+k\downarrow} - f_{p\uparrow})}{\omega_{p,k}}.
\end{eqnarray}
We use the following notation 
\begin{eqnarray}  \label{e159}
\omega^{s}_{p,k} = (\omega + \epsilon_{p }  - \epsilon_{p+k }  - \Delta_{I} ); \,\,
 \Delta_{I}  = 2I S_{z},   \nonumber
\end{eqnarray}
$$n_{\sigma} = \frac{1}{N}\sum_{q}  \langle a^{\dagger}_{q\sigma}a_{q\sigma}\rangle  =
\frac{1}{N} \sum_{q} f_{q\sigma}
=\sum_{q}(\exp(\beta \varepsilon(q\sigma)) + 1), $$ 
$$ \varepsilon(q \sigma) = \epsilon_{q }  -z_{\sigma}I  S_{z},  $$

$$ \bar n = \sum  ( n_{\uparrow} + n_{\downarrow}); \quad 0 \le
\bar n \le 2;  \,\, S_{z} = N^{-1/2}\langle S^{z}_{0}\rangle.  $$.
The magnetic polaron's spectrum is given by
\begin{equation} \label{e160}
E_{k \sigma} = \varepsilon(k\sigma) + I^{2} N^{-1} \chi^{b}_{k \sigma} (E_{k \sigma} ).
\end{equation}
One can show that for any value of the electron's spin
projection the polaron spectrum of the bound electron-magnon's state contains two branches. In the so-called
atomic limit ($\epsilon_{k} = 0$ ), when $k \rightarrow 0$, $\omega \rightarrow 0$,
we obtain
\begin{equation} \label{e161}
\langle \langle a_{k \sigma}  \vert a^{\dagger}_{k \sigma} \rangle \rangle^{0} =
\frac{S + z_{\sigma} S_{z}}{2S + 1} ( \omega + IS)^{-1} +
\frac{S - z_{\sigma} S_{z}}{2S + 1} ( \omega - I(S + 1))^{-1}.
\end{equation}
Here, $S$ and $S_{z} = \langle S^{z}_{0}\rangle/\sqrt{N}$ denote the spin magnitude
and the magnetization, respectively. The obtained
result, Eq. (\ref{e160}), is in perfect agreement with the result
of Mattis and Shastry~\cite{mattis81}, who investigated the magnetic
polaron's problem for $ \textrm{T} = 0$
\begin{equation} \label{e162}
\langle \langle a_{k \sigma}  \vert a^{\dagger}_{k \sigma} \rangle \rangle^{0}\vert_{T=0} =
\{ \omega - \varepsilon ( k\sigma) -
\delta_{\sigma\downarrow} 2I^{2}S \frac{\Lambda_{k\sigma}(\omega)}{(1 - I \Lambda_{k\sigma}(\omega))}\}^{-1}
\end{equation}
Thus, the magnetic polaron is formed in the case of
antiferromagnetic $s$-$d$ interaction ($ I < 0$). In order to get
a clear idea of the spectrum character let us now consider
two limiting cases:
\begin{itemize}
\item[(i)]
 a wide-band semiconductor ($|I|S \ll W $) \\
\begin{eqnarray} \label{e163}
E_{k  \downarrow}  \simeq \epsilon _{k} +
I \frac{S (S + S_{z} +1)   + S_{z} (S - S_{z} +1 )}{2S }   + \\
\frac{(-I)}{N}\sum_{q}   \frac{( \epsilon_{k-q} - \epsilon _{k} +
2I (S - S_{z} ))}{(\epsilon_{k-q} - \epsilon _{k} + 2IS_{z}  )}   \frac{ \langle S^{+}_{q}S^{-}_{-q} \rangle}{2S}, \nonumber
\end{eqnarray}
\item[(ii)]
a narrow-band semiconductor ($|I|S \gg W $) \\
\begin{eqnarray} \label{e164}
E_{k  \downarrow}  \simeq I ( S + 1 ) + \frac{2 ( S + 1 )(S + S_{z})}{( 2S + 1 )(S + S_{z} +1)}
\epsilon _{k} + \nonumber \\
\frac{1}{N}\sum_{q} \frac{(\epsilon_{k-q} - \epsilon _{k})}{( 2S + 1 )}\frac{ \langle S^{+}_{q}S^{-}_{-q} \rangle}{(S + S_{z} +1) }.
\end{eqnarray}
\end{itemize}
Here, $W$ is the band width for $\textrm{I} = 0$. Note that in order
to make expressions more compact we omitted the correlation
function $K^{zz}_{q}$ in the above formulae.\\
Consider now the low-temperature spin-wave
regime, where one can assume that $S_{z} \simeq S$. In this case
we have
$$\langle S^{+}_{q}S^{-}_{-q}\rangle \simeq 2S( 1 + N(\omega (q))). $$
One can show that for
\begin{itemize}
\item[(i)]
a wide-band semiconductor ($|I|S \ll W $) \\
\begin{eqnarray} \label{e165}
E_{k  \downarrow}  \simeq \epsilon _{k} +  IS + \frac{2I^{2}S}{N}
\sum_{q} \frac{1}{(\epsilon _{k} - \epsilon_{k-q} + 2IS )} + \nonumber \\
\frac{(-I)}{N}\sum_{q}   \frac{( \epsilon_{k-q} - \epsilon _{k} )}{(\epsilon_{k-q} - \epsilon _{k} - 2IS )}
N(\omega (q)),
\end{eqnarray}
\item[(ii)]
a narrow-band semiconductor ($|I|S \gg W $) \\
\begin{eqnarray} \label{e166}
E_{k  \downarrow}  \simeq I ( S + 1 ) + \frac{2S }{( 2S + 1 )}\epsilon _{k} +
\frac{1}{N}\sum_{q} \frac{2S }{( 2S + 1 )} \frac{(\epsilon_{k-q} - \epsilon _{k})}{( 2S + 1 )}
N(\omega (q)).
\end{eqnarray}
\end{itemize}
Let us now estimate the energy of the bound state of the
magnetic polaron
\begin{equation} \label{e167}
\varepsilon_{B} = \varepsilon_{k\downarrow} - E_{k  \downarrow}.
\end{equation}
Taking into account that
$$ \varepsilon_{k\downarrow} =  \epsilon _{k} + IS.$$
we obtain the following expressions for the binding
energy $\varepsilon_{B}$:
\begin{itemize}
\item[(i)]
a wide-band semiconductor ($|I|S \ll W $) \\
\begin{eqnarray} \label{e168}
\varepsilon_{B} = \varepsilon^{0}_{B1} -
\frac{(-I)}{N}\sum_{q}   \frac{( \epsilon_{k-q} - \epsilon _{k} )}{(\epsilon_{k-q} - \epsilon _{k} - 2IS )}
N(\omega (q)),
\end{eqnarray}
\item[(ii)]
a narrow-band semiconductor ($|I|S \gg W $) \\
\begin{eqnarray} \label{e169}
\varepsilon_{B} =  \varepsilon^{0}_{B2} -
\frac{1}{N}\sum_{q} \frac{2S }{( 2S + 1 )} \frac{(\epsilon_{k-q} - \epsilon _{k})}{( 2S + 1 )}
N(\omega (q)),
\end{eqnarray}
\end{itemize}
где
\begin{eqnarray} \label{e170}
\varepsilon^{0}_{B1} = \frac{(2I^{2}S)}{N}\sum_{q} \frac{1}{(\epsilon_{k-q} - \epsilon _{k} - 2IS )}
\simeq \frac{|I|S}{W}|I|, \nonumber \\
 \varepsilon^{0}_{B2} = - I + \frac{\epsilon _{k}}{( 2S + 1 )} \simeq |I|.
\end{eqnarray}
The outlined theory gives a complete description of
the magnetic polaron for finite temperatures~\cite{b81,b160},
revealing the fundamental importance of the complicated
structure of generalized mean-fields, which cannot
be reduced to simple functionals of mean spin and
particle densities.
%
\section{Broken Symmetry, Quasiaverages, and Physics of Magnetic Materials}
%
%
It is well known that the concept of spontaneously
broken symmetry~\cite{ander58,namb60,namb61,gold61,blud63,pander63,hig66,wagn66,dfors,dak78} is one of the most important
notions in the quantum field theory and elementary
particle physics. This is especially so as far as creating
a unified field theory, uniting all the different forces of
nature~\cite{nnb85}, is concerned. One should stress that the
notion of spontaneously broken symmetry came to the
quantum field theory from solid-state physics. It was
originated in quantum theory of magnetism, and later
was substantially developed and found wide applications
in the gauge theory of elementary particle physics~\cite{grib,fstr05}.
It was in the quantum field theory where the
ideas related to that concept were quite substantially
developed and generalized. The analogy between the
Higgs mechanism giving mass to elementary particles
and the Meissner effect in the Ginzburg-Landau superconductivity
theory is well known~\cite{ander58,namb60,blud63,pander63,hig66,dak78,du05}.
Both effects are consequences of spontaneously
broken symmetry in a system containing two
interacting subsystems. \\ A similar situation is encountered
in the quantum solid-state theory~\cite{pwa84}. Analogies
between the elementary particle and the solid-state theories
have both cognitive and practical importance for
their development~\cite{heis69}. We have already mentioned the
analogies with the Higgs effect playing an important
role in these theories~\cite{ewkk01}. However, we have every
reason to also consider analogies with the Meissner
effect in the Ginzburg-Landau superconductivity
model, because the Higgs model is, in fact, only a relativistic
analogue of that model~\cite{ander58,blud63,pander63,hig66,dak78}. On
the same ground one can consider the existence of magnons
in spin systems at low temperatures~\cite{umez84}, acoustic
and optical vibration modes in regular lattices or in
multi-sublattice magnets, as well as the vibration spectra
of interacting electron and nuclear spins in magnetically-ordered crystals~\cite{tur83}. \\  The isotropic Heisenberg
ferromagnet (\ref{e12}) is often used as an example of a system
with spontaneously broken symmetry~\cite{dfors}. This
means that the Hamiltonian symmetry, the invariance
with respect to rotations, is no longer the symmetry of
the equilibrium-state. Indeed the ferromagnetic states
of the model are characterized by an axis of the preferred
spin alignment, and, hence, they have a lower
symmetry than the Hamiltonian itself. However, as was
stressed by Anderson~\cite{pwa84,ander75,ander90}, the ground state
of the Heisenberg ferromagnet is an eigenstate of the
relevant transformation of continuous symmetry (spin
rotation). Therefore, \textbf{the symmetry is not broken} and the
low-energy excitations do not have novel properties.
The symmetry breaking takes place when the ground
state is no longer an eigenstate of a particular symmetry
group, as in antiferromagnets or in superconductors.
Only in this case the concepts of quasi-degeneracy,
Goldstone bosons, and Higgs phenomenon can be
applied~\cite{pwa84,ander75,ander90}.\\  The essential role of the physics
of magnetism in the development of symmetry ideas
was noted in the paper~\cite{namb07} by the 2008 Nobel Prize
Winner Y. Nambu, devoted to the development of the
elementary particle physics and the origin of the concept
of spontaneous symmetry breakdown. Nambu
points out that back at the end of the 19th century
P. Curie~\cite{pcu94,pcu66} used symmetry principles in the
physics of condensed matter. P. Curie~\cite{pcu94} used symmetry
ideas in order to obtain analogues of selection
rules for various physical effects, for instance, for the
Wiedemann effect~\cite{pcu94,pcu66} (see the books~\cite{pcu66,pcu68,nye}). Nambu also notes:\\
"More relevant examples for us, however, came
after Curie. The ferromagnetism is the prototype of
today's spontaneous symmetry breaking, as was
explained by the works of Weiss~\cite{pweis}, Heisenberg~\cite{wh28},
and others. Ferromagnetism has since served us as a
standard mathematical model of spontaneous symmetry
breaking".\\
This statement by Nambu should be understood in
light of the clarification made by Anderson~\cite{pwa84,ander75,ander90}  
(see also the paper~\cite{cmp07}). P. Curie was indeed a
forerunner of the modern concepts of the quantum theory
of magnetism. He formulated the \textbf{Curie principle}:
"Dissymmetry creates the phenomenon". According to
this principle~\cite{pcu94,pcu66}:
\begin{quote}
"A phenomenon can exist in a medium possessing
a characteristic symmetry $ (G_{1})$ or the symmetry of one of
that characteristic symmetry subgroups $ (G  \subseteq   G_{i})$".
\end{quote}
In other words, some symmetry elements may coexist
with some phenomena, but this is not necessarily the
case. What is required is that some symmetry elements
are absent. This is that dissymmetry, which creates the
phenomenon. One of the formulations of the dissymmetry
principle has the following form~\cite{shub72}
\begin{equation}
\label{e176}
G_{i}^{phenomena} \supseteq G_{media} = \bigcap  G_{i}^{phenomena},
\end{equation}
or, alternatively
\begin{equation}
\label{e177}
G_{i}^{properties} \supseteq G_{object} = \bigcap  G_{i}^{properties}.
\end{equation}
Note that the concepts of symmetry, dissymmetry,
and broken symmetry became very widespread in various
branches of science and art~\cite{shub72,cag98,chat08}.\\
Essential progress in the understanding of the spontaneously
broken symmetry concept is connected with
\textbf{Bogoliubov's ideas about quasiaverages}~\cite{nnb73,nnb75}.\\
Indeed, as was noticed in the book~\cite{dfors}:
"\ldots the canonical ensemble $\rho \sim \exp (- \beta H)$ is no
longer a  good ensemble for the spontaneously
ordered systems. Averaging over this ensemble
would be to average, among other properties, over all  directions of
the total spin. That is fine  in a
paramagnet, and passes for a number of purposes  in the 
ferromagnetic regime as well,  but for other purposes,   such as the calculation
of $\langle \vec{S}^{\textrm{tot}} \rangle,$ it would be a foolish thing to do.
One could use $\exp (- \beta H)$ to weight states of different energy, but in addition
one should specify that the trace is to be taken only over those states for which $ \vec{S}^{\textrm{tot}} $
points in the $z$-direction. Formally, one would then have something like
$$ \rho = \textrm{const} \, {\mathcal P} \vec{S}^{\textrm{tot}} \exp (- \beta H),$$
where the projection operator ${\mathcal P} \vec{S}^{\textrm{tot}}$ eliminates all but those states for which
$ \vec{S}^{\textrm{tot}} $ points along $z.$"\\
As we see, this statement written in 1975 contains in
a concise form an argumentation in favor of using the
ideas of quasiaverages~\cite{nnb73,nnb75}, but it does not mention
them explicitly. However, the notion of quasiaverages~\cite{nnb73,nnb75}
was formulated by N.N. Bogoliubov back in
1960-1961 (see also the paper~\cite{wagn66}). \\ It is necessary to stress,
that the starting point for Bogoliubov's paper~\cite{nnb73,nnb75} was
an investigation of \textbf{additive conservation laws and
selection rules}, continuing and developing the already
mentioned above approach by P. Curie for derivation of
selection rules for physical effects. Bogoliubov demonstrated
that in the cases when the state of statistical
equilibrium is degenerate, as in the case of a ferromagnet,
one can remove the degeneracy of equilibrium
states with respect to the group of spin rotations by
including in the Hamiltonian $H$ an additional noninvariant
term $\nu M_{z} V$ with an infinitely small $\nu$. This replaces
the ordinary averages by quasiaverages~\cite{nnb73,nnb75} of
the form
\begin{equation}
\label{e178}
\curlyeqprec A \curlyeqsucc = \lim_{\nu \rightarrow 0} \langle A \rangle_{\nu \vec{e}},
\end{equation}
where $\langle A \rangle_{\nu \vec{e}}$ is the ordinary average of the quantity $A$
with respect to the Hamiltonian $H_{\nu \vec{e}} = H + \nu (\vec{e} \cdot \vec{M}) V.$
Thus, the presence of degeneracy is directly reflected
on quasiaverages via their dependence on the arbitrary
vector $\vec{e}.$ The ordinary averages can be obtained from
the quasiaverages by integrating over all possible
directions of $\vec{e}$
\begin{equation}
\label{e179}
\langle A \rangle = \int \curlyeqprec A \curlyeqsucc d \vec{e}.
\end{equation}
The question of symmetry breaking within the
localized and band models of antiferromagnets was
studied by the author of this review in the papers~\cite{kuz02,b100,b142}.
It has been found there that the concept of
symmetry breaking in the band model of
magnetism~\cite{b142} is much more complicated than in the
localized model. In the framework of the band model of
magnetism one has to additionally consider the so-called
\textbf{anomalous propagators} of the form
\begin{eqnarray} \textrm{FM}: G_{fm} \sim \langle \langle a_{k\sigma};a^{\dag}_{k-\sigma} \rangle \rangle,  \nonumber \\
\nonumber \textrm{AFM}: G_{afm} \sim \langle \langle a_{k+Q\sigma};a^{\dag}_{k+Q'\sigma'} \rangle \rangle. \nonumber
\end{eqnarray}
In the case of the band antiferromagnet the ground
state of the system corresponds to a spin-density wave
(SDW), where a particle scattered on the internal inhomogeneous
periodic field gains the momentum $Q - Q'$
and changes its spin:  $\sigma \rightarrow \sigma'$. The long-range order
parameters are defined as follows
\begin{eqnarray} \label{e180}
\textrm{FM}: m =
1/N\sum_{k\sigma} \langle a^{\dag}_{k\sigma}a_{k-\sigma} \rangle,\\  \textrm{AFM}: M_{Q}
= \sum_{k\sigma} \langle a^{\dag}_{k\sigma}a_{k+Q-\sigma} \rangle. \label{e181}  \end{eqnarray}
It is important to stress, that the long-range order
parameters here are functionals of the internal field, which
in turn is a function of the order parameter. Thus, in the
cases of rotation and translation invariant Hamiltonians of
band ferro- and antiferromagnetics one has to add the following
infinitesimal sources removing the degeneracy:
\begin{eqnarray} \label{e182}
\textrm{FM}:  \nu\mu_{B}
H_{x}\sum_{k\sigma}a^{\dag}_{k\sigma}a_{k-\sigma},\\   \textrm{AFM}:
\nu \mu_{B} H \sum_{kQ} a^{\dag}_{k\sigma}a_{k+Q-\sigma}.
\label{e183}
\end{eqnarray}
Here,  $\nu \rightarrow 0$ after the usual in statistical mechanics
infinite-volume limit $V \rightarrow \infty$. The ground state in the
form of a spin-density wave was obtained for the first
time by Overhauser in investigations of nuclear matter~\cite{over28}.
There, the vector  $\vec{Q}$ is a measure of inhomogeneity
or translation symmetry breaking in the system. It
was written in the paper~\cite{tas46} (see also~\cite{koma93,bach94,tas48}) that
\begin{quote}
''\ldots  \emph{ in antiferromagnets a staggered magnetic field plays the role of a
symmetry-breaking field. No mechanism can generate a real staggered magnetic field in antiferromagnetic
materials}''.
\end{quote}
The analysis performed in the papers by Penn~\cite{penn66,penn67}
showed (see also~\cite{fal80}) that the antiferromagnetic
and more complicated states (for instance, ferrimagnetic)
can be described in the framework of a generalized
mean-field approximation. In doing that we have to
take into account both the normal averages
$\langle a^{\dag}_{i\sigma}a_{i\sigma}\rangle$,
and the anomalous averages $\langle a^{\dag}_{i\sigma}a_{i-\sigma}\rangle.$\\
It is clear that
the anomalous terms (\ref{e182}) and (\ref{e183})   break the original
rotational symmetry of the Hubbard Hamiltonian.
Thus, the generalized mean-field's approximation has
the following form 
$n_{i-\sigma}a_{i\sigma} \simeq \langle n_{i-\sigma}\rangle a_{i\sigma} -
\langle a^{\dag}_{i-\sigma}a_{i\sigma}\rangle a_{i-\sigma}.$  
A self-consistent theory of band antiferromagnetism
was developed by the author of this review in the papers~\cite{kuz02,b142}
using the method of the irreducible GF. The
following definition was used:
\begin{eqnarray} \label{e184}
^{ir}\langle \langle a_{k+p\sigma}a^{\dag}_{p+q-\sigma}a_{q-\sigma} \vert
a^{\dag}_{k\sigma} \rangle \rangle_ {\omega} =
\langle \langle a_{k+p\sigma}a^{\dag}_{p+q-\sigma}a_{q-\sigma}\vert
a^{\dag}_{k\sigma} \rangle \rangle_{\omega} - \nonumber\\ \delta_{p,
0}\langle n_{q-\sigma}\rangle G_{k\sigma} - \langle a_{k+p\sigma}a^{\dag}_{p+q-\sigma}\rangle
\langle \langle a_{q-\sigma} \vert a^{\dag}_{k\sigma} \rangle \rangle_{\omega}. \end{eqnarray} 
The algebra of relevant operators must be chosen as follows
 $(a_{i\sigma}$,
$a^{\dag}_{i\sigma}$, $n_{i\sigma}$, $a^{\dag}_{i\sigma}a_{i-\sigma})$. The corresponding
initial GF will have the following matrix structure
$$
G^{AFM} \sim
\begin{pmatrix}
\langle \langle a_{i\sigma}\vert a^{\dag}_{j\sigma}\rangle \rangle & \langle \langle a_{i\sigma}\vert
a^{\dag}_{j-\sigma}\rangle \rangle \cr
\langle \langle a_{i-\sigma}\vert a^{\dag}_{j\sigma}\rangle \rangle & \langle \langle a_{i-\sigma}\vert
a^{\dag}_{j-\sigma} \rangle \rangle \cr   \end{pmatrix}.$$
The off-diagonal terms select the vacuum state of
the band's antiferromagnet in the form of a spin-density
wave. It is necessary to stress that the problem of the
band's antiferromagnetism~\cite{mizia,cal94} is quite involved,
and the construction of a consistent microscopic theory
of this phenomenon remains a topical problem.
%
\subsection{Quantum Protectorate and Microscopic Models of Magnetism}
%
%
The "\textbf{quantum protectorate}" concept was formulated
in the paper~\cite{pnas00}. Its authors, R. Laughlin and
D. Pines, discussed the most fundamental principles of
matter description in the widest sense of this word:\\
"It is possible to perform approximate calculations
for larger systems, and it is through such calculations that
we have learned why atoms have the size they do, why chemical
bonds have the length and strength they do, why solid matter has
the elastic properties it does, why some things are transparent
while others reflect or absorb light. With a little more
experimental input for guidance it is even possible to predict
atomic conformations of small molecules, simple chemical reaction
rates, structural phase transitions, ferromagnetism, and
sometimes even superconducting transition temperatures.
But the schemes for approximating are not first-principles
deductions but are rather art keyed to experiment, and thus tend
to be the least reliable precisely when reliability is most needed,
i.e., when experimental information is scarce, the physical behavior
has no precedent, and the key questions have not yet been
identified. \ldots We have succeeded in reducing all of ordinary
physical behavior to a simple, correct Theory of Everything only
to discover that it has revealed exactly nothing about many things
of great importance."~\cite{pnas00}\\
R. Laughlin and D. Pines show that there are facts
that are clearly true, (for instance, the value $e2/hc$) yet
they cannot be deduced by direct calculation from the
Theory of Everything, for exact results cannot be predicted by
approximate calculations. Thus, the
existence of these effects is profoundly important, for it
shows us that for at least some fundamental things in
nature the Theory of Everything is irrelevant.  Next,
the authors formulate their main thesis: emergent physical
phenomena, which are regulated by higher physical
principles, have a certain property, typical for these
phenomena only. This property is their insensitivity to
microscopic description. Thus, here, in essence, a most
broad question is posed:\\ 
"\textbf{what is knowable in the deepest sense of the term}?" \\

For
instance, the low-energy excitation spectrum of ordinary
crystal dielectrics contains a transversal and longitudinal
sound wave and nothing else, irrespective of
microscopic details (see also~\cite{pines05}). Therefore, in the
opinion of R. Laughlin and D. Pines, there is no need
"to prove" the existence of sound in solid bodies; this is
a consequence of the existence of elastic modules in the
long-wave scale, which in turn follows from the spontaneous
breaking of translation and rotation symmetries,
typical for the crystal state. This implies the converse
statement: very little one can learn about the
atomic structure of the solid bodies of crystal by investigating
their acoustic properties. Therefore, the authors
summarize, the crystal state is the simplest known
example of \textbf{the quantum protectorate, a stable state of
matter with low-energy properties determined by
higher physical principles and by nothing else}.\\
The existence of two scales, the low-energy and
high-energy scales, relevant to the description of magnetic
phenomena was stressed by the author of this
review in the papers~\cite{b41,b151,b155} devoted to comparative
analysis of models of localized and band models of
quantum theory of magnetism. It was shown there, that
the low-energy spectrum of magnetic excitations in the
magnetically-ordered solid bodies corresponds to a
hydrodynamic pole ($\vec{k}, \omega \rightarrow 0 $) in the generalized
spin susceptibility, which is present in the Heisenberg,
Hubbard, and the combined $s-d$ model (see Fig. 1).\\ 
%
%
%
%
\begin{figure}[h]
\centerline{\includegraphics[width=7cm]{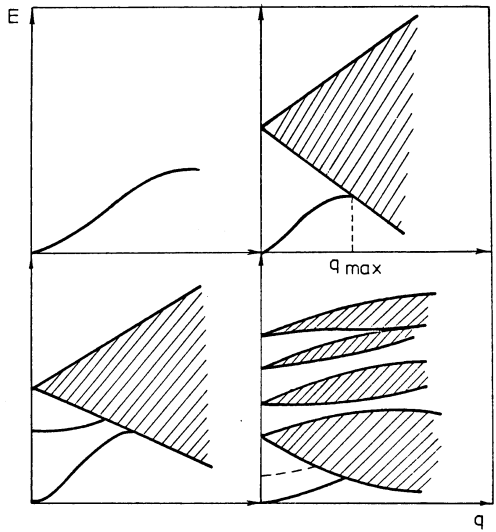}}
\caption{Schematic diagrams of excitation spectra in four
microscopic models of theory of magnetism. Upper left: the
Heisenberg model; upper right: the Hubbard model; lower
left: the Zener model; lower right: the multiband Hubbard
model.}
\end{figure}
In the Stoner band model the hydrodynamic pole is
absent, there are no spin waves there. At the same time,
the Stoner single-particle's excitations are absent in the
Heisenberg model's spectrum. The Hubbard model~\cite{b41,b151,b155} 
with narrow energy bands contains both types
of excitations: the collective spin waves (the low energy
spectrum) and Stoner single-particle's excitations
(the high-energy spectrum). This is a big advantage
and flexibility of the Hubbard model in comparison
to the Heisenberg model. The latter, nevertheless, is
a very good approximation to the realistic behavior in
the domain where the hydrodynamic description is
applicable, that is, for long wavelengths and low energies.
The quantum protectorate concept was applied to
the quantum theory of magnetism by the author of this
review in the paper~\cite{b155}, where a criterion of models
of the quantum theory of magnetism applicability to
description of concrete substances was formulated. The
criterion is based on the analysis of the model's low energy
and high-energy spectra.
%
\section{The Lawrence-Doniach Model}
%
The Ginzburg-Landau model~\cite{du05,kett} is a special
form of the mean-field theory. This model operates with
a pseudo-wave function $\Psi(\vec{r})$, which plays the role of a
parameter of complex order, while the square of this
function modulus $|\Psi(\vec{r})|^{2}$ should describe the local density
of superconducting electrons. It is well known, that
the Ginzburg-Landau theory is applicable if the temperature
of the system is sufficiently close to its critical
value $T_{c}$, and if the spatial variations of the functions $\Psi$
and of the vector potential $\vec{A}$  are not too large. The
main assumption of the Ginzburg-Landau approach is
the possibility to expand the free-energy density $f$ in a
series under the condition, that the values of $\Psi$ are
small, and its spatial variations are sufficiently slow.
Then, we have
\begin{eqnarray}
 f = f_{n0}  +  \alpha |\Psi|^{2}  +   \frac{\beta}{2} |\Psi|^{4} + \frac{1}{2m^{*}}  
\Bigg|\bigg(-i\hbar\nabla + \frac{2e\vec{A}}{c}\bigg)\Psi\Bigg|^2 + \frac{\hbar^{2}}{8 \pi}.
\label{eq1} \end{eqnarray} 
The Ginzburg-Landau equations follow from an applications
of the variational method to the proposed
expansion of the free energy density in powers of $|\Psi|^{2}$ and $|\nabla  \Psi|^{2},$
which leads to a pair of coupled differential
equations for $\Psi(\vec{r})$ and the vector potential $\vec{A}$.\\
The Lawrence-Doniach model was formulated in
the paper~\cite{don} for analysis of the role played by layered
structures in superconducting materials~\cite{kkc00,kuku02,kuku03}.
The model considers a stack of parallel two dimensional
superconducting layers separated by an
insulated material (or vacuum), with a nonlinear
interaction between the layers. It is also assumed that
an external magnetic field is applied to the system. In
some sense, the Lawrence-Doniach model can be
considered as an anisotropic version of the Ginzburg-
Landau model~\cite{du05,kett}. More specifically,
an anisotropic Ginzburg-Landau model can be considered
as a continuous limit approximation to the
Lawrence-Doniach model. However, when the
coherence length in the direction perpendicular to
the layers is less than the distance between the layers, these models are difficult to compare. In the
framework of the approach used by Lawrence and
Doniach the superconducting properties of the layered
structure were considered under the assumption
that in the superconducting state the free energy per
cell relative to its value in the zero external field can
be written in the following form
\begin{eqnarray}
 f(\vec{r}) = \sum\limits_i^n \Bigg[\alpha_i(T)|\Psi_i(\vec{r})|^2
     +\beta|\Psi_i(\vec{r})|^4 + \frac{1}{2m_{ab}}\Bigg|\bigg(-i\hbar\nabla+
     \frac{2e\vec{A}}{c}\bigg)\Psi_i(\vec{r})\Bigg|^2\Bigg] +
\nonumber \\
        + \sum\limits_{<ij>} \eta_{ij}\Big|\Psi_i(\vec{r})-
           \Psi_j(\vec{r})\Big|^2.
\label{eq2} \end{eqnarray} 
Here, $\Psi_i(\vec{r})$ is the order parameter of the Ginzburg-Landau
order of the layer number $i,$ ($\Psi_i(x,y)$ is a function of
two variables), the operator $\nabla$ acts in the $x$-$y$ plane; $\vec{A}$ is
the corresponding vector's potential, $\alpha$ and $\beta$  are the
usual Ginzburg-Landau parameters, $\eta_{ij}$ describes a
positive Josephson interaction between the layers; and
$<ij>$ denotes summation over neighboring layers. It is
assumed that the layers correspond to planes $ab$, and
the $c$ axis is perpendicular to these planes. Accordingly,
the $z$ axis is aligned with $c$, and the coordinates $x$-$y$
belong to the plane $ab$. The quantities $\eta_{ij}$ are usually
written as follows
\begin{equation}\label{eq2a}
 \eta_{ij} = \frac{\hbar^{2}}{2m_{c} s^{2}}.
\end{equation} 
Here, $s$ is the distance between the layers. As one can
see, for a rigorous treatment of the problem one has to
take into account the anisotropy of the effective mass
at the planes $ab$  and between them, $m_{ab}$ and $m_{c},$  respectively.
Frequently, the distinction between these two
types of anisotropy is ignored, and a quasi-isotropic
case is considered. If we write down $\Psi_i$ in the form
$\Psi_i = |\Psi_i| \exp(i \varphi_i)$ and assume that all $|\Psi_i|$ are equal, then
$\eta_{ij}$ is given by
\begin{equation}\label{eq2b}
 \eta_{ij} = \frac{\hbar^{2}}{2m_{c} s^{2}} |\Psi_i|^{2}[1 - \cos (\varphi_i - \varphi_{i-1})].
\end{equation} 
The coefficient $\alpha_i(T)$ for the layer number $i$ is given by
\begin{equation}
\alpha_i(T)=\alpha_i^{\prime}\frac{(T-T^0_i)}{T^0_i} \label{eq3}, 
\end{equation}
where $T^0_i$  denotes the critical temperature for the layer
number $i$. Next, one can consider the situation where
$\Psi_i(\vec{r}) = \Psi_i(r)$ and $\vec{A}=0.$  In the vicinity of $T_c$ the contribution
from $\beta|\Psi_i|^4$ is small. Taking into account all
these simplifications one can write down the free
energy's density in the following form
\begin{equation} f=\sum\limits_i^n \alpha_i(T)|\Psi_i|^2+
   \sum\limits_{<ij>} \eta_{ij}|\Psi_i-\Psi_j|^2.
\label{eq4} \end{equation} 
This is the quasi-isotropic approximation with single
mass parameter $\alpha$. The Ginzburg-Landau equations
follow from the free-energy extremum conditions with
respect to variations of $\Psi_i$
\begin{equation}
  \frac{\delta{}f}{\delta\Psi_i^\ast}=(\alpha_i+\eta_{i-1 \; i}+
        \eta_{i \; i-1})\Psi_i-(\eta_{i-1 \; i}\Psi_{i-1}+
        \eta_{i \; i+1}\Psi_{i+1})=0.
\label{eq5} \end{equation} 
The corresponding secular equation is given by
\begin{equation}
 \Big|(\alpha_i(T)+\eta_{i-1 \; i}+\eta_{i \; i+1})\delta_{ij}-
    \eta_{ij}\delta_{i \; j\pm1}\Big|=0.
\label{eq6} \end{equation} 
It is assumed in the framework of the Lawrence-Doniach model~\cite{don} that the transition temperature
corresponds to the largest root of the secular equation.
In other words, one has to investigate solutions of the
equation
\begin{equation}
  \left|(T-T^0_i+\frac{\eta_{i-1 \; i}}{\alpha_i^\prime}T^0_i+
              \frac{\eta_{i \; i+1}}{\alpha_i^\prime}T^0_i)\delta_{ij}-
             \frac{\eta_{ij}}{\alpha_i^\prime}T^0_i\delta_{i \; j\pm1}\right|=0,
\label{eq7} \end{equation}
or, in other form 
\begin{equation}
 \textrm{det} (TI-M)=0 , \quad \textrm{where} \quad
 M_{ij}=\left(T^0_i-\frac{\eta_{i-1 \; i}}{\alpha_i^\prime}T^0_i-
              \frac{\eta_{i \; i+1}}{\alpha_i^\prime}T^0_i\right)\delta_{ij}+
             \frac{\eta_{ij}}{\alpha_i^\prime}T^0_i\delta_{i \; j\pm1}.
\label{eq8} \end{equation}
Thus, the problem is reduced to finding the maximal
eigenvalue of the matrix $M$. If we take into account the
external field, then the complete form of the Lawrence-Doniach equation~\cite{don} is given by
\begin{eqnarray}
\alpha  \Psi_i 
+ \beta |\Psi_i|^2 \Psi_i  -  \frac{\hbar^{2}}{2m_{ab}} \bigg(\nabla + i \frac{2e}{\hbar c} \vec{A} \bigg)^{2}\Psi_i \nonumber \\
-  \,  \frac{\hbar^{2}}{2m_{c} s^{2}} \bigg(  \Psi_{i+1} e^{2ie A_{z}s/ \hbar c} - 2\Psi_{i} -  
\Psi_{i-1} e^{2ie A_{z}s/ \hbar c} \bigg) = 0.
\label{eq9} \end{eqnarray} 
A large number of papers are devoted to investigations
of the Lawrence-Doniach model and to development
of various methods for its solution~\cite{kkc00,kuku02,kuku03,chap95,chen97,bau05}. In
many respects this model corresponds to layered structures
of high-temperature superconductors~\cite{lev95}, and
in particular to mercurocuprates~\cite{kkc00,kuku02,kuku03}. A relativistic
version of the Lawrence-Doniach model was studied
in the paper~\cite{ewkk01}, where violation of the local $U(1)$
gauge's symmetry was considered by analogy with
Higgs mechanism~\cite{hig66}. A spontaneous breaking of the
global U(1) invariance is taking place through the
superconducting condensate. The paper~\cite{ewkk01} also
studied in detail the consequences of spontaneous symmetry
breaking in connection with the Anderson-Higgs
phenomenon~\cite{hig66}. As was mentioned already, the concept
of spontaneous symmetry breaking corresponds to
situations with symmetric action, but asymmetric realization
(the vacuum condensate) in the low-energy
regime. As a result the realization has a lower symmetry
than the causing action~\cite{pander63,grib}. \\  In essence, the
Higgs mechanism~\cite{hig66} follows from the Anderson
idea~\cite{pander63} on the connection between the gauge's
invariance breaking and appearance of the zero-mass
collective mode in superconductors. Difference-differential
equations for the order parameter, as well as for
the vector potential at the plane and between the planes
were also derived in the paper~\cite{ewkk01}. These equations
correspond to the Klein-Gordon, Proca and sine-Gordon
equations. The paper also contains a comparison of
the superconducting phase shift $(\varphi_i - \varphi_{i-1})$  between the
layers in the London limit with the standard sine-Gordon
equation. A possible application of this approach to
description of the high-temperature superconductivity
in layered cuprates with a single plane in the elementary
cell and with a weak Josephson interaction
between the layers was also considered. \\ Thus, a systematic
scheme for a phenomenological description of the
macroscopic behavior of layered superconductors can
be constructed by applying the covariance and gauge invariance
principles to a four-dimensional generalization
of the Lawrence-Doniach model. The Higgs
mechanism~\cite{hig66} plays the role of a guiding idea,
which allows one to place this approach on a deep and
nontrivial foundation. The surprising formal simplicity
of the Lawrence-Doniach model once again stresses
the R. Peierls idea~\cite{pei80} on the efficiency of physical
model creating.
%
\section{Nonequilibrium Statistical Operators and Quasiaverages in the Theory of Irreversible Processes}
%
%
It has been mentioned above that Bogoliubov's
quasiaverages concept~\cite{nnb73,nnb75} plays an important
role in equilibrium statistical mechanics. According to
that concept, infinitely small perturbations can trigger
macroscopic responses in the system if they break some
symmetry and remove the related degeneracy (or quasidegeneracy)
of the equilibrium state. As a result, they
can produce macroscopic effects even when the perturbation
magnitude is tend to zero, provided that happens
after passing to the thermodynamic limit. D.N. Zubarev
showed~\cite{dnz70,dnz701} that the concepts of symmetry
breaking perturbations and quasiaverages play an
important role in the theory of irreversible processes as
well~\cite{zub}. The method of the construction of a nonequilibrium
statistical operator~\cite{zub} becomes especially
deep and transparent when it is applied in the framework
of the quasiaverage concept. The main idea of
the papers~\cite{dnz70,dnz701} was to consider infinitesimally
small sources breaking the time-reversal symmetry of
the Liouville equation
\begin{equation}\label{e186}
 \frac{\partial \rho (t,0)}{\partial t} + \frac{1}{i \hbar}[\rho(t,0), H ] = 0
\end{equation}
which become vanishingly small after a thermodynamic
limiting transition.  \\The main idea of the method
of a nonequilibrium statistical operator (NESO)~\cite{zub}
can be summarized as follows. In the scale of sufficiently
large times the nonequilibrium state of the system
can be described by some set of parameters $F_{m}(t)$,
and one can find such a particular solution of the Liouville
equation (\ref{e186}) which depends on time only
through $F_{m}(t)$. The first argument of the operator $\rho (t,0)$
refers to an implicit time dependence. It is assumed that
the nonequilibrium statistical ensemble can be characterized
by a small set of \textbf{relevant operators} $P_{m}(t)$ (quasi-integrals
of motion). The corresponding NESO is a
functional of $P_{m}(t)$:
\begin{equation}\label{e187}
  \rho(t ) = \rho \{\ldots P_{m}(t)  \ldots \}.
\end{equation}
One can show, see~\cite{zub}, that if the statistical operator
$\rho (t,0)$ satisfies the Liouville equation, then it is
given by
\begin{equation}\label{e188}
 \rho =   \exp \bigl (  \Lambda - \int^{0}_{-\infty} dt_{1}  \sum_{m}G_{m}(t_{1}) P_{m}(t_{1})  \bigr); \quad   \Lambda = 1 - \lambda,
\end{equation}
where
\begin{equation}\label{e189}
 G_{m}(t_{1}) = \varepsilon  e^{\varepsilon t_{1}}F_{m}(t + t_{1}),
\end{equation}
\begin{equation}\label{e190}
\Lambda = \varepsilon \int^{0}_{-\infty}  dt_{1}e^{\varepsilon t_{1}}\lambda(t + t_{1}) = \lambda(t) - 
\int^{0}_{-\infty}  dt_{1}e^{\varepsilon t_{1}}\dot{\lambda}(t + t_{1}).
\end{equation}
Alternatively, it can be written as follows
\begin{eqnarray}\label{e191}
 \rho =  \exp (\overline{\ln \rho_{q}}) = \exp \bigl (  \varepsilon \int^{0}_{-\infty}  dt_{1}e^{\varepsilon t_{1}} 
e^(\frac{iHt_{1}}{\hbar}) \ln \rho_{q}(t+t_{1}) e^(\frac{- iHt_{1}}{\hbar}) \bigr ) = \\ \nonumber
\exp (- \overline{S(t,0)}) = \exp (- \varepsilon \int^{0}_{-\infty}  dt_{1}e^{\varepsilon t_{1}}S(t + t_{1},t_{1})) = \\ \nonumber
\exp \bigl ( - S(t,0) +  \int^{0}_{-\infty}  dt_{1}e^{\varepsilon t_{1}} \dot{S}(t + t_{1},t_{1}) \bigr ),
\end{eqnarray}
where
\begin{align}\label{e192}
\rho_{q}   =   \exp \left( \Omega - \sum_{m}F_{m}(t)P_{m}\right) \equiv \exp (- S(t,0)),\quad
\Omega = \ln \textrm{Tr}  \exp \left( - \sum_{m}F_{m}(t)P_{m}\right),
\end{align}
\begin{eqnarray}\label{e193}
 \dot{S}(t,0) = \frac{\partial S(t,0)  }{\partial t} + \frac{1}{i \hbar}[S(t,0) , H ]; \\ \nonumber
 \dot{S}(t,t_{1}) = \exp \left(\frac{iHt_{1}}{\hbar} \right) \dot{S}(t,0) \exp \left(\frac{- iHt_{1}}{\hbar} \right).
\end{eqnarray}
Here, $\rho_{q}$ is the quasi-equilibrium statistical operator,
which corresponds to the extremum value of the information's
entropy
\begin{equation}\label{e194}
  S = - \textrm{Tr} ( \rho \ln \rho),
\end{equation}
under the additional conditions that $\textrm{Tr} ( \rho P_{m} ) = \langle P_{m}\rangle_{q}; 
\quad  \textrm{Tr}   \rho  = 1.$  In this case
\begin{equation}\label{e195}
 \frac{\delta \Phi}{\delta F_{m}}  =  - \langle P_{m}\rangle_{q};   
 \quad    \langle  \ldots  \rangle_{q} =  \textrm{Tr}( \rho_{q}  \ldots),
\end{equation} 
\begin{equation}\label{e196}
  \Phi ( \rho ) = - \textrm{Tr} ( \rho \ln \rho) - \sum_{m}F_{m} \textrm{Tr}(\rho P_{m}) +    \lambda \textrm{Tr}\rho,
\end{equation}
\begin{equation}\label{e197}
 \langle P_{m} \rangle^{t} =  \langle P_{m} \rangle^{t}_{q}. 
\end{equation}
The quantum Liouville equation (\ref{e186}) (as well as
the classical one) is invariant with respect to the time
reversal. One can show~\cite{dnz70,dnz701} that $\rho (t,0)$ satisfies
the Liouville equation with an additional infinitesimally
small (proportional to $\varepsilon$) source-term in the right
hand side, and we send $\varepsilon$ to zero after the thermodynamic
limit. Indeed, let us consider the equation
\begin{equation}\label{e198}
 \frac{\partial \rho_{\varepsilon}  }{\partial t} + \frac{1}{i \hbar}[\rho_{\varepsilon}, H ] = 
 - \varepsilon ( \rho_{\varepsilon}  -  \rho_{q})
\end{equation}
or, equivalently,
\begin{equation}\label{e199}
 \frac{\partial \ln \rho_{\varepsilon}  }{\partial t} + \frac{1}{i \hbar}[ \ln \rho_{\varepsilon}, H ] = 
 - \varepsilon ( \ln \rho_{\varepsilon}  - \ln \rho_{q}),
\end{equation}
where $\varepsilon \rightarrow 0$ after passage to the thermodynamic
limit. Equation (\ref{e198}) is an analogue of the corresponding
equation in the quantum scattering theory~\cite{gell,watson}.
The introduction of infinitely small sources in the
Liouville equation corresponds to imposing the following
boundary conditions
\begin{equation}\label{e200}
  \exp \left( \frac{iHt_{1}}{\hbar} \right) \left( \rho(t + t_{1})  -  
  \rho_{q}(t + t_{1})  \right)  \exp \left( \frac{-iHt_{1}}{\hbar} \right) \rightarrow 0.
\end{equation}
Here,$t_{1} \rightarrow -\infty$ after the thermodynamic limit. It was
shown in the papers~\cite{zub,dnz70,dnz701} that the operator$\rho_{\varepsilon}$
is given by
\begin{equation}\label{e201}
\rho_{\varepsilon}(t,t) = \varepsilon \int^{t}_{-\infty} dt_{1} 
e^{\varepsilon (t_{1} - t)} \rho_{q}( t_{1},t_{1})  =
\varepsilon \int^{0}_{-\infty} dt_{1} e^{\varepsilon t_{1}}\rho_{q}(t + t_{1}, t + t_{1}). 
\end{equation}
Here, the first argument in $\rho(t,t)$ refers to the implicit
time dependence via the parameters $F_{m}(t)$, while the
second argument refers to the time dependence via the
Heisenberg representation. The desired statistical operator
is given by
\begin{equation}\label{e202}
\rho_{\varepsilon}  = \rho_{\varepsilon}(t,0)   =  \overline {\rho_{q}(t,0)} = 
\varepsilon \int^{0}_{-\infty} dt_{1} 
e^{\varepsilon t_{1}} \rho_{q}(t + t_{1}, t_{1}).   
\end{equation}
Hence, the nonequilibrium statistical operator is
given by
\begin{align}\label{e203}
 \rho = Q^{-1} \exp \left(  - \sum_{m} B_{m} \right)   =
 Q^{-1} \exp \left(  - \sum_{m} \varepsilon \int^{0}_{-\infty}  dt_{1}e^{\varepsilon t_{1}} 
 \left( F_{m}(t+ t_{1})P_{m}(t_{1}) \right) \right)  =    
\\ Q^{-1} \exp \left(  - \sum_{m} F_{m}(t) P_{m} +
\sum_{m} \int^{0}_{-\infty}  dt_{1}e^{\varepsilon t_{1}} [ \dot{F}_{m}(t+ t_{1})P_{m}(t_{1}) + 
  F_{m}(t+ t_{1}) \dot{P}_{m}(t_{1})] \right).  \nonumber
\end{align}
One can rewrite Eq. (\ref{e199}) in the following form
\begin{equation}\label{e204}
  \frac{d}{dt}\left( e^{\varepsilon t} \ln  \rho (t,t) \right) = \varepsilon e^{\varepsilon t}\ln  \rho_{q} (t,t),
\end{equation}
where
\begin{equation}\label{e205}
  \ln  \rho (t,t) = U^{\dagger}(t,0)  \ln  \rho (t,0) U(t,0); \quad U(t,0) = \exp \left( \frac{i H t}{\hbar} \right).
\end{equation}
Integrating Eq. (\ref{e204}) over the interval $(-\infty, 0)$ we
obtain
\begin{equation}\label{206}
  \ln  \rho (t,t) =  
  \varepsilon \int^{0}_{-\infty} dt_{1} e^{\varepsilon t_{1}} \ln \rho_{q}(t + t_{1}, t + t_{1}).  
\end{equation}
It is assumed that $ \lim_{\varepsilon \rightarrow 0^{+}} \ln  \rho (t,t) = 0$.  Therefore,
\begin{equation}\label{e207}
    \rho (t,0) =  \exp \left(
 - \varepsilon \int^{0}_{-\infty} dt_{1} e^{\varepsilon t_{1}} \ln \rho_{q}(t + t_{1},  t_{1})  \right) = 
 \exp \overline{\left( \ln \rho_{q}(t,0)\right)} \equiv \exp \overline{\left( -S(t,0)\right)}.
\end{equation}
The average value of any dynamic variable $A$ is now
given by
\begin{equation}\label{e208}
  \langle A \rangle = \lim_{\varepsilon \rightarrow 0^{+}} \textrm{Tr} ( \rho (t,0) A ).
\end{equation}
We see that the above average is in fact nothing else
but a quasiaverage. The normalization of the quasi-equilibrium
distribution $\rho_{q}$ is preserved if
\begin{equation}\label{e209}
  \textrm{Tr} ( \rho (t,0) P_{m} )= \langle P_{m}\rangle = \langle P_{m}\rangle_{q} ; \quad  \textrm{Tr}   \rho  = 1.  
\end{equation}
Thus, one can assert that the origin of the irreversibility
effect is closely related to the violation of the
time-reversal symmetry~\cite{petr}, as well as to the notion
of quasiaverages from statistical mechanics~\cite{zub,dnz70,dnz701}.
%
%
 \subsection{Generalized Kinetic Equations}   
%
%
The NESO method~\cite{zub} found wide applications in
various problems of statistical mechanics. An important
contribution to the development of the kinetic equations' 
theory in the framework of NESO method was
made by L.A. Pokrovskii~\cite{pok67,pok,pok1}. 
Generalized kinetic and transport equations describing the time evolution
of the variables $\langle P_{m}\rangle$ and $F_{m}(t)$   are obtained by
averaging the equation of motion for $ P_{m}$ over the
derived NESO
\begin{equation}\label{e210}
 \langle P_{m}\rangle = - \frac{\delta \Omega}{\delta F_{m}(t)};
\quad F_{m}(t)  = \frac{\delta S}{\delta \langle P_{m}\rangle }.   
\end{equation}
The generalized transport equations are given by
\begin{equation}\label{e211}
 \langle \dot {P}_{m}\rangle = - \sum_{n} \frac{\delta^{2} \Omega}{\delta F_{m}(t)\delta F_{n}(t)}\dot{F}_{n}(t);
\quad \dot{F}_{m}(t)  =  \sum_{n} \frac{\delta^{2} S}{\delta \langle P_{m}\rangle \delta \langle P_{n}\rangle} \langle\dot {P}_{n}\rangle.  
\end{equation}
The corresponding entropy production can be written
down in the following form
\begin{equation}\label{e212}
\dot {S}(t) = \langle \dot {S}(t,0)\rangle = - \sum_{m}\langle \dot {P}_{m}\rangle F_{m}(t) = -
\sum_{n,m}\frac{\delta^{2} \Omega }{\delta F_{m}(t)\delta F_{n}(t)}\dot{F}_{n}(t)F_{m}(t).
\end{equation}
The two equations in (\ref{e211}) are mutually conjugate,
and together with Eq. (\ref{e212}) they form a complete system
of equations for calculation of the quantities $\langle P_{m}\rangle$ and $F_{m}$.\\
Now, following the paper~\cite{pok1}, we are going to
write down kinetic equations for a system with weak
interaction. The corresponding Hamiltonian is given by
\begin{equation}\label{e213}
 H = H_{0} + V.
\end{equation}
Here, $H_{0}$ is the Hamiltonian of noninteracting particles
(or quasiparticles) and $V$ is the interaction operator. As
the set of relevant operators we choose the operators
$ P_{m} = P_{k}$ of the form $a^{\dagger}_{k}a_{k}$ or $a^{\dagger}_{k}a_{k+q}.$  Here, $a^{\dagger}_{k}$ and $a_{k}$ are
the usual creation and the annihilation operators (either
Fermi or Bose). We begin with the following equations
of motion:
\begin{equation}\label{e214}
 \dot{P}_{k} = \frac{1}{i \hbar}[P_{k} , H ].
\end{equation}
It is usually assumed that
\begin{equation}\label{e215}
 [P_{k} , H_{0} ] = \sum_{l} c_{kl}P_{l},
\end{equation}
where $c_{kl}$ are some coefficients ({\em c}-numbers).\\
According to Eq. (\ref{e203}) we have
\begin{align}\label{e216}
 \rho =  Q^{-1} \exp \left(  - \sum_{k} F_{k}(t) P_{k} +
\sum_{k} \int^{0}_{-\infty}  dt_{1}e^{\varepsilon t_{1}} [ \dot{F}_{k}(t+ t_{1})P_{k}(t_{1}) + 
  F_{k}(t+ t_{1}) \dot{P}_{k}(t_{1})] \right).   
\end{align}
Keeping in mind that $\langle P_{k}\rangle = \langle P_{k}\rangle_{q},$ we can write
down the generalized kinetic equations~\cite{pok1} for $\langle P_{k}\rangle$ as
follows
\begin{equation}\label{e217}
 \frac{d \langle P_{k}\rangle  }{d t} = \frac{1}{i \hbar}\langle [  P_{k} , H ]\rangle =
 \frac{1}{i \hbar} \sum_{l} c_{kl}\langle P_{l}\rangle +  \frac{1}{i \hbar}\langle [  P_{k} , V ]\rangle.
\end{equation}
The right hand side of Eq. (\ref{e217}) contains the generalized
\textbf{collision integral}, which, using an expansion in
powers of $V$, can be written as follows
\begin{equation}\label{e218}
 \frac{d \langle P_{k}\rangle  }{d t} = L^{0}_{k} + L^{1}_{k} + L^{21}_{k} + L^{22}_{k},
 \end{equation}
where
\begin{equation}\label{e219}
 L^{0}_{k} =  \frac{1}{i \hbar} \sum_{l} c_{kl}\langle P_{l}\rangle _{q},
\end{equation}
\begin{equation}\label{e220}
 L^{1}_{k} =  \frac{1}{i \hbar} \langle [  P_{k} , V ]\rangle_{q},
\end{equation}
\begin{equation}\label{e221}
 L^{21}_{k} =  \frac{1}{ \hbar^{2}}  \int^{0}_{-\infty} dt_{1} e^{\varepsilon t_{1}} \langle [ V(t_{1}), [ P_{k} , V ]]\rangle_{q},
\end{equation}
\begin{equation}\label{e222}
 L^{22}_{k} =  \frac{1}{ \hbar^{2}}  \int^{0}_{-\infty} dt_{1} 
 e^{\varepsilon t_{1}} \langle [ V(t_{1}),  i \hbar \sum_{l} P_{l}   
 \frac{\partial   L^{1}_{k}(\ldots \langle P_{l}\rangle \ldots)  }{\partial  \langle P_{l}\rangle } ]\rangle_{q}.
\end{equation}
Analogously one can find the higher-order terms $V^{3}$, $V^{4}$ and so on.
%
\subsection{Generalized Kinetic Equations for a System in a Thermal Bath}
%
%
The papers~\cite{wk70,wzk70,kuz05} (see also~\cite{kuz07}) generalize the
equations in (\ref{e217}) for the case of a system interacting
with a thermal bath. The concept of a thermal bath or
heat reservoir is fairly complicated and has certain specific
features~\cite{koz08}. According to the standard definition,
a thermal bath is a system with, effectively, an infinite
number of degrees of freedom. A thermal bath is a
heat reservoir maintaining the investigated system
under a particular temperature. Following Bogoliubov~\cite{nnb78},
we will assume that a thermal bath is a source of
stochasticity for a small subsystem (which in an
extreme situation can be just a single particle). Such a
small subsystem can be, for example, an atomic or a
molecular system interacting with an electromagnetic
field, or a system of nuclear or electron spins interacting
with the crystal lattice, etc. We will describe the entire
system by the Hamiltonian
\begin{equation}\label{e223}
 H = H_{1} + H_{2} + V,
\end{equation}
where
\begin{equation}\label{e224}
 H_{1} = \sum_{\alpha} E_{\alpha}a^{\dagger}_{\alpha}a_{\alpha}; 
 \quad V = \sum_{\alpha,\beta}\Phi_{\alpha \beta}a^{\dagger}_{\alpha}a_{ \beta}, \quad \Phi_{\alpha \beta} = 
 \Phi^{\dagger}_{ \beta \alpha}.
\end{equation}
Here, $H_{1}$  is the Hamiltonian of the small subsystem;
$a^{\dagger}_{\alpha}$, $a_{\alpha}$ are the creation and annihilation operators of
quasiparticles with the energies $ E_{\alpha}$ in the small subsystem;
$V$ is the operator describing the interaction
between the small subsystem and the thermal bath; and
$H_{2}$ is the thermal bath's Hamiltonian, which we do not
write down explicitly. The quantities $\Phi_{\alpha \beta}$ are operators
acting on the thermal bath's degrees of freedom. We
assume that the state of the system can be characterized
by a set of operators %
$\langle P_{\alpha \beta}\rangle = \langle a^{\dagger}_{\alpha}a_{ \beta}\rangle$, and the state of the
thermal bath by the operator $\langle H_{2}\rangle$. Here, $ \langle \ldots \rangle$ denotes
the averaging with respect to the NESO, which is
defined as follows:
\begin{eqnarray}\label{e225}
\rho_{q} (t) =  \exp (- S(t,0)), \quad  S(t,0) = \Omega(t) + 
\sum_{\alpha \beta }P_{\alpha \beta }F_{\alpha\beta }(t) + \beta H_{2},\\ 
\Omega = \ln \textrm{Tr} \exp (- \sum_{\alpha \beta }P_{\alpha \beta }F_{\alpha\beta }(t) - \beta H_{2} ). \nonumber
\end{eqnarray}
Here, $F_{\alpha\beta }(t)$ are the thermodynamic parameters conjugate
to $P_{\alpha \beta },$ $\beta$ is the inverse temperature of the thermal
bath. All operators are considered in the Heisenberg
representation. We write down the nonequilibrium statistical
operator as follows
\begin{eqnarray}\label{e226}
\rho (t) =  \exp (- \overline{S(t,0)}),\\
\overline{S(t,0)} = \varepsilon \int^{0}_{-\infty} dt_{1} e^{\varepsilon t_{1}} 
\left( \Omega(t + t_{1}) +  \sum_{\alpha \beta }P_{\alpha \beta }F_{\alpha\beta }(t) + \beta H_{2}  \right).
\nonumber
\end{eqnarray}
The parameters $F_{\alpha\beta }(t)$ are determined by the condition
$\langle P_{\alpha \beta }\rangle = \langle P_{\alpha \beta }\rangle_{q}$. To derive the kinetic equations we will use
an expansion over the small parameter in the interaction $V$.
It is also assumed that the equation  
$ \langle \Phi_{\alpha \beta}\rangle_{q} = 0$. holds. It is
convenient to rewrite $\rho_{q}$ as follows
\begin{equation}\label{e227}
 \rho_{q}  = \rho _{1}\rho_{2} = Q^{-1}_{q} \exp (- L_{0}(t)),
\end{equation}
where
\begin{eqnarray}\label{e228}
\rho _{1} = Q^{-1}_{1}\exp \left(- \sum_{\alpha \beta } P_{\alpha \beta } F_{\alpha \beta }(t)\right); \quad
Q_{1} = \textrm{Tr} \exp \left( - \sum_{\alpha \beta } P_{\alpha \beta } F_{\alpha \beta }(t) \right),\\
\rho _{2} = Q^{-1}_{2} e^{- \beta H_{2} }; \quad Q_{2} = \textrm{Tr} \exp (- \beta H_{2}),\\
Q_{q} = Q_{1}Q_{2}; \quad L_{0} = \sum_{\alpha \beta } P_{\alpha \beta } F_{\alpha \beta }(t) + \beta H_{2}.
\end{eqnarray}
We begin from the following relationship:
\begin{equation}\label{e229}
 \frac{d \langle P_{\alpha \beta }\rangle }{d t} =  \frac{1}{i \hbar}\langle [  P_{\alpha \beta  } , H ]\rangle =
 \frac{1}{i \hbar}(E_{\beta} - E_{\alpha})\langle P_{\alpha \beta  }\rangle +  \frac{1}{i \hbar}\langle [  P_{\alpha \beta} , V ]\rangle.
\end{equation}
We terminate the expansions at the second order
terms in $V$. The kinetic equations for the quantities $\langle P_{\alpha \beta }\rangle$
of a system in a thermal bath are given by
\begin{equation}\label{e230}
 \frac{d \langle P_{\alpha \beta }\rangle }{d t} =  \frac{1}{i \hbar}(E_{\beta} - E_{\alpha})\langle P_{\alpha \beta  }\rangle -
 \frac{1}{\hbar^{2}} \int^{0}_{-\infty} dt_{1} e^{\varepsilon t_{1}} 
 \langle \left[[P_{\alpha \beta}, V], V(t_{1}) \right]\rangle_{q}.
\end{equation}
These equations generalize the results of the paper~\cite{pok1}
for a system in a thermal bath. One can show that
the choice of the concrete model's form for the Hamiltonian
(\ref{e223}) is not essential. For arbitrary $H _{1}$ and $V$, and
for some set of variables $\langle P_{k}\rangle$ satisfying the condition
$[H _{1}, P_{k}] =  \sum_{l} c_{kl}P_{l}$, one can construct a quasi-equilibrium
statistical operator $\rho_{q}$ in the following form
\begin{eqnarray}\label{e231}
\rho_{q}  
 = Q^{-1}_{q} \exp \left(  - \sum_{k} P_{k}F_{k}(t)  
-  \beta H _{2} \right). 
\end{eqnarray}
Here, $F_{k}(t)$ are the parameters conjugate to $\langle P_{k}\rangle$. The
kinetic equations for $\langle P_{k}\rangle$ are given by
\begin{equation}\label{e232}
 \frac{d \langle P_{k}\rangle }{d t} =  \frac{i}{ \hbar} \sum_{l} c_{kl}\langle P_{l}\rangle -
 \frac{1}{\hbar^{2}} \int^{0}_{-\infty} dt_{1} e^{\varepsilon t_{1}} 
 \langle \left[[P_{k}, V], V(t_{1}) \right]\rangle_{q}.
\end{equation}
%
%
\subsection{A Schr\"{o}dinger-Type Equation for a Dynamic System in a Thermal Bath}
%
%
Following the papers~\cite{kuz07,wk70,wzk70,kuz05}, we consider now
the behavior of a small dynamical subsystem with a
Hamiltonian $H_{1}$, which interacts with a thermal bath
described by the Hamiltonian $H_{2}$. As the operators characterizing
the state of the small subsystem we choose
the operators $a^{\dagger}_{\alpha}, a_{\alpha}$, 
and $n_{\alpha} = a^{\dagger}_{\alpha}a_{\alpha}$.  In this case the
quasi-equilibrium statistical operator $\rho_{q}$  is given by
\begin{eqnarray}\label{e233}
\rho_{q}   =   \exp \left( \Omega - \sum_{\alpha} ( f_{\alpha}(t)a_{\alpha} + f^{\dagger}_{\alpha}(t)a^{\dagger}_{\alpha} + 
F_{\alpha}(t)n_{\alpha} ) - \beta H_{2} \right)  \equiv \exp ( - S(t,0)),\\
\Omega = \ln \textrm{Tr}  \exp \left( - \sum_{\alpha}( f_{\alpha}(t)a_{\alpha} + f^{\dagger}_{\alpha}(t)a^{\dagger}_{\alpha} + 
F_{\alpha}(t)n_{\alpha} ) - \beta H_{2}  \right).\nonumber
\end{eqnarray}
Here, $f_{\alpha}, f^{\dagger}_{\alpha}$ and $F_{\alpha}$,  play the role of Lagrange multipliers.
They are the parameters conjugate to  $\langle a_{\alpha}\rangle_{q}, \langle a^{\dagger}_{\alpha}\rangle_{q}$  
and $\langle n_{\alpha}\rangle_{q}$:
\begin{equation}\label{e234}
\langle a_{\alpha}\rangle_{q} =  - \frac{\delta \Omega}{\delta f_{\alpha}(t)},  \quad \langle n_{\alpha}\rangle_{q} =
- \frac{\delta \Omega}{\delta F_{\alpha}(t)},
\quad \frac{\delta S}{\delta \langle a_{\alpha}\rangle_{q} }  =  f_{\alpha}(t),  \quad
\frac{\delta S}{\delta \langle n_{\alpha}\rangle_{q} } = F_{\alpha}(t). 
\end{equation}  
The quantities $ a_{\alpha}, \, a^{\dagger}_{\alpha}$  in the statistical operator
can be interpreted as sources of \emph{quantum noise} (those terms break the spin conservation law in the case of fermions; see the
papers~\cite{kuz07,kuz05}). Let us write down the quasi-equilibrium
statistical operator as follows
\begin{equation}\label{e235}
\rho_{q} = \rho_{1} \rho_{2},    
\end{equation}
where
\begin{eqnarray}\label{e236}
\rho_{1}   =   \exp \left( \Omega_{1} - \sum_{\alpha} ( f_{\alpha}(t)a_{\alpha} + f^{\dagger}_{\alpha}(t)a^{\dagger}_{\alpha} + 
F_{\alpha}(t)n_{\alpha} ) \right) \\
\Omega_{1} = \ln \textrm{Tr}  \exp \left( - \sum_{\alpha}( f_{\alpha}(t)a_{\alpha} + f^{\dagger}_{\alpha}(t)a^{\dagger}_{\alpha} + 
F_{\alpha}(t)n_{\alpha} ) \right)\nonumber \\
\rho_{2}   =   \exp \left( \Omega_{2}  - \beta H_{2} \right), \quad
\Omega_{2} = \ln \textrm{Tr}  \exp \left( - \beta H_{2}  \right)
\label{eq86} 
\end{eqnarray}
As a result we obtain the expression (\ref{e226}) for the
NESO $\rho$. We assume that the following conditions are
satisfied:
\begin{equation}\label{e237}
\langle a_{\alpha}\rangle_{q} = \langle a_{\alpha}\rangle, \quad \langle a^{\dagger}_{\alpha}\rangle_{q} = \langle a^{\dagger}_{\alpha}\rangle, \quad
\langle n_{\alpha}\rangle_{q} = \langle n_{\alpha}\rangle.
\end{equation}  
We start from the equations of motion
\begin{eqnarray}\label{e238}
i\hbar \frac{d \langle a_{\alpha}\rangle}{dt} = \langle [a_{\alpha}, H_{1} ]\rangle + \langle [a_{\alpha}, V ]\rangle,\\
i\hbar \frac{d \langle n_{\alpha}\rangle}{dt} = \langle [n_{\alpha}, H_{1} ]\rangle + \langle [n_{\alpha}, V ]\rangle.
\label{e239} 
\end{eqnarray}
In the second order in $V$ we obtain
\begin{eqnarray}\label{e240}
i\hbar \frac{d \langle a_{\alpha} \rangle}{dt} = E_{\alpha}\langle a_{\alpha}\rangle + \frac{1}{i\hbar }
\int^{0}_{-\infty} dt_{1} e^{\varepsilon t_{1}} 
 \langle \left[[a_{\alpha}, V], V(t_{1}) \right]\rangle_{q},\\
i\hbar \frac{d \langle n_{\alpha}\rangle}{dt} = 
\frac{1}{i\hbar }
\int^{0}_{-\infty} dt_{1} e^{\varepsilon t_{1}} 
 \langle \left[[n_{\alpha}, V], V(t_{1}) \right]\rangle_{q}.
\label{e241} 
\end{eqnarray}
Here, $V(t_{1})$ denotes the operator $V$ in the interaction
representation. The expansion yields
\begin{align}\label{e242}
i\hbar \frac{d \langle a_{\alpha}\rangle}{dt} = E_{\alpha}\langle a_{\alpha}\rangle +  \nonumber \\  \frac{1}{i\hbar }
\int^{0}_{-\infty} dt_{1} e^{\varepsilon t_{1}} \left( \sum_{\beta \mu \nu} 
\langle \Phi_{\alpha \beta}\phi_{\mu \nu}(t_{1})\rangle_{q}\langle a_{\beta}a^{\dagger}_{\mu} a_{\nu}\rangle_{q}   -     
\langle \phi_{\mu \nu}(t_{1})\Phi_{\alpha \beta}\rangle_{q}\langle a^{\dagger}_{\mu} a_{\nu} a_{\beta}\rangle_{q} \right),  
\end{align}
where $\phi_{\mu \nu}(t_{1}) = \Phi_{\mu \nu}(t_{1}) \exp ( \frac{i}{\hbar }(E_{\mu} - E_{\nu})t_{1} )$,
or, equivalently
\begin{eqnarray}\label{e243}
i\hbar \frac{d \langle a_{\alpha}\rangle}{dt} = E_{\alpha}\langle a_{\alpha}\rangle + \frac{1}{i\hbar }\sum_{\beta \mu }
\int^{0}_{-\infty} dt_{1} e^{\varepsilon t_{1}} \langle \Phi_{\alpha \mu}\phi_{\mu \beta}(t_{1})\rangle_{q}\langle a_{\beta}\rangle + \\
\frac{1}{i\hbar }\sum_{\beta \mu \nu}
\int^{0}_{-\infty} dt_{1} e^{\varepsilon t_{1}} \langle [\Phi_{\alpha \nu}, \phi_{\mu \nu}(t_{1})] \rangle_{q}
\langle a^{\dagger}_{\mu} a_{\nu} a_{\beta}\rangle_{q}. \nonumber
\end{eqnarray}
Therefore, we obtain
\begin{equation}\label{e244}
  i\hbar \frac{d \langle a_{\alpha} \rangle}{dt} = E_{\alpha}\langle a_{\alpha}\rangle + \frac{1}{i\hbar }\sum_{\beta \mu }
\int^{0}_{-\infty} dt_{1} e^{\varepsilon t_{1}} \langle \Phi_{\alpha \mu}\phi_{\mu \beta}(t_{1})\rangle_{q}\langle a_{\beta}\rangle.
\end{equation}
Using the spectral representations for the correlation
functions we can write down
\begin{equation}\label{e245}
  i\hbar \frac{d \langle a_{\alpha}\rangle}{dt} = E_{\alpha}\langle a_{\alpha}\rangle + \sum_{\beta }
K_{\alpha \beta} \langle a_{\beta}\rangle,
\end{equation}
where $K_{\alpha \beta} $ are defined as follows:
\begin{eqnarray}
\label{e246}
 \frac{1}{i\hbar}\sum_{\mu} 
 \int^{ 0}_{ - \infty} dt_{1}  e^{\varepsilon t_{1}}   \langle \Phi_{ \beta \mu} \phi_{\mu \nu}(t_{1})\rangle_{q} =  \\
\frac{1}{2\pi}  \sum_{\mu}  \int^{ + \infty}_{ - \infty} d\omega
 \frac{J_{\mu \nu, \beta \mu} (\omega)}{\hbar\omega - E_{\mu} - E_{\nu} + i\varepsilon} = K_{\beta\nu}. \nonumber
\end{eqnarray}
Thus, we have obtained a Schr\"{o}dinger-type equation
for the \textbf{mean amplitudes} $\langle a_{\alpha}\rangle$. In a certain sense
this equation is an analogue (or a generalization) of the
Schr\"{o}dinger equation for the case of a particle moving
in a medium. Let us consider this analogy in a more
detail. First, we write down the \emph{analogue} of the wave
function in the following form
\begin{equation}\label{e247}
\psi ( \vec{r}) = \sum_{\alpha} \chi_{\alpha}( \vec{r})\langle a_{\alpha}\rangle.
\end{equation}
Here, $\{ \chi_{\alpha}( \vec{r}) \}$ is a complete orthonormal set of single-particle
eigenfunctions of the operator
$\left( - \frac{\hbar^{2}}{2m}\nabla^{2}  + v(\vec{r}) \right)$,
where $v(\vec{r})$ is the potential energy,  
\begin{equation}\label{e248}
\left(- \frac{\hbar^{2}}{2m}\nabla^{2}  + v(\vec{r}) \right)\chi_{\alpha}( \vec{r}) = 
E_{\alpha} \chi_{\alpha}( \vec{r}).
\end{equation}
Thus, the quantity $\psi ( \vec{r})$  plays the role of a wave
function describing a particle moving in a medium.
Equation (\ref{e245}) can be rewritten in the following form
\begin{equation}\label{e249}
 i\hbar \frac{\partial \psi ( \vec{r}) }{\partial t} = 
 \left(- \frac{\hbar^{2}}{2m} \nabla^{2}  + v(\vec{r}) \right) \psi( \vec{r})  +
 \int K (\vec{r},\vec{r'})\psi ( \vec{r'})d \vec{r'}.
\end{equation}
The kernel $K (\vec{r},\vec{r'})$  of the integral equation (\ref{e249}) is
given by
\begin{equation}\label{e250}
  K (\vec{r},\vec{r'}) = \sum_{\alpha \beta} K_{\alpha \beta}\chi_{\alpha}( \vec{r})\chi^{\dag}_{\beta}( \vec{r'}) = 
\frac{1}{i\hbar} \sum_{\alpha, \beta, \mu } \int^{0}_{-\infty} dt_{1} e^{\varepsilon t_{1}}
\langle \Phi_{\alpha \mu}\phi_{\mu \beta}(t_{1})\rangle_{q}\chi_{\alpha}( \vec{r})\chi^{\dag}_{\beta}( \vec{r'}).
\end{equation}
We see that Eq. (\ref{e249}) can indeed be classified as a
Schr\"{o}dinger-type equation for a dynamical system in
a thermal bath. It is interesting to note that very similar
equations of the Schr\"{o}dinger-type with a nonlocal
interaction were used in the collision theory~\cite{mot} for
description of particle scattering on a cluster of many
scattering centers.\\
In order to make clear some special features of
Eq. (\ref{e249}) let us consider the translation operator
$ \exp (i\vec{q} \vec{p} / \hbar)$, где $\vec{q} = \vec{r'} - \vec{r},$  where $ \vec{p} = -i\hbar \nabla_{r},$
Then, Eq. (\ref{e249}) can be rewritten in the following form
\begin{equation}\label{e251}
 i\hbar \frac{\partial \psi ( \vec{r}) }{\partial t} = 
 \left(- \frac{\hbar^{2}}{2m} \nabla^{2}  + v(\vec{r}) \right) \psi( \vec{r})  +
\sum_{p} D(\vec{r},\vec{p})\psi ( \vec{r}), 
\end{equation}
where
\begin{equation}\label{e252}
D(\vec{r},\vec{p}) = \int d^{3}q K (\vec{r},\vec{r} + \vec{q})e^{\frac{i\vec{q}\vec{p}}{\hbar} }.
\end{equation}
It is reasonable to assume that the wave function $\psi ( \vec{r})$
does not change very rapidly over distances comparable
to the characteristic correlation length of the
kernel $K (\vec{r},\vec{r'}).$ Then, using the series expansion for $ \exp (i\vec{q} \vec{p} / \hbar)$
in Eq. (\ref{e251}) we obtain in the zeroth order
\begin{equation}\label{e253}
 i\hbar \frac{\partial \psi ( \vec{r}) }{\partial t} = 
 \left(- \frac{\hbar^{2}}{2m} \nabla^{2}  + v(\vec{r}) + \textrm{Re} \,  U( \vec{r}) \right) \psi( \vec{r})  +
 i  \textrm{Im} \,  U( \vec{r}) \psi( \vec{r}),
\end{equation}
where
\begin{equation}\label{e254}
U( \vec{r}) =  \textrm{Re} \, U( \vec{r}) +  i  \textrm{Im} \,  U( \vec{r}) =
 \int d^{3}q K (\vec{r},\vec{r} + \vec{q}). 
\end{equation}
Equation (\ref{e253}) has the exact functional form of a
Schr\"{o}dinger equation with a complex potential well
known in the collision theory~\cite{mot}. Note that the introduction
of the quantity  $\psi ( \vec{r})$ does not mean that the
state of the small dynamical subsystem becomes pure.
The state remains mixed because it is described by a
statistical operator. The dynamics of the system is
described by a system of coupled evolution equations
for the quantities $f_{\alpha}, f^{\dagger}_{\alpha}$  and $F_{\alpha}$. Note, that there were
many attempts to derive a Schr\"{o}dinger-type equation
for a particle in a medium~\cite{korr,kostin,tzan}. Korringa~\cite{korr}
tried to obtain such an equation in the form of an evolution
equation with a non-hermitian Hamiltonian.
However, his equation (cf. Eq. (29) from~\cite{korr})
\begin{equation}\label{e255}
 i  \frac{\partial W' }{\partial t} =  \left ( H'(t) + h'(t)) + \frac{i}{2\theta} \frac{d h'}{dt} + \ldots \right)  W'(t),
\end{equation}
where $ W'(t)$ is a statistical density matrix describing the
original system, is rather a modified Bloch equation. An
attempt to derive a Schroedinger-type equation for a
Brownian particle interacting with a thermal environment
was made in the paper~\cite{kostin}. The evolution equation
obtained there is given by
\begin{equation}\label{e256}
 i\hbar  \frac{\partial \psi }{\partial t} =  - \frac{\hbar^{2}}{2 m} \nabla^{2} \psi + V \psi + V_{R} \psi + 
 [\frac{\hbar f}{2i m} \ln (\frac{\psi}{\psi^{*}} ) + W(t) ] \psi (\vec{r},t),
\end{equation}
where
\begin{equation}\label{e257}
 W(t) =  - (\frac{\hbar f}{2i m} ) \int  \psi^{*}\ln (\frac{\psi}{\psi^{*}} ) \psi dr. 
\end{equation}
Here, $f$ is the friction coefficient, $V_{R}$ is a random potential,
and $V_{R}(\vec{r},t) = - \vec{r} \vec{F}_{R}(t),$  where $\vec{F}_{R}(t)$  is a random
vector-function of time. Excluding the function $W(t)$
with the help of the transformation
\begin{equation}\label{e258}
 \psi (\vec{r},t) = \exp [i \theta (t)] \phi (\vec{r},t),
\end{equation}
where
$$\theta (t) = - \hbar^{-1} \exp ( - \frac{t f}{ m}) \int_{0^{t}}^{t}\exp (\frac{s f}{ m}) W(s)d s,$$
one obtains the equation for $\phi (\vec{r},t)$  in the following form
\begin{eqnarray}\label{e259}
 i\hbar  \frac{\partial \phi }{\partial t} =  - \frac{\hbar^{2}}{2 m} \nabla^{2} \phi(\vec{r},t) + V(\vec{r}) \phi(\vec{r},t) + 
 V_{R}(\vec{r},t) \phi(\vec{r},t) + \\ \nonumber
 \frac{\hbar f}{2i m} \phi(\vec{r},t)\ln [\frac{\phi(\vec{r},t)}{\phi^{*}(\vec{r},t)} ].  
\end{eqnarray}
It is clear that the dynamic behavior of a particle in
a dissipative environment is most accurately described
by a Schr\"{o}dinger-type equation with damping, see Eq.(\ref{e249}) above. This is actually the reason for applications
of this equation in numerous problems of physics,
physical chemistry, biophysics, and other areas~\cite{lnc71,ivic91,vasco93,luz94,mesq96,mesq98,luz00,mesq04,luz04}. 
A more detailed discussion of various aspects of
dissipative behavior and of stochastic process in complex
systems is given in the reviews~\cite{kuz07,kuz06,rama08,tan06,tan08}.
%
\section{Conclusion}
%
%
In the present paper we have shown the determining
role played by correlation effects in systematic microscopic
descriptions of magnetic, electrical, and superconducting
properties of complex substances. We have
stressed that the approximation of tight-binding electrons
and the method of model Hamiltonians are very
effective tools for description of these substances. In
many cases the methods of quantum statistical mechanics,
many of which were formulated and developed by
N.N. Bogoliubov, allow one to develop efficient approaches
for solution of complicated problems from microscopic
theory of correlation effects, especially in the case of
strong electron correlations. The method of two-time
temperature Green's functions allows one to efficiently
investigate the quasiparticle dynamics generated by
the main model Hamiltonians from the quantum solid state
theory and the quantum theory of magnetism. The
method of quasiaverages allows one to take a deeper
look at the problems of spontaneous symmetry breaking,
as well as at the problems of symmetry and dissymmetry
in the physics of condensed matter. Further
development of the theory describing many-particle
effects and investigations of more realistic models will
allow one to gain more precise ideas on the effective
interactions in the systems, which determine various
phenomena, the main features of electron states, and
therefore, the physical properties of real substances.
The methods developed by N.N. Bogoliubov are and
will remain the important core of a theoretician's
toolbox, and of the ideological basis behind this
development.
%
\section{Acknowledgements}
%
%
The author recollects with gratefulness discussions
of this review topics with N.N. Bogoliubov
(21.08.1909-13.02.1992) and D.N. Zubarev (30.11.1917-16.07.1992). He is also grateful to Professor
N.N. Bogoliubov, Jr., for valuable discussions and
bringing the reference~\cite{bog90} to his attention.
%
%
%

%
%

%

\end{document}